\documentclass{lmcs}
\pdfoutput=1

\usepackage[utf8]{inputenc}
\usepackage{lastpage}
\lmcsdoi{17}{1}{11}
\lmcsheading{}{\pageref{LastPage}}{}{}%
{Apr.~22,~2020}{Feb.~03,~2021}{}

\pdfoutput=1


\ACMCCS{[{\bf Theory of computation}]: Logic ; 
[{\bf Mathematics of
  computing}]: Trees ; Model theory ;}

\usepackage{amsmath}
\usepackage{amsfonts}
\usepackage{amssymb}
\usepackage{graphicx}

\usepackage{xparse}
\ExplSyntaxOn
\NewDocumentEnvironment{ThmEnv} { m m m o}
 {
   \leavevmode
   \smallskip
  \par\noindent
  \textbf{#1~#2}
  \IfNoValueTF{#4}{
  }{
   {~\upshape(#4)}
  }
  \textbf{.}\hspace{0.3em plus 0.3em minus 0.2em}#3
 \ignorespaces
 }
{\upshape\par\smallskip}

\ExplSyntaxOff
\def\NewCustomTheorem#1#2{
  \NewDocumentEnvironment{#1}{m o}{%
    \IfNoValueTF{##2}{\begin{ThmEnv}{#1}{##1}{#2}}{\begin{ThmEnv}{#1}{##1}{#2}[##2]}
    }{\end{ThmEnv}}
}
\NewCustomTheorem{Example}{\upshape}
\NewCustomTheorem{Examples}{\upshape}
\NewCustomTheorem{Remark}{\upshape}
\NewCustomTheorem{Remarks}{\upshape}
\NewCustomTheorem{Definition}{\upshape}
\NewCustomTheorem{Definitions}{\upshape}

\NewCustomTheorem{Proposition}{\itshape}
\NewCustomTheorem{Lemma}{\itshape}
\NewCustomTheorem{Theorem}{\itshape}
\NewCustomTheorem{Corollary}{\itshape}
\NewCustomTheorem{Conjecture}{\itshape}

\begin{document}

\title{Axiomatizations of betweenness\\ in order-theoretic trees}

\author{Bruno COURCELLE} 
\address{LaBRI, CNRS\\and Bordeaux University, France}

\email{courcell@labri.fr}

\begin{abstract}

\noindent The ternary \emph{betweenness relation} of a tree, $B(x,y,z),$
expresses that the node $y$ is on the unique path between nodes $x$ and
$z$.\ This notion can be extended to \emph{order-theoretic trees }defined as
partial orders such that the set of nodes larger than any node is linearly
ordered.\ In such generalized trees, the unique "path" between two nodes is
linearly ordered and can be infinite.\ 

\qquad We generalize some results obtained in a previous article for the
betweenness relation of \emph{join-trees}.\ Join-trees are order-theoretic
trees such that any two nodes have a least upper-bound. The motivation was to
define conveniently the rank-width of a countable graph.\ We called
\emph{quasi-tree} the structure $(N,B)$ based on the betweenness relation
$B$\ of a join-tree with vertex set $N$.\ We proved that quasi-trees are
axiomatized by a first-order sentence.

\qquad Here, we obtain a monadic second-order axiomatization of betweenness in
order-theoretic trees.\ We also define and compare several \emph{induced
betweenness relations}, \emph{i.e.}, restrictions to sets of nodes of the
betweenness relations in countable generalized trees of different kinds.\ We
prove that induced betweenness in quasi-trees is characterized by a
first-order sentence. The proof uses order-theoretic trees.
\end{abstract}

\maketitle

\section*{Introduction}

\bigskip

The \emph{rank-width} $rwd(G$) of a finite graph $G,$\ defined by Oum and
Seymour in \cite{OumSey},\ is a complexity measure based on ternary trees
whose leaves hold the vertices.\ If $H$ is an induced subgraph of $G$, then
$rwd(H)\leq rwd(G)$. In order to define the rank-width of a countable graph in
such a way that it be the least upper-bound of those of its finite induced
subgraphs, we have defined in \cite{Cou14}\ certain generalized (undirected)
trees called \emph{quasi-trees} (forming the class \textbf{QT}), such that the
unique "path" between any two nodes is linearly ordered and can be
infinite.\ In particular, it can have the order-type of an interval of the set
$\mathbb{Q}$ of rational numbers. As no notion of adjacency can be used, we
have defined quasi-trees in terms of a notion of betweenness.\ 

The \emph{betweenness relation} of a tree is the ternary relation $B$ such
that $B(x,y,z)$ holds if and only if $x,y,z$ are distinct and $y$ is on the
unique path between $x$ and $z$.\ It can be extended to \emph{order-theoretic
trees }defined as partial orders such that the set of elements larger than any
element is linearly ordered.\ A \emph{join-tree} is an order-theoretic tree
such that any two nodes have a \emph{least upper-bound}, equivalently in this
case, a \emph{least common ancestor}. A join-tree may have no root,\emph{
i.e.}, no largest element. A quasi-tree is defined abstractly as a ternary
structure $S=(N,B)$\ satisfying finitely many first-order \emph{betweenness
axioms.\ }But quasi-trees are equivalently characterized as the betweenness
relations of join-trees \cite{Cou14}.

In the present article we axiomatize in monadic second-order logic betweenness
in order-theoretic trees\footnote{All trees and related structures (except
lines in the plane in the definition of topological trees) are finite or
countably infinite.}.\ We also define and study several \emph{induced
betweenness relations}, \emph{i.e.}, restrictions to sets of nodes of
betweenness relations in generalized trees of different kinds. An induced
betweenness relation in a quasi-tree need not be that of a
quasi-tree.\ However, induced betweenness relations in quasi-trees, forming
the class \textbf{IBQT}, are also axiomatized by a single \emph{first-order
sentence}.\ This fact does not follow immediately by a general logical
argument from the first-order characterization of quasi-trees.\ The proof that
this axiomatization is valid uses order-theoretic trees.

We define actually four types of betweenness structures $S=(N,B)$ for which we
prove that the inclusions following from the definitions are proper.\ For each
type of betweenness, a structure $S$ is defined from an order-theoretic tree
$T$.\ Except for the case of induced betweenness in order-theoretic trees,
some defining tree $T$ can be described in $S$ by monadic second-order
formulas.\ In technical words, $T$ is defined from $S$ by a \emph{monadic
second-order transduction}, a notion thoroughly studied in \cite{CouEng}. The
construction of a monadic second-order transduction for induced betweenness in
quasi-trees is not straighforward.\ It is based on a notion of
\emph{structuring }of order-theoretic trees already used in
\cite{Cou15,Cou14,CouLMCS}, that consists in decompositing them into pairwise
disjoint "branches", that are convex and linearly ordered.\ Monadic
second-order formulas can identify structurings of order-theoretic trees.\ In
these articles, we also obtained algebraic characterizations of the join-trees
and quasi-trees that are the unique countable models of monadic-second order
sentences\footnote{This type of characterization will be extended to
order-theoretic trees in a work in progress.}.\ 

In order to provide a concrete view of our generalized trees, we embed them
into \emph{topological trees}, defined as connected unions of possibly
unbounded segments of straight lines in the plane that have no subset
homeomorphic to a circle. Countable induced betweenness relations in
topological trees and in quasi-trees are the same.

Our main results are the following ones:
\begin{itemize}
\item this class \textbf{IBQT} is first-order axiomatizable (Theorem 3.1),

\item a join-tree witnessing that a ternary structure $S$\ is in \textbf{IBQT}  can be specified in $S$  by monadic second-order formulas (Theorem 3.25),

\item induced betweenness relations in topological trees and in quasi-trees are
the same (Theorem 4.4).
\end{itemize}

\paragraph{About motivations}
This article arises from three research directions of theoretical nature. The
first one concerns \emph{Model Theory}.\ A general goal is to understand the
power of logical languages, here first-order (FO in short) and monadic
second-order (MSO in short) logic, for expressing properties of trees, graphs
and related relational structures, and of transformations of such structures.
For finite structures, monadic second-order logic yields tractable algorithms
parameterized by appropriate \emph{widths}, based on hierarchical
decompositions \cite{CouEng,Hli+}. For countably infinite structures described
in appropriate finitary ways, it yields decidability results\footnote{Of high
complexity, so that these results do not provide usable algorithms.\ However,
they contribute to the theory of calculability.}.\ The relevant graphs and
trees belong to Caucal's hierarchy (see \cite{Car,Par,Tho}). On both aspects
the literature is enormous. When a property is proved to be MSO expressible,
we try to answer the natural question of asking whether it is FO expressible.

The second research direction concerns \emph{order-theoretic trees}
(\emph{O-trees} in short), a classical notion in the \emph{Theory of
Relations}, studied in particular by Fra\"{\i}ss\'{e} in \cite{Fra}. He
defined a countable\emph{ universal }O-tree, in which every countable O-tree
embeds.\ We used O-trees for defining rank-width and modular decomposition of
countable graphs \cite{Cou14,CouDel}.\ Infinite words based on countable
linear orders (of any type) are studied with the concepts of the \emph{Theory
of Automata} and monadic second-order logic \cite{Car+}.\ Hence, our study of
order-theoretic trees with such tools aims at completing this theory of
countable structures \cite{Cou15,CouLMCS}.

The third research direction concerns \emph{Combinatorial Geometry} and, in
particular, the natural notion of \emph{betweenness}. The betweenness of a
linear order describes it \emph{up to reversal}.\ This notion is FO
axiomatizable, but offers difficult problems and open questions.\ It is
NP-complete to decide if a finite ternary relation is included in the
betweenness relation of a linear order\footnote{On the contrary, one can
decide in polynomial time if a finite binary relation is included in a linear
order.} (see Chapter 9 of \cite{CouEng}).\ Betweenness has also been studied
in \emph{partial orders}.\ It is axiomatized by an infinite set of first-order
sentences in \cite{Lih}, that cannot be replaced by a finite one
\cite{CouBetPO}. In the latter article, we axiomatize betweenness in partial
orders by an MSO\ sentence.\ Several notions of betweenness in \emph{graphs}
have also been investigated and axiomatized.\ We only refer to the survey
\cite{Cha+} that contains a rich bibliography. Another reference is \cite{Chv}
about the betweenness in graphs relative to induced paths: $y$ is between $x$
and $z$ if it is an intermediate vertex on a chordless path between $x$ and
$z$.

\bigskip

\paragraph{Summary} We review definitions and notation in Section 1. We define
four different notions of betweenness in order-theoretic trees in Section
2.\ We establish in Section 3 the first-order and monadic second-order
axiomatizations presented above.\ The case of induced betweenness in
order-theoretic trees is left as a conjecture. We also examine whether monadic
second-order transductions can produce witnessing trees from given betweenness
structures.\ In Section 4, we describe embeddings of join-trees into
topological trees. In an appendix (Section 6), we give an example of a
first-order class of relational structures (actually of labelled graphs) whose
induced substructures do not form a first-order (and even a monadic
second-order) axiomatizable class.

\bigskip

\section{Definitions and basic facts}

All trees, graphs and logical structures are countable, which means, finite or
countably infinite. We will not repeat this hypothesis in our statements.

In some cases, we denote by $X\uplus Y$ the union of sets $X$ and $Y$ to
insist that they are disjoint.\ Isomorphism of ordered sets, trees, graphs and
other logical structures is denoted by $\simeq$. We denote by $[n]$ the set of
integers $\{1,...,n\}$.

The \emph{arity} of a relation $R$ is $\rho(R).$ The \emph{restriction} of a
relation $R$ defined on a set $V$\ to a subset $X$ of $V$, \emph{i.e.}, $R\cap
X^{\rho(R)},$ is denoted by $R[X]$. If $S$ is an \{$R_{1},..,R_{k}%
\}$-structure $(V,R_{1},..,R_{k})$, then $S[X]:=(V,R_{1}[X],..,R_{k}[X])$.

The \emph{Gaifman graph} of $S=(V,R_{1},..,R_{k})$ is the graph $\mathit{Gf}%
(S)$ with vertex set $V$ and an edge between $x$ and $y\neq x$ if and only if
$x$ and $y$ belong to a tuple of some relation $R_{i}$. We say that $S$ is
\emph{connected} if its Gaifman graph is connected. If it is not, $S$ is the
disjoint union of connected structures, each of them corresponding to a
connected component of the Gaifman graph of $S$.

A family of sets is \emph{overlapping} if it contains two sets $X$ and $Y$
such that $X\cap Y$, $X-Y$ and $Y-X$ are all not empty.

\subsection{Partial orders}

For partial orders $\leq,\preceq,\sqsubseteq$, ... we denote respectively by
$<,\prec,\sqsubset$, ... the corresponding strict partial orders. We write
$x\bot y$ if $x$ and $y$ are incomparable for the considered order.

Let $(V,\leq)$ be a partial order.\ For $X,Y\subseteq V$, the notation $X<Y$
means that $x<y$ for every $x\in X$ and $y\in Y$.\ We write $X<y$ instead of
$X<\{y\}$ and similarly for $x<Y$. We use similar notation for $\leq$\ and
$\bot$. The least upper-bound of $x$ and $y$ is denoted by $x\sqcup y$ if it
exists and is called their \emph{join}.\ 

If $X\subseteq V$, then we define $N_{\leq}(X):=\{y\in V\mid y\leq X\}$\ and
similarly for $N_{<}$.\ We define $\downarrow(X):=\{y\in V\mid y\leq x$ for
some $x\in X\}.\ $We have $N_{\leq}(X)\leq X$, $N_{\leq}(\emptyset)=V$, and
$\downarrow(\emptyset)=\emptyset.$ We also define $L_{\geq}(X):=\{y\in V\mid
y\geq X\}$, and similarly $L_{>}(X).$ We write $L_{\geq}(x)$ (resp. $L_{\geq
}(x,y))$ if $X=\{x\}$ (resp. $X=\{x,y\}$) and similarly for $L_{>}.$ Note that
$L_{\geq}(X)$ is $N_{\geq^{\prime}}(X)$ for the opposite order $\leq^{\prime}$
of $\leq.$

An \emph{interval} $X$\ of $(V,\leq)$ is a \emph{convex subset}, \emph{i.e.},
$y\in X$ if $x<y<z$ and $x,z\in X$.

Let $(V,\leq)$\ and $(V^{\prime},\leq^{\prime})$ be partial orders.\ An
\emph{embedding} $j:(V,\leq)\rightarrow(V^{\prime},$ $\leq^{\prime})$\ is an
injective mapping such that $x\leq y$ if and only if $j(x)\leq^{\prime}j(y)$;
in this case,\ $(V,\leq)$\ is isomorphic by $j$ to $(j(V),\leq^{\prime\prime
})$, where $\leq^{\prime\prime}$ is the restriction of $\leq^{\prime}$\ to
$j(V)$ (\emph{i.e.}, is $\leq^{\prime}\ [j(V)]$).\ We will write more simply
$(j(V),\leq^{\prime})$.\ 

We say that $j$ is a \emph{join-embedding} if, furthermore, $j(x)\sqcup
^{\prime}j(y)$ is defined and equal to $j(x\sqcup y)$ whenever $x\sqcup y$ is defined.

Here is an example of an embedding that is not a join-embedding: $j$ is the
inclusion mapping $(X,\leq)\rightarrow(V,\leq)$\ where $V:=\{a,b,c,d\}$,
$a<c<d$, $b<c$ , $a\bot b$ and $X=\{a,b,d\}$. We have $a\sqcup b=d$ in
$(X,\leq)$ but $a\sqcup b=c\neq j(d)$ in $(V,\leq).$

\subsection{Trees}

A \emph{forest} is a possibly empty, undirected graph $F$ that has no cycles.
Hence, it has neither loops nor multiple edges\footnote{No two edges with same
ends.}. We call \emph{nodes} its vertices. Their set is denoted by $N_{F}$. A
\emph{tree} is a connected forest.\ 

A \emph{rooted tree} $R=(T,r)$\ is a tree $T$ equipped with a distinguished
node $r$ called its \emph{root}. We define on $N_{R}:=N_{T}$\ the partial
order $\leq_{R}$ such that $x\leq_{R}y$ if and only if $y$ is on the unique
path in $T$ between $x$ and the root $r$. The minimal nodes are the
\emph{leaves} and the root is the largest node.\ The least upper-bound of $x$
and $y$, denoted by $x\sqcup_{R}y$ is their least common ancestor in $R$.

We will specify a rooted tree $R$\ by $(N_{R},\leq_{R})$ and we will omit the
index $R$ when the considered tree is clear.

A partial order $(N,\leq)$ is $(N_{R},\leq_{R})$ for some rooted tree $R$ if
and only if\ it has a largest element and, for each $x\in N$, the set
$L_{\geq}(x)$ is finite and linearly ordered. These conditions imply that any
two nodes have a join.

\bigskip

\subsection{Order-theoretic forests and trees}

\begin{Definition}{1.1}[O-forests and O-trees]
In order to have a simple terminology, we will use the prefix \emph{O-} to
mean \emph{order-theoretic.}

\begin{enumerate}[label=(\alph*)]
\item An \emph{O-forest} is a pair $F=(N,\leq)$ such that:

\begin{enumerate}[label=(\arabic*)]
\item $N$\ is a possibly empty set called the set of \emph{nodes},

\item $\leq$ is a partial order on $N$\ such that, for every node $x,$ the set
$L_{\geq}(x)$ is linearly ordered.
\end{enumerate}

It is called an \emph{O-tree} if furthermore:

\begin{enumerate}[resume*]
\item every two nodes $x$ and $y$ have an upper-bound.
\end{enumerate}

An O-forest $F$ is the union of disjoint O-trees $T_{1},T_{2},...$\ such that
the Gaifman graphs $\mathit{Gf}(T_{i})$ are the connected components of
$\mathit{Gf}(F)$. Two nodes of $F$\ are in a same O-tree $T_{i}$ if and only
if they have an upper-bound.

The \emph{leaves} are the minimal elements. If $N$ has a largest element $r$
(\emph{i.e.}, $x\leq r$ for all $x\in N$) then $F$\ is a \emph{rooted }O-tree
and $r$ is its \emph{root}.

\item A \emph{line} in an O-forest $(N,\leq)$ is a linearly ordered subset
$L$\ of $N$\ that is \emph{convex}, \emph{i.e.}, such that $y\in L$ if $x,z\in
L$ and $x<y<z$. A subset $X$ of $N$\ is \emph{upwards closed}
(resp.\ \emph{downwards closed}) if $y\in X$\ whenever $y>x$ (resp.\ $y<x$)
for some $x\in X$. In an O-forest, the set $L_{\geq}(X)$ of
\emph{upper-bounds} of a nonempty set $X\subseteq N$ is an upwards closed line.\ 

\item An O-tree $T=(N,\leq)$ is a \emph{join-tree}\footnote{An \emph{ordered
tree} is a rooted tree such that the set of sons of any node is linearly
ordered. This notion is extended in \cite{CouLMCS} \ to join-trees.\ Ordered
join-trees should not be confused with order-theoretic trees, that we call
O-trees for simplicity.} if every two nodes $x$ and $y$ have a least
upper-bound (for $\leq$) denoted by $x\sqcup y$ and called their \emph{join}
(cf.\ Section 1.1).\ In a join-tree, every finite set has a least upper-bound,
but an infinite one may have none.

\item Let $J=(N,\leq)$ be an O-forest and $X\subseteq N$. Then $J[X]:=(X,\leq)$
is an O-forest\footnote{We recall from Subsection 1.1 that the notation $\leq$
is also used for the restriction of $\leq$ to $X$.}.\ It is the
\emph{sub-O-forest} of $J$ \emph{induced on} $X$. Two elements $x,y$ having a
join $z$ in $J$\ may have no join in $J[X]$ or they may have a join different
from $z$. If $J$ is an O-tree, then $J[X]$ may not be an O-tree.
\end{enumerate}
\end{Definition}

\begin{Examples}{1.2}
  \begin{enumerate}
  
\item If $R$ is a rooted tree, then $(N_{R},\leq_{R})$ is a join-tree.\ Every
finite O-tree is a join-tree of this form.

\item Every linear order is a join-tree.\ 

\item Let $S:=\mathbb{N}\cup\{a,b,c\}$ be strictly partially ordered by $<_{S}%
$\ such that $a<_{S}b,c<_{S}b$ and $b<_{S}i<_{S}j$ for all $i,j\in\mathbb{N}$
such that\footnote{The standard\ strict order on $\mathbb{N}$ is  $<$  .} $j<i$, and $a$ and $c$ are incomparable.\ Then 
$T:=(S,\leq_{S})$ is a
join-tree, see the left part of Figure 1.\ In particular $a\sqcup_{S}%
c=b$.\ The relation $\leq_{S}$ is not the partial order associated with any
rooted tree (by the remark at the end of Section 1.2).

We can consider $\mathbb{N}\cup\{a,b\}$ as forming a "path" between $a$ and 0
in the join-tree $T$ (where 0 is the largest element). A formal definition of
such "paths" will be given. Let $S^{\prime}:=S-\{b\}.$ The O-tree
$T[S^{\prime}]:=(S^{\prime},\leq_{S})$ is not a join-tree because $a$ and $c$
have no join.%

\begin{figure}
[ptb]
\begin{center}
\includegraphics[
height=2.322in,
width=2.9118in
]{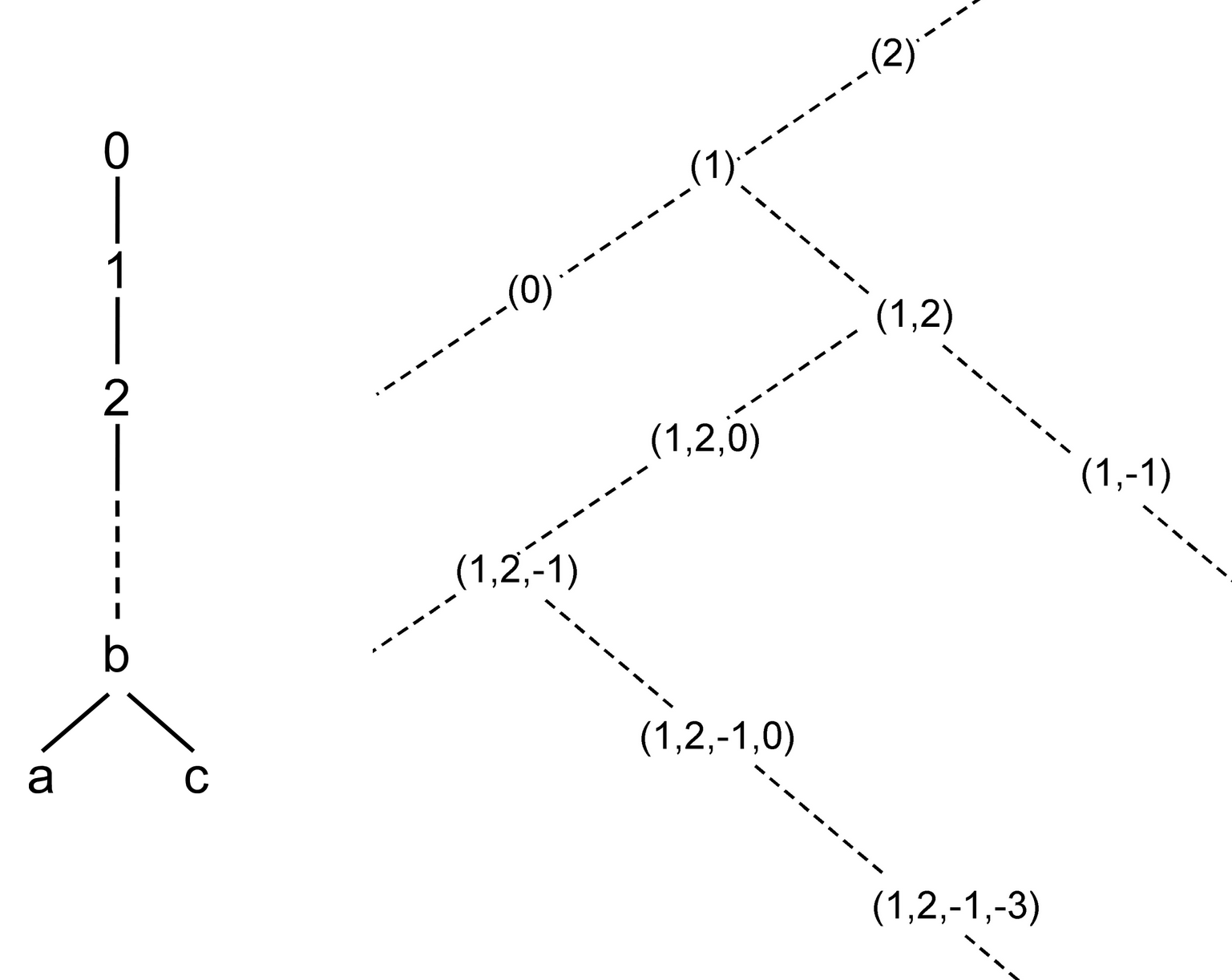}%

\caption{The join-tree of Examples 1.2(3) and 3.4(a). A part of the universal
O-tree of Example 1.2(4).}%
\end{center}
\end{figure}

\item Fra\"{\i}ss\'{e} has defined in \cite{Fra}\ (Section 10.5.3) a join-tree
$T:=(Seq_{+}(\mathbb{Q}),\preceq)$\ where $Seq_{+}(\mathbb{Q})$ is the set of
finite nonempty sequences of rational numbers, such that every O-tree
$(N,\leq)$ is isomorphic to $T[X]$ for some subset $X$ of $Seq_{+}%
(\mathbb{Q}).$ The strict partial order $\prec$ is defined as follows. For two
sequences $\boldsymbol{x}=(x_{1},...,x_{n})$ and $\boldsymbol{y}%
=(y_{1},...,y_{m})$ we have $\boldsymbol{x}\prec\boldsymbol{y}$ if and only if:

$\qquad$(i) $n\geq m$,\ $(x_{1},...,x_{m-1})=(y_{1},...,y_{m-1})$ and
$x_{m}<y_{m},$ or

\qquad(ii) $n>m$ and\ $(x_{1},...,x_{m})=(y_{1},...,y_{m}).$

In particular, for all $x_{1},...,x_{n}$ and $z<x_{n}.$\ we have
$(x_{1},...,x_{n-1},z)\prec(x_{1},...,x_{n-1},x_{n})$ by (i) and
$(x_{1},...,x_{n})\prec(x_{1},...,x_{n-1})$ by (ii).\ The strict partial order
$\prec$ is generated by transitivity from these particular relations.

Two sequences $\boldsymbol{x}$ and $\boldsymbol{y}$ as above are incomparable
if and only if there is a sequence $(z_{1},...,z_{p})$ such that either $p\leq
n$, $\boldsymbol{x}=(z_{1},...,z_{p-1},x_{p},...,x_{n}),z_{p}>x_{p}$, and
$\boldsymbol{y}=(z_{1},...,z_{p-1},z_{p},y_{p+1},...,y_{m})$ or vice-versa by
exchanging $\boldsymbol{x}$ and $\boldsymbol{y}$. Their join is
$\boldsymbol{z}=(z_{1},...,z_{p}).$

Examples of lines in $T$\ are $\{(x)\mid x\in\mathbb{Q}\}$ and, for each
$x_{1},x_{2}\in\mathbb{Q}$, the sets $\{(x_{1},x_{2},x)\mid x\in\mathbb{Q}\}$
and $\ \{(x_{1},x\,_{2},x),(x_{1},z),(y)\mid x,y,z\in\mathbb{Q},y\geq
x_{1},z\geq x_{2}\}.$

The right part of Figure 1 sketches some parts of this join-tree. We have
$(1,-1)\prec$ $(1,2)\prec(2)$ and $(1,2,-1,0)\prec(1,2,0)$ by Case (i), and
$(1,2,-1,-3)\prec(1,2,-1)\prec(1,2)\prec(1)$ by Case (ii).

Examples of joins are $(0)\sqcup(1,2)=(1)$ (with $\boldsymbol{x}=(0)$,
$\boldsymbol{y}=(1,2)$ and $\boldsymbol{z}=(1)$),

and $(1,-1)\sqcup(1,2,-1,0)=(1,2)$ (with $\boldsymbol{x}=(1,-1)$,
$\boldsymbol{y}=(1,2,-1,0)$ and $\boldsymbol{z}=(1,2)$).

Examples of lines are $\{(1,2,x)\mid x\in\mathbb{Q}\},$ $\{(1,x)\mid
x\in\mathbb{Q},x\geq2\}$ and $\ \{(1,2,x),(1,z),(y)\mid x,y,z\in
\mathbb{Q},y\geq1,z\geq2\}.$ We will also consider this tree in Examples 3.6
and 3.28.$\ \square$
  \end{enumerate}
\end{Examples}

\bigskip

\begin{Definitions}{1.3}[The join-completion of an O-forest]
Let \ $J=(N,\leq)$ be an O-forest.\ We let $\mathcal{K}$\ be the set of
upwards closed lines $L_{\geq}(x,y)$\ for all (possibly equal) nodes
$x,y.\ $If $x$ and $y$ have no upper-bound, then $L_{\geq}(x,y)$\ is empty. If
$x\sqcup y$ is defined, then $L_{\geq}(x,y)=L_{\geq}(x\sqcup y).$

The family $\mathcal{K}$ is countable.\ We let $j:N\rightarrow\mathcal{K}%
$\ map $x$ to $L_{\geq}(x)$ and $\widehat{J}:=(\mathcal{K},\supseteq)$. We
call $\widehat{J}$ the \emph{join-completion of }$J$ because of the following
proposition, stated with these hypotheses and notation.
\end{Definitions}

\begin{Proposition}{1.4} The partially ordered set $\widehat{J}%
:=(\mathcal{K},\supseteq)$ is a join-tree and $j$ is a join-embedding
$J\rightarrow\widehat{J}$.
\end{Proposition}

\begin{proof}[Proof Sketch] We indicate the main steps.\ First,
$\widehat{J}:=(\mathcal{K},\supseteq)$ is an O-tree: if $L,L^{\prime
},L^{\prime\prime}\in\mathcal{K},L^{\prime}\subseteq L$ and $L^{\prime\prime
}\subseteq L,$ then $L^{\prime}\subseteq L^{\prime\prime}$ or $L^{\prime
\prime}\subseteq L^{\prime}$ because $L,L^{\prime},L^{\prime\prime}$ are
upwards closed lines.

\emph{Claim}: $\widehat{J}$ is a join-tree.

\emph{Proof }: Let $L_{\geq}(x,y)$\ and $L_{\geq}(z,u)$\ be incomparable.\ We
have $w\in L_{\geq}(x,y)-L_{\geq}(z,u)$\ and $w^{\prime}\in L_{\geq
}(z,u)-L_{\geq}(x,y).$\ We claim that $L_{\geq}(w,w^{\prime})=L_{\geq
}(x,y)\cap L_{\geq}(z,u),$ hence that it is the join of $L_{\geq}(x,y)$\ and
$L_{\geq}(z,u)$\ in $\widehat{J}$. To prove the claim, we note that $L_{\geq
}(w,w^{\prime})\subseteq L_{\geq}(w)\subseteq L_{\geq}(x,y)$ and similarly,
$L_{\geq}(w,w^{\prime})\subseteq L_{\geq}(z,u)$, hence $L_{\geq}(w,w^{\prime
})\subseteq L_{\geq}(x,y)\cap L_{\geq}(z,u).$ Conversely, assume we have
$t\in(L_{\geq}(x,y)\cap L_{\geq}(z,u))-L_{\geq}(w,w^{\prime}).$ As
$x\leq\{w,t\}$, we have $w\leq t$ or $t\leq w$. Assume $t\leq w$. Then since
$t\in L_{\geq}(z,u),$ we have $w\in L_{\geq}(z,u)$, contradicting its
definition.\ So we should have $w\leq t$ and similarly, $w^{\prime}\leq
t$.\ Hence $t\in L_{\geq}(w,w^{\prime})$, contradicting its definition.\ This
proves the claim.\ Note that $L_{\geq}(x,y)\cap L_{\geq}(z,u)=L_{\geq}(x,z).$
\qed
\smallskip

Then we have $x\leq y$ if and only if $L_{\geq}(y)\subseteq L_{\geq}(x)$,
hence $j$ is an embedding.\ Since $L_{\geq}(x\sqcup y)=L_{\geq}(x)\cap
L_{\geq}(y)$ that is the join of $L_{\geq}(x)$ and $L_{\geq}(y)$ in
$\widehat{J}$, $j$ is a join-embedding.
\end{proof}

\bigskip

\ Its construction adds to $J$\ the "missing joins".\ The existing joins are
preserved. It follows that every O-forest $J$\ with set of nodes $N$\ is
$T[N]$ for some join-tree $T$, in particular for $T:=\widehat{J}.$

\subsection{Monadic second-order logic}

We will express properties of relational structures by first-order (FO in
short) and monadic second-order (MSO) formulas and sentences.\ Logical
structures are relational\ (they have only relation symbols) and countable.

\bigskip

\begin{Definitions}{1.5}[Quick review of terminology and notation]
\emph{Monadic second-order logic} extends first-order logic by the use of
\emph{set variables} $X,Y,Z$ ...\ denoting subsets of the domain of the
considered logical structure.\ The atomic formula $x\in X$ expresses the
membership of $x$ in $X$. We call \emph{first-order} a formula where set
variables are not quantified. For example, a first-order formula can express
that $X\subseteq Y$. A\emph{ sentence} is a\ formula without free variables.

A property $P$ of $\mathcal{R}$-structures where $\mathcal{R}$ is a finite set
of relation symbols, is \emph{first-order }or \emph{monadic second-order
expressible} (\emph{FO} or \emph{MSO expressible}) if it is equivalent to the
validity, in every $\mathcal{R}$-structure $S$, of a first-order or monadic
second-order sentence $\varphi$. The validity of $\varphi$ in $S$ is denoted
by $S\models\varphi$. We say that a property of tuples of subsets
$X_{1},...,X_{n}$ of the domains of structures in a class $\mathcal{C}$\ is
\emph{FO} or \emph{MSO definable} if it is equivalent to $S\models
\varphi(X_{1},...,X_{n}$ ) in every $\mathcal{R}$-structure $S$ in
$\mathcal{C}$, where $\varphi$ is a fixed FO or MSO formula with $n$ free set
variables.\ A class of structures is \emph{FO} or \emph{MSO definable} or
\emph{axiomatizable} if it is characterized by an FO or MSO sentence.
\end{Definitions}

Transitive closures and choices of sets, typically in graph coloring problems,
are MSO\ but not FO expressible.\ See \cite{CouEng} for a detailed study of
MSO expressible graph properties. Other comprehensive books are \cite{Hod,Lib}.

\bigskip

\begin{Examples}{1.6 }[Partial orders and graphs]
  \begin{enumerate}

\item A simple undirected graph $G$\ can be identified with the \{$edg$%
\}-structure $(V_{G},edg_{G})$ where $V_{G}$\ is its vertex set and
$edg_{G}(x,y)$ means that there is an edge between $x$ and $y$ if $G$. For
example, 3-colorability is expressed by the MSO sentence :

\begin{quote}
$\exists X,Y[X\cap Y=\emptyset\wedge\lnot\exists u,v(edg(u,v)\wedge
\lbrack(u\in X\wedge v\in X)\vee$

$\qquad\qquad\qquad(u\in Y\wedge v\in Y)\wedge(u\notin X\cup Y\wedge v\notin
X\cup Y)])].$
\end{quote}

\item We now consider partial orders $(N,\leq)$.\ The FO\ formula $Lin(X)$
defined as $\forall x,y[(x\in X\wedge y\in X)\Longrightarrow(x\leq y\vee y\leq
x)]$ expresses that a subset $X$\ of $N$ is linearly ordered. The MSO\ formula

\begin{quote}
$Lin(X)\wedge\exists a,b[Min(X,a)\wedge Max(X,b)\wedge\theta(X,a,b)]$
\end{quote}

expresses that $X$ is linearly ordered and finite, where $Min(X,a)$ and
$Max(X,b)$ are FO formulas expressing respectively that $X$\ has a least
element $a$ and a largest one $b$, and $\theta(X,a,b)$\ is an MSO formula
expressing that :

\begin{quote}
(i) each element $x$ of $X$ except $b$ has a \emph{successor} $c$ in $X$
(\emph{i.e.}, $c$ is the least element of $L_{>}(x)\cap X$), and

(ii) $(a,b)\in Suc^{\ast},$ where $Suc$ is the above defined successor
relation (depending on $X$) and $Suc^{\ast}$\ is its reflexive and transitive closure.
\end{quote}

Assertion (ii) is expressed by the MSO\ formula with free variables $a,b,X$ :

$\forall U[U\subseteq X\wedge a\in U\wedge\forall x,y((x\in U\wedge(x,y)\in
Suc)\Longrightarrow y\in U)\Longrightarrow b\in U].$
\end{enumerate}
\end{Examples}

First-order formulas expressing $U\subseteq X$, $(x,y)\in Suc$ and Property
(i) are easy to write. The finiteness of a linear order is not FO
expressible\footnote{Follows from the Compactness Theorem for FO\ logic
\cite{Hod}.}. Without a linear order, the finiteness of a set $X$\ is not MSO expressible.\ 

\begin{Definitions}{1.7}[Transformations of relational structures]
As in \cite{CouEng}, we call \emph{transduction} a transformation of
relational structures specified by logical formulas\footnote{The usual
terminology of \emph{interpretation} is inconvenient as it is frequently
unclear what is defined from what.\ The term\emph{ transduction} is borrowed
to formal language theory that is concerned with transformations of words,
trees and terms. There are deep links between monadic second-order definable
transductions and tree transducers \cite{CouEng}.}.\ We will try to be not too
formal but nevertheless precise.
\end{Definitions}

\begin{enumerate}[label=(\alph*)]
\item The basic type of transduction $\tau$\ is as follows.\ A structure
$S^{\prime}=(D^{\prime},R_{1}^{\prime},..,$ $R_{m}^{\prime})$ is\ defined from
a structure $S=(D,R_{1},..,R_{n})$ and a $p$-tuple ($X_{1},..,X_{p})$ of
subsets of $D$ called \emph{parameters} by means of formulas $\chi
,\delta,\theta_{R_{1}^{\prime}},...,\theta_{R_{m}^{\prime}}$ used as follows:

\begin{quote}
$\tau(S,(X_{1},..,X_{p}))=S^{\prime}$ is defined if and only if $S\models
\chi(X_{1},...,X_{p}),$

$S^{\prime}=(D^{\prime},R_{1}^{\prime},..,R_{m}^{\prime})$ has domain
$D^{\prime}\subseteq D$ such that $d\in D^{\prime}$ if and only if
$S\models\delta(X_{1},...,X_{p},d),$

$R_{i}^{\prime}$ is the set of tuples ($d_{1},...,d_{s})\in D^{\prime s}$,
$s=\rho(R_{i}^{\prime})$, such that $S\models\theta_{R_{i}^{\prime}}%
(X_{1},\ldots,X_{n}, d_{1},$ $\ldots,d_{s}).$
\end{quote}

We call $\tau$\ an FO or an MSO transduction if the formulas that define it
are, respectively, first-order or monadic second-order ones.

As an example, the mapping from a graph $G=(V,edg)$ to the connected component
$(V^{\prime},edg[V^{\prime}])$ containing a vertex $u$ is defined by
$\chi,\delta$ and $\theta_{edg}$ where $\chi(X)$ expresses that $X$ is a
singleton $\{u\}$, $\delta(X,d)$ expresses that there is a path between $d$
and the vertex in $X$, and $\theta_{edg}(x,y)$ is the formula always
$\boldsymbol{true}$, say, $x=x$. It is an MSO transduction as path properties
are expressible by monadic second-order formulas.

\bigskip

\item Transductions of the general type may enlarge the domain of the input
structure. A structure $S^{\prime}=(D^{\prime},R_{1}^{\prime},..,R_{m}%
^{\prime})$ is defined from $S=(D,R_{1},..,R_{n})$ and a $p$-tuple
($X_{1},..,X_{p})$ of parameters as above by means of formulas $\chi
,\delta_{1},...,\delta_{k}$ and others, $\theta_{R_{i}^{\prime},i_{1}%
,...,i_{s}}$, used as follows:

\begin{quote}
$\tau(S,(X_{1},..,X_{p}))=S^{\prime}$ is defined if and only if $S\models
\chi(X_{1},...,X_{p}),$

$S^{\prime}=(D^{\prime},R_{1}^{\prime},..,R_{m}^{\prime})$ has domain
$D^{\prime}\subseteq(D\times\{1\})\uplus...\uplus(D\times\{k\})$ such that
$(d,i)\in D^{\prime}$ if and only if $S\models\delta_{i}(X_{1},...,X_{p},d),$

$R_{i}^{\prime}$ is the set of tuples (($d_{1},i_{1}),...,(d_{s},i_{s}))\in
D^{\prime s}$, $s=\rho(R_{i}^{\prime})$, such that

$\qquad\qquad\qquad S\models\theta_{R_{i}^{\prime},i_{1},...,i_{s}}%
(X_{1},...,X_{p},d_{1},...,d_{s}).$
\end{quote}

If $D$ is finite, then $\left\vert D^{\prime}\right\vert \leq k\left\vert
D\right\vert $.
\end{enumerate}

An easy example consists in the \emph{duplication} of a graph $G=(V,edg)$ into
the graph $H:=G\oplus G^{\prime}$, that is $G$ together with a disjoint copy
$G^{\prime}$ of it.\ We get a graph $H$ up to isomorphism, because of the use
of disjoint isomorphic copies.\ To define a transduction, we take $k=2$, $p=0$
(no parameter is needed), $\chi,\delta_{1},\delta_{2}$ always
$\boldsymbol{true}$, $\theta_{edg,i,j}(x,y)$ always $\boldsymbol{false}$ if
$i\neq j$, and equal to $edg(x,y)$ if $i=j$, where $i,j\in\lbrack2]$.

Another more complicated example is the transformation of an O-forest
$J=(N,\leq)$ into its join-completion $\widehat{J}$.\ We define concretely the
set of nodes of $\widehat{J}$\ as ($N\times\{1\})\uplus(M\times\{2\})$ where
$M$ is a subset of $N$\ in bijection with the set of sets $L_{\geq}(x,y)$ such
that $x$ and $y$ have no join, cf.\ Definition 1.3.\ This bijection can be
made MSO definable, and so is the order relation of $\widehat{J}$. Defining
$M$ is not straightforward because the sets $L_{\geq}(x,y)$ are not pairwise
disjoint.\ We can use the notion of \emph{structuring of an O-tree}: see
Remark 3.35.

\section{Quasi-trees and betweenness in O-trees}

In this section, we define a \emph{betweenness relation }in O-trees, and
compare it with the \emph{betweenness relation induced} by sets of nodes in
join-trees or O-trees.\ We generalize the notion of quasi-tree defined and
studied in \cite{Cou14} and \cite{CouLMCS}.

For a ternary relation $B$ on a set $N$ and $x,y\in N$, we define
$[x,y]_{B}:=\{x,y\}\cup\{z\in N\mid(x,z,y)\in B\}.$ If $n>2$, then the
notation $\neq(x_{1},x_{2},...,x_{n})$ means that $x_{1},x_{2},...,x_{n}$ are
pairwise distinct (hence abreviates an FO formula).

\subsection{Betweenness in trees and quasi-trees}

\begin{Definition}{2.1}[Betweenness in linear orders and in trees]
  \begin{enumerate}[label=(\alph*)]
\item Let $L=(X,\leq)$ be a linear order.\ Its \emph{betweenness
relation\footnote{This definition can be used for partial orders.\ The
corresponding notion of betweenness is axiomatized in \cite{CouBetPO,Lih}.\ We
will \emph{not} use it for defining betweenness in order-theoretic trees,
although these trees are partial orders, because it would not yield the
desired generalization of quasi-trees. See Example 2.2.}} $B_{L}$ is the
ternary relation on $X$ defined by :

\begin{quote}
$B_{L}(x,y,z):\Longleftrightarrow x<y<z$ or $z<y<x.\ $
\end{quote}

\item If $F$ is a forest, its \emph{betweenness relation} $B_{F}$ is the ternary
relation on $N_{F}$ defined by:
\[
\text{$B_{F}(x,y,z):\Longleftrightarrow x,y,z$ are pairwise distinct and $y$ is on a
path between $x$ and $z$.}
\]
Such a path is unique if it does exist.

\item If $R=(N_{R},\leq_{R})$ is a rooted tree, we define its \emph{betweenness
relation} $B_{R}$\ as $B_{\mathit{Und}(R)}$ where $\mathit{Und}(R)$ is the
tree obtained from $R$ by forgetting its root.
\end{enumerate}
\end{Definition}

For all $x,y,z\in N$, we have the following characterization of $B_{R}%
=B_{\mathit{Und}(R)}$:

\begin{quote}
$B_{R}(x,y,z)\Longleftrightarrow x,y,z$ are pairwise distinct, $x$ and $z$
have a join $x\sqcup_{R}z$ and $x<_{R}y\leq_{R}x\sqcup_{R}z$ or $z<_{R}%
y\leq_{R}x\sqcup_{R}z$.
\end{quote}

It follows that the betweenness relation of a rooted tree is invariant under a
change of root: $B_{R}=B_{R^{\prime}}$\ if $\mathit{Und}(R)=\mathit{Und}%
(R^{\prime})$.

\begin{Example}{2.2} Figure 2 shows a rooted tree $R$ with root 0. For
illustrating the above description of $B_{R}$, we note that $B_{R}(b,a,0)$ and
$b<a<0=b\sqcup0$, and also that $B_{R}(b,a,c)$ and $b<a<1=b\sqcup c.$ The
betweenness of the partial order $(N_{R},\leq_{R})$ in the sense of
\cite{CouBetPO,Lih} does not contain the triple $(b,a,c)$. It is only the
union of those of the four paths from the leaves $b,d,f,h$ to the root
0.
\end{Example}

\begin{figure}
[ptb]
\begin{center}
\includegraphics[
height=1.6259in,
width=1.0352in ]%
{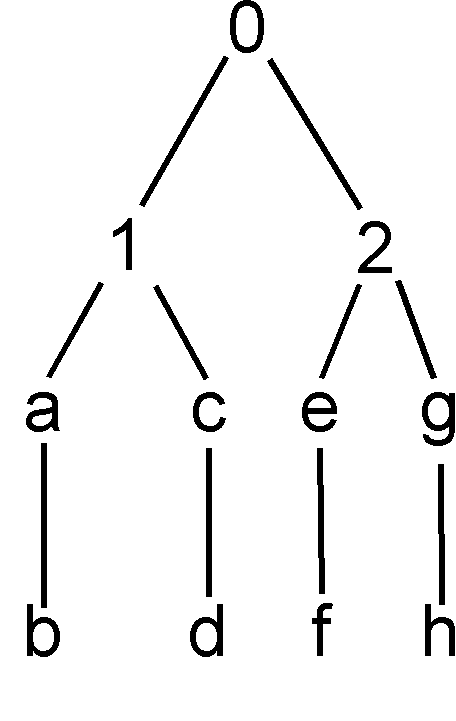}%
\caption{The rooted tree $R$ of\ Example 2.2.}%
\end{center}
\end{figure}

\bigskip

With a ternary relation $B$ on a set $X$, we associate the ternary relation
$A$ on $X$\ : $A(x,y,z):\Longleftrightarrow B(x,y,z)\vee B(x,z,y)\vee
B(y,x,z)$, to be read : $x,y,z$ are \emph{aligned}.\ If $n\geq3$, then
$B^{+}(x_{1},x_{2},...,x_{n})$ stands for the conjunction of the conditions
$B(x_{i},x_{j},x_{k})$ for all $1\leq i<j<k\leq n.$ They imply that
$x_{1},x_{2},...,x_{n}$ are pairwise distinct.

\bigskip

The following is Proposition 5.2\ in \cite{CouLMCS} or Proposition 9.1\ in
\cite{CouEng}.

\begin{Proposition}{2.3}
  \begin{enumerate}[label=(\alph*)]
  \item The betweenness relation $B$ of a linear order
$(X,\leq)$ satisfies the following properties for all $x,y,z,u\in X$.

\begin{quote}
A1 : $B(x,y,z)\Rightarrow\neq(x,y,z).$

A2 : $B(x,y,z)\Rightarrow B(z,y,x).$

A3 : $B(x,y,z)\Rightarrow\lnot B(x,z,y).$

A4 : $B(x,y,z)\wedge B(y,z,u)\Rightarrow B^{+}(x,y,z,u).$

A5 : $B(x,y,z)\wedge B(x,u,y)\Rightarrow B^{+}(x,u,y,z).$

A6 : $B(x,y,z)\wedge B(x,u,z)\Rightarrow y=u\vee B^{+}(x,u,y,z)\vee
B^{+}(x,y,u,z).$

A7' : $\neq(x,y,z)\Rightarrow A(x,y,z).$
\end{quote}

\item The betweenness relation $B$ of a tree $T$ satisfies the properties A1-A6
for all $x,y,z,u$ in $N_{T}$ together with the following weakening of A7':

\begin{quote}
A7 : $\neq(x,y,z)\Rightarrow A(x,y,z)\vee\exists w[B(x,w,y)\wedge
B(y,w,z)\wedge B(x,w,z)].$
\end{quote}
\end{enumerate}
\end{Proposition}

\begin{Remarks}{2.4}
  \begin{enumerate}

\item Property A4 could be written equivalently : $B(x,y,z)\wedge
B(y,z,u)\Rightarrow B(x,y,u)\wedge B(x,z,u).$ Property A5 could be written
$B(x,y,z)\wedge B(x,u,y)\Rightarrow B(x,u,z)\wedge B(u,y,z).$

\item Property A7' says that if $x,y,z$ are three elements in a linear order,
then, one of them is between the two others. Properties A1-A5 belong to the
axiomatization of betweenness in partial orders given in \cite{CouBetPO,Lih}%
.\ Property A6 is actually a consequence of Properties A1-A5 and A7', as one
proves easily.

\item Property A7 says that, in a tree $T$, if $x,y,z$ are three nodes not on a
same path, some node $w$ is between any two of them. In this case, we have :

\begin{quote}
$\{w\}=P_{x,y}\cap P_{y,z}\cap P_{x,z}$ where $P_{x,y}$ is the set of nodes on
the path between $x$ and $y$,
\end{quote}

so that we have $B(x,w,y)\wedge B(y,w,z)\wedge B(x,w,z)$.

If $T$ is a rooted tree, and $x,y,z$ are not on a path from a leaf to the
root, then $w$ is the join (the least common ancestor) of two nodes among
$x,y,z$. In the rooted\ tree $R$ of Figure 2, if $x\ =a,y=d$ and $z=e$, we
have $w=1=x\sqcup y$.

Property A6 is a consequence of Properties A1-A5 and A7.

\item Properties A1-A6 (for an arbitrary structure $S=(N,B)$) imply that the two
cases of the conclusion of A7 are exclusive\footnote{The three cases of
$A(x,y,z)$ are exclusive by A2 and A3.}\emph{ }and that, in the second one,
there is a unique node $w$ satisfying $B(x,w,y)\wedge B(y,w,z)\wedge B(x,w,z)$
(by Lemma 11 of \cite{Cou14}), that is denoted by $M_{S}(x,y,z)$.
\end{enumerate}
\end{Remarks}

\paragraph*{Convention} The letter $B$\ and its variants, $B_{T},B_{1}$, etc.\ will
always denote ternary relations.\ We will only consider ternary relations
satisfying Properties A1 and A2.\ In other words, we will consider $B(x,y,z)$
as identical to $B(z,y,x)$ and $\neq(x,y,z)$ as an immediate consequence of
$B(x,y,z).$ This is similar to the standard usage of considering $x=y$ as
identical to $y=x$ and $x\neq y$ as an immediate consequence of $x<y.$ It
follows that $B^{+}(x_{1},x_{2},...,x_{n})$ stands also for the conjunction of
the conditions $B(x_{k},x_{j},x_{i})$ for $1\leq i<j<k\leq n.$ In the proofs
and discussions about structures $(N,B)$, we \emph{will not make explicit} the
uses of A1 and A2 .

\begin{Definitions}{2.5}[Another betweenness property]
We define the following property of a structure $S=(N,B)$:

\begin{quote}
A8 : $\forall u,x,y,z[\neq(u,x,y,z)\wedge B(x,y,z)\wedge\lnot
A(y,z,u)\Rightarrow B(x,y,u)]$.\ 
\end{quote}
\end{Definitions}

\begin{ThmEnv}{Example and Remark}{2.6}{\upshape}
  \begin{enumerate}

\item Properties A1-A6 do not imply A8. Consider $S:=([5],B)$ where
$B$\ satisfies (only) $B^{+}(1,2,3,4)\wedge B(4,3,5)$ illustrated in Figure 3.
(There is no curve line going through 1,2,5 because $B(1,2,5)$ is not assumed
to be valid).\ Conditions A1-A6 hold but A8 does not, because we have $\lnot
A(2,3,5)\wedge B(1,2,3)$. Then, A8\ would imply $B(1,2,5)$ that is not assumed.

\item Properties A1-A5 and A8 imply A6. Assume we have $B(x,y,z)\wedge
B(x,u,z)\wedge y\neq u$.$\ $

If $\lnot A(y,z,u),$ we have $B(x,y,u)$ by A8 and then $B^{+}(x,y,u,z)$ by A5,
which implies $B(y,u,z)$ and $A(y,z,u)$ by the definitions, which contradicts
the assumption.

Hence, we have $A(y,z,u),$ that is, $B(y,z,u)$ or $B(z,y,u)$ or $B(y,u,z)$%
.\ If $B(y,z,u)$ holds, then we have \ $B^{+}(x,y,z,u)$ by A4, hence
$B(x,z,u)$ which contradicts $B(x,u,z)$ by A3.\ If $B(z,y,u)$ holds, we have
\ $B^{+}(x,u,y,z)$ by A5 (since $B(x,u,z)$ holds), that is one case of the
desired conclusion.\ The last case is $B(y,u,z),$ that yields by A5 (since
$B(x,y,z)$ holds) the other case of the conclusion.\ We will keep Property
A6\ in our axiomatization for its clarity and to shorten proofs. \hfill $\qed$%

\begin{figure}
[ptb]
\begin{center}
\includegraphics[
height=1.4313in,
width=1.254in
]%
{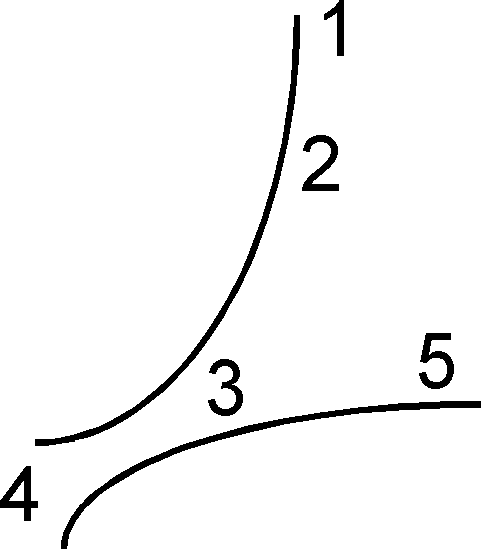}%
\caption{Structure $S$ of Example 2.6(1)}%
\end{center}
\end{figure}

\end{enumerate}
\end{ThmEnv}

\bigskip\noindent
We say that $(N,B)$ is \emph{trivial} if $B=\emptyset$. In this case,
Properties A1-A6, and A8 hold.

\begin{Lemma}{2.7} Let $S=(N,B)$ satisfy A1-A6.
  \begin{enumerate}

\item A7 implies A8.

\item If A8 holds, then the Gaifman graph\footnote{Defined in Section 1.} of $S$
is either edgeless (if $B=\emptyset$) or connected.
\end{enumerate}
\end{Lemma}

\begin{proof}
  \begin{enumerate}
  \item
    Let us assume $\neq(u,x,y,z)\wedge B(x,y,z)\wedge\lnot
A(u,y,z)$ and prove $B(x,y,u).$ There is $w$ such that $B(u,w,y)\wedge
B(y,w,z)\wedge B(u,w,z).$\ From $B(x,y,z)$, we get $B^{+}(x,y,w,z)$ by A5,
hence, $B(x,y,w)$ by the definitions. Then, from $B(y,w,u)$ and \ $B(x,y,w),$
we get $B^{+}(x,y,w,u)$ by A4,\ whence $B(x,y,u)$ by the definitions, as desired.

\item Assume that the Gaifman graph $\mathit{Gf}(S)$ is not edgeless.\ We have
$B(x,y,z)$ for some $x,y,z$. Consider $u$ different from them. Either
$A(y,z,u)\ $or $B(x,y,u)$ (or both) hold by A8.\ Hence, $u$ is in the same
connected component as $x,y,z$.
\qedhere
\end{enumerate}
\end{proof}

\begin{Definition}{2.8}[Quasi-trees and betweenness in join-trees \cite{Cou14}]
  \begin{enumerate}[label=(\alph*),beginpenalty=99]

\item A \emph{quasi-tree} is a structure $S=(N,B)$ such that $B$ is a ternary
relation on a set $N$, called the set of \emph{nodes}, that satisfies
conditions A1-A7. To avoid uninteresting special cases, we also require that
$\left\vert N\right\vert \geq$3. We say that $S$ is \emph{discrete} if
$[x,y]_{B}:=\{x,y\}\cup\{z\in N\mid B(x,z,y\}$ is finite for all $x,y$.

\item From a join-tree $J=(N,\leq),$ we define a ternary relation $B_{J}$ on $N$
by :

\begin{quote}
$B_{J}(x,y,z):\Longleftrightarrow\neq(x,y,z)\wedge([x<y\leq x\sqcup
z]\vee\lbrack z<y\leq x\sqcup z]),$
\end{quote}

called its \emph{betweenness relation}. As a definition, we use here the
observation made for rooted trees in Definition 2.1(c).\ The join $x\sqcup z$
is always defined.

\item In a quasi-tree $S=(N,B)$, we define \emph{the path} that links $x$ and
$y$ as the set $[x,y]_{B}$. It is linearly ordered with least element $x$ and
largest one $y$ in such a way that $u<v$ if and only if $x=u\wedge y=v$ or
$B(x,u,v)$\ or $B(u,v,y)$. An element may have no successor or no predecessor
(hence it may not be a path in the usual sense). However, this set is
connected in the Gaifman graph $\mathit{Gf}(S)$.
\end{enumerate}
\end{Definition}

\bigskip

Figure 4 shows a quasi-tree, where the dashed lines represent infinite paths
in the above sense.\ In such a structure, no adjacency notion is
available.\ The ternary relation of betweenness replaces it.%

\begin{figure}
[ptb]
\begin{center}
\includegraphics[
height=1.1614in,
width=3.493in ]%
{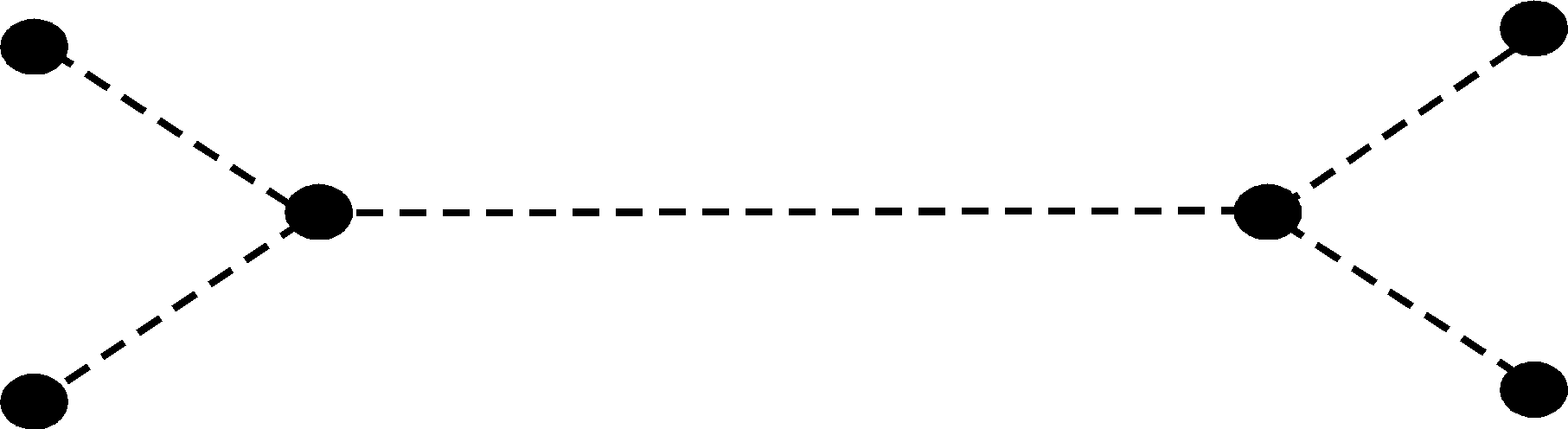}%
\caption{A quasi-tree.}%
\end{center}
\end{figure}

The following theorem is Proposition 5.6 of \cite{CouLMCS}.

\begin{Theorem}{2.9}
  \begin{enumerate}
  \item The structure $\mathit{qt}(J):=(N,B_{J})$
associated with a join-tree $J=(N,\leq)$ with at least 3 nodes is a
quasi-tree.\ Conversely, every quasi-tree $S$ is $\mathit{qt}(J)$ for some
join-tree $J$.

\item A quasi-tree is discrete if and only if it is $\mathit{qt}(J)$ for the
join-tree $J:=(N_{R},\leq_{R})$ where $R$ is a rooted tree.

  \end{enumerate}
\end{Theorem}

This theorem shows that one can specify a quasi-tree by a binary relation,
actually a partial order.\ However, this is inconvenient because choosing a
partial order breaks the symmetry.\ This motivates our use of a ternary
relation.\ Similarily, betweenness can formalize the notion of a linear order,
\emph{up to reversal}.

\subsection{Other betweenness structures}

\begin{Definition}{2.10}[Induced betweenness in a quasi-tree]
If $Q=(N,B)$ is a quasi-tree, $X\subseteq N$, we say that $Q[X]:=(X,B[X])$\ is
an \emph{induced betweenness relation in} $Q$.\ It is \emph{induced on}
$X$.
\end{Definition}

\begin{ThmEnv}{Remark and example}{2.11}{\upshape} The structure $Q[X]$ need not be a
quasi-tree because A7 does not hold for a triple $(x,y,z)\in X^{3}$ such that
$M_{Q}(x,y,z)$ is not in $X$ (cf.\ Proposition 2.3).

Figure 5\ shows a tree $T$ to the left with $N_{T}=[7]$.\ Its betweenness
relation $B_{T}$\ is expressed in a short way by the properties $B_{T}%
^{+}(1,2,7,3,4)$, $B_{T}^{+}(1,2,7,5,6)$ and $B_{T}^{+}(6,5,7,3,4).$ Let
$Q:=(N_{T},B_{T})$ and $N_{1}:=[6].$ The induced betweenness $S_{1}:=Q[N_{1}]$
is illustrated on the right, where the curve lines represent the facts
$B_{T}^{+}(1,2,3,4)$, $B_{T}^{+}(1,2,5,6)$ and $B_{T}^{+}(6,5,3,4).$ It is not
a quasi-tree because $7=M_{Q}(1,4,6)$ is not in $N_{1}$.$\square$%
\end{ThmEnv}

\begin{figure}
[ptb]
\begin{center}
\includegraphics[
height=1.6163in,
width=3.6772in ]%
{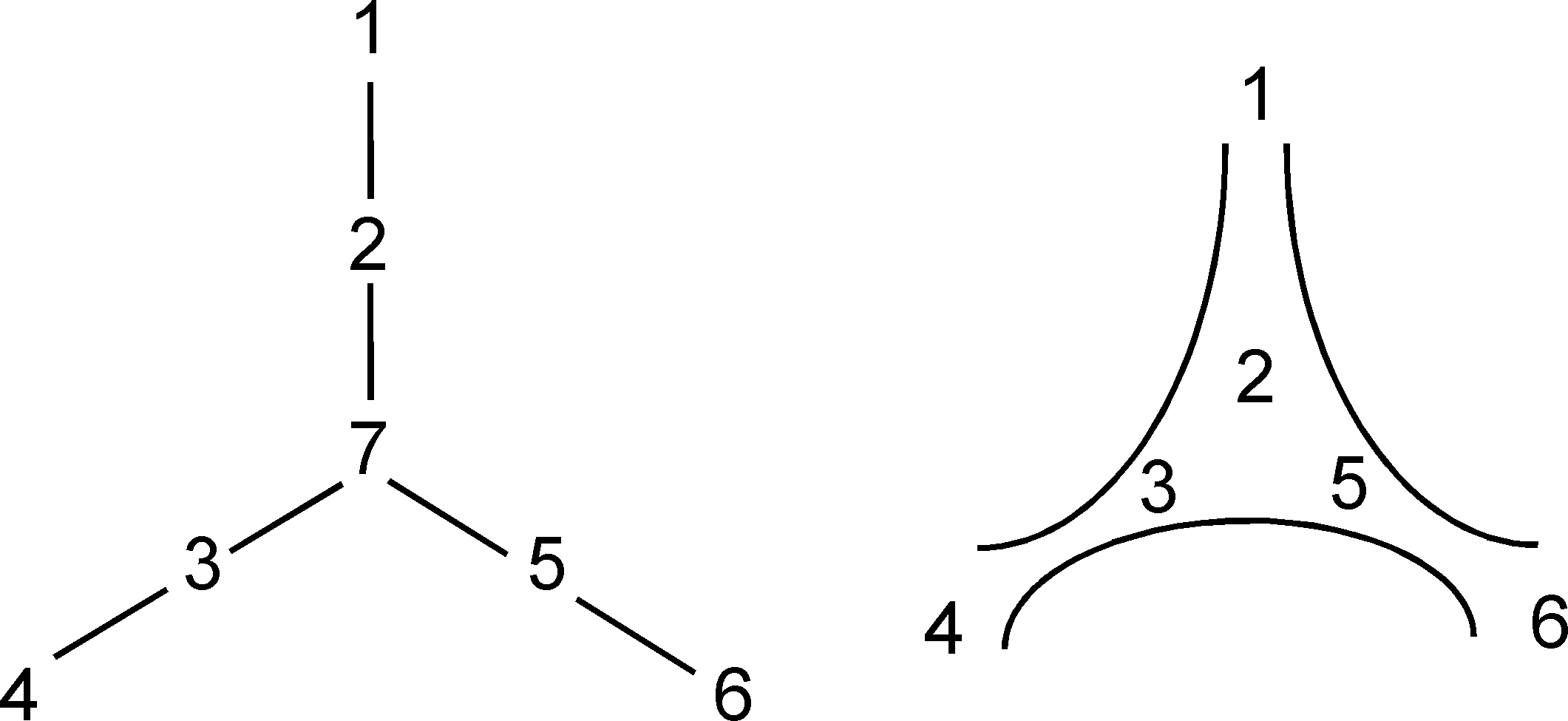}%
\caption{An induced betweenness in a quasi-tree, cf.\ Example 2.11.}%
\end{center}
\end{figure}

Our objective is to axiomatize induced betweenness relations in quasi-trees
(equivalently in join-trees), similarly as betweenness relations in
join-trees\footnote{As in \cite{Cou14}, we have defined quasi-trees
(Definition 2.8) as the ternary structures that satisfy A1-A7.\ In the sequel,
we will rather consider them as the betweenness relations of join-trees, and
A1-A7\ as their axiomatization.} are by A1-A7 in Theorem 2.9(1).

\bigskip

\begin{Proposition}{2.12} An induced betweenness relation in a quasi-tree
satisfies properties A1-A6 and A8.
\end{Proposition}

\begin{proof} The FO sentences expressing A1-A6 and A8 are universal, that
is, are of the form $\forall x,y,...,z.\varphi(x,y,...,z)$ where $\varphi$\ is
quantifier-free.\ The validity of such sentences is preserved under taking
induced substructures (we are dealing with relational structures). The result
follows from Theorem 2.9\ and Lemma 2.7(1) showing that a quasi-tree satisfies
A8.
\end{proof}

\bigskip

Our objective is to prove that a ternary relation is an induced betweenness in
a quasi-tree if and only if it satisfies Properties A1-A6 and A8. Our proof
will use O-trees.

\bigskip

Figure 6 illustrates Property A8 which says: $B(x,y,z)\wedge\lnot
A(y,z,u)\Rightarrow B(x,y,u).$ The white circle between $y$ and $z$ represents
the node $M_{Q}(y,z,u)$ of a quasi-tree $Q$ that has been deleted, so that
Property A7\ does not hold in the structure $Q[N-\{M_{Q}(y,z,u)\}]$.%

\begin{figure}
[ptb]
\begin{center}
\includegraphics[
width=2.4743in
]%
{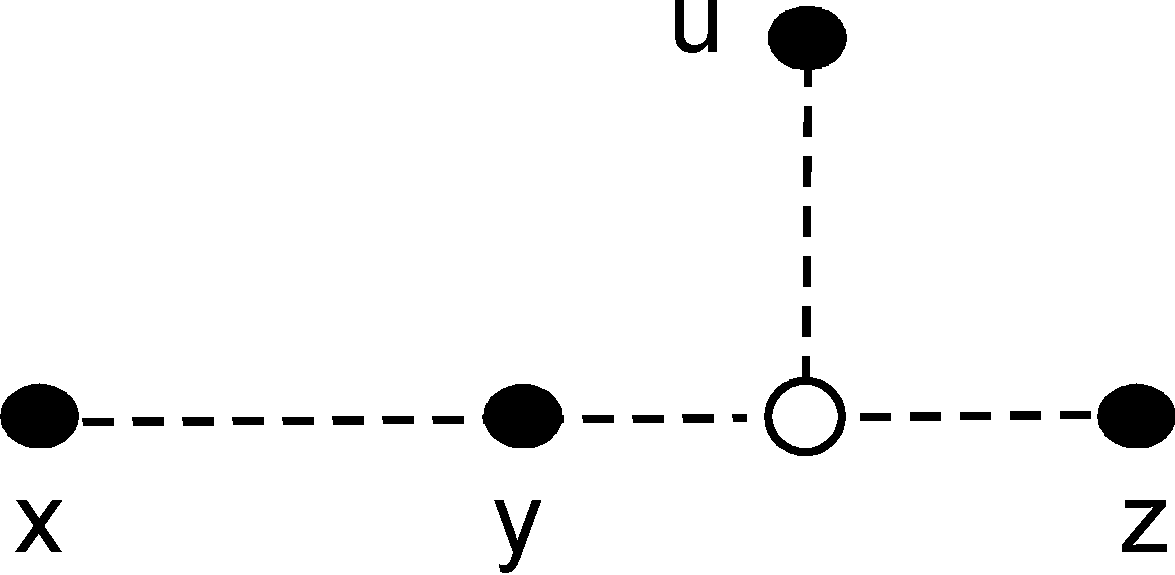}%
\caption{Illustration of Property A8.}%
\end{center}
\end{figure}

\bigskip

\begin{Definition}{2.13 }[Betweenness in O-forests]
  \begin{enumerate}[label=(\alph*),beginpenalty=99]

\item The \emph{betweenness relation} of an O-forest $F=(N,\leq)$ is the ternary
relation $B_{F}$ on $N$ such that :

\begin{quote}
$B_{F}(x,y,z):\Longleftrightarrow\neq(x,y,z)\wedge\lbrack(x<y\leq x\sqcup
z)\vee(z<y\leq x\sqcup z)].$
\end{quote}

The validity of the right handside needs that $x\sqcup z$\ be defined.\ 

\item If $F=(N,\leq)$ is an O-forest and $X\subseteq N$, then $B_{F}[X]$ is an
\emph{induced betweenness relation} in $F$ and $(X,B_{F}[X])$ is an
\emph{induced betweenness structure.}

The difference with Definition 2.8(b)\ is that if $x$ and $z$ have no least
upper-bound (\emph{i.e.}, if $x\sqcup z$\ is undefined, which implies that $x$
and $z$ are incomparable, denoted by $x\bot z$), then $B_{F}$\ contains no
triple of the form $(x,y,z).$\ 

If $F$ is a finite O-tree, it is a join-tree and thus, $(N,B_{F})$ is a
quasi-tree.$\square$
\end{enumerate}
\end{Definition}

\bigskip

We have four classes of betweenness structures $S=(N,B)$ : quasi-trees,
induced betweenness structures in quasi-trees, betweenness and induced
betweenness structures in O-forests, denoted respectively by \textbf{QT},
\textbf{IBQT}, \textbf{BO}\ and \textbf{IBO}.

\begin{Remarks}{2.14}
  \begin{enumerate}
\item Let $T$ be a tree and $X$ a set of leaves. The
induced betweenness relation $B_{T}[X]$ is trivial.

\item The Gaifman graph of a betweenness structure $S$ is connected in the
following cases : $S\in$\textbf{IBQT} and is not trivial or $S$ is the
betweenness structure of an O-tree.\ It may be not connected in the other cases.

\item If $S$ is an induced betweenness in an O-forest consisting of several
disjoint O-trees, then two nodes in the different O-trees cannot belong to a
same triple.\ It follows that they cannot be linked by a path in the graph
$\mathit{Gf}(S)$.\ Hence, a structure $(N,B)$ is the betweenness of an
O-forest, or an induced betweenness in an O-forest if and only if each of its
connected components is so in an O-tree. We will only consider betweenness of
O-trees (class \textbf{BO}) and induced betweenness in O-trees (class
\textbf{IBO}).
\end{enumerate}
\end{Remarks}

\begin{Proposition}{2.15} We have the following proper inclusions:
\begin{center}
\textbf{QT} $\subset$ \textbf{IBQT} $\cap$ \textbf{BO}, \textbf{IBQT}$\ \subset
$\textbf{ IBO} and \textbf{BO} $\subset$ \textbf{IBO}.
\end{center}
The classes \textbf{IBQT} and \textbf{BO} are incomparable. For finite
structures, we have \textbf{QT} $=$ \textbf{BO}.
\end{Proposition}

\bigskip

These inclusions are illustrated in Figure 7. Structures $S_{1},S_{2},S_{4}$
and $S_{5}$\ witnessing proper inclusions are described in the proof.

\begin{proof}All inclusions are clear from the definitions. We give
examples to prove that the inclusions are proper. We recall that
$S[X]:=(X,B[X])$ if $S=(N,B)$ and $X\subseteq N$.

\begin{enumerate}

\item The structure $S_{1}$ of Example 2.11, shown in the right part of Figure
5, is in \textbf{IBQT} but not in \textbf{QT}. It is not in \textbf{BO}%
\ either, because otherwise, it would be a quasi-tree as it is finite.%

\begin{figure}
[ptb]
\begin{center}
\includegraphics[
height=2.0833in,
width=3.1116in
]%
{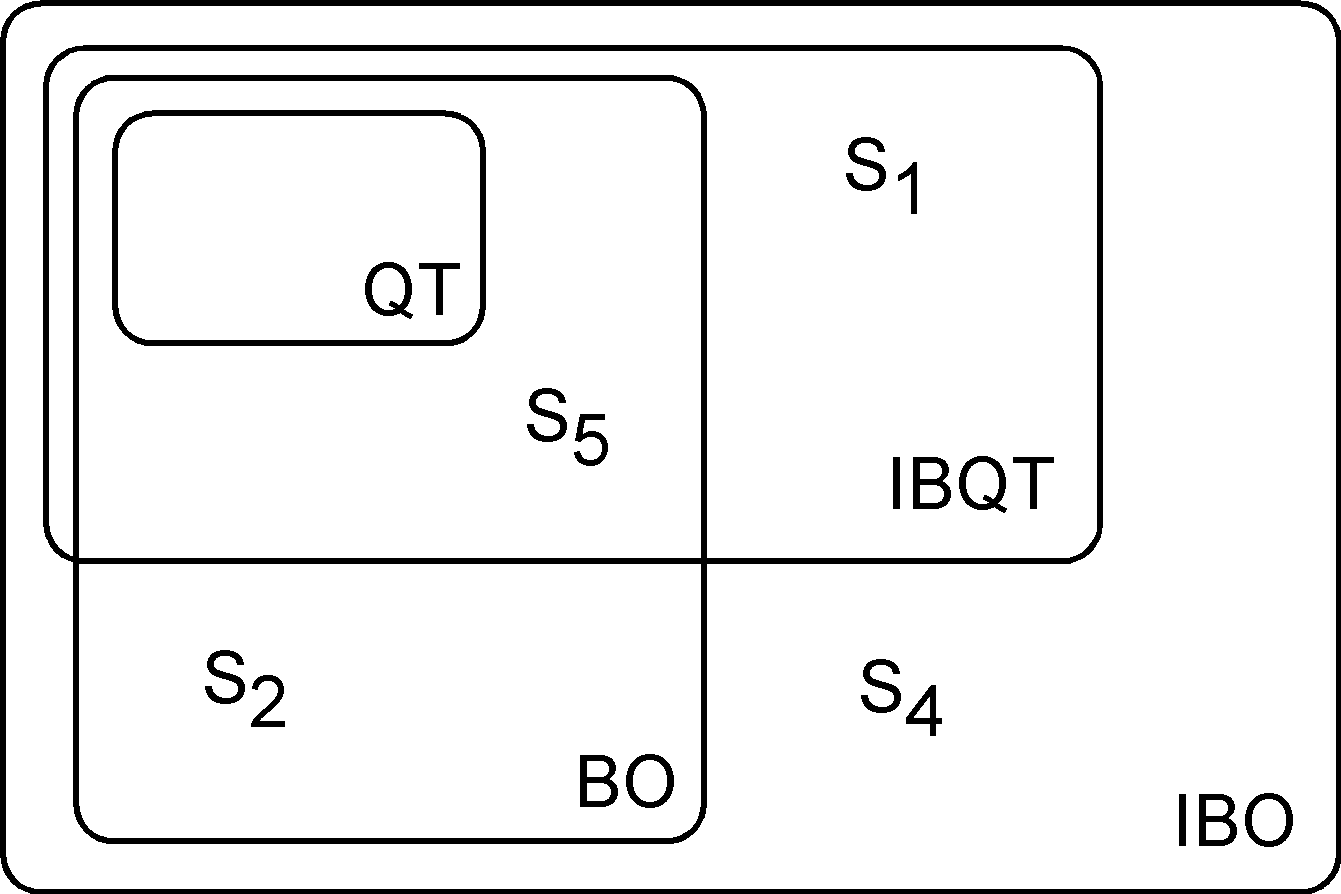}%
\caption{Proper inclusions of classes proved in Proposition 2.15.}%
\end{center}
\end{figure}

\begin{figure}[ptb]
\begin{center}
\includegraphics[
height=1.5489in,
width=0.6832in
]%
{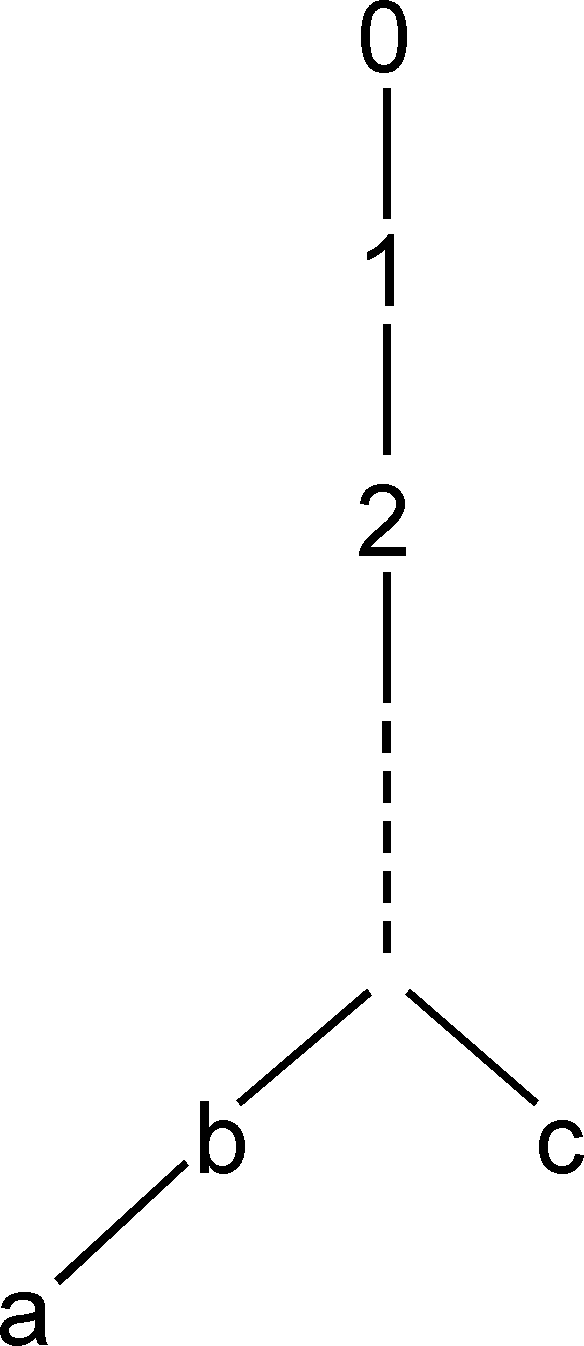}%
\caption{The O-tree $T_{2}$\ used in the proof of Proposition 2.15, Parts (2)
and (4).}%
\end{center}
\end{figure}

\item We consider $N_{2}:=\mathbb{N}\cup\{a,b,c\}$ and the O-tree $T_{2}%
:=(N_{2},\preceq)$ in Figure 8\ such that $a\prec b\prec i\prec j$ and $c\prec
i\prec j$ for all $i,j$ in $\mathbb{N}$ such that $j<i$.\ Its betweenness
structure $S_{2}:=(N_{2},B_{2})$ is described by the properties $B_{2}%
^{+}(a,b,i,j,k)$ and $B_{2}^{+}(c,i,j,k)$ for all $i,j,k$ in $\mathbb{N}$ such
that $k<j<i$. Since $b$ and $c$ have no least upper-bound in $T_{2}$, we do
not have $B_{T_{2}}(a,b,c)$. Hence, $S_{2}$ is in \textbf{BO} but not in
\textbf{IBQT}, as it does not satisfy A8: we have\ $\lnot A_{T_{2}%
}(0,b,c)\wedge B_{T_{2}}(a,b,0)$ but not $B_{T_{2}}(a,b,c)$.\ The classes
\textbf{IBQT} and \textbf{BO}\ are incomparable.

If we take $c$ as new root, we obtain a join-tree $U=(N_{2},\preceq^{\prime})$
where $a\prec^{\prime}b\prec^{\prime}c$ and $0\prec^{\prime}1\prec^{\prime
}2...\prec^{\prime}i\prec^{\prime}...\prec^{\prime}c.$ and $\{a,b\}\bot
^{\prime}\mathbb{N}$.\ Clearly $B_{U}\neq B_{T_{2}}$.\ 

Hence, betweenness in O-trees depends on some kind of orientation, that can be
specified either by a root or by an upwards closed line (cf. the notion of
structuring in Definition 3.27 below).\ To the opposite, in the case of
quasi-trees and induced betweenness in quasi-trees, any node can be taken as
root in the constructions of the relevant join-trees (cf.\ \cite{CouLMCS} for
quasi-trees, and the proof of Theorem 3.1\ and Remark 3.4(d) for induced
betweenness in quasi-trees).

\bigskip

\item To prove that the inclusion of \textbf{BO} in \textbf{IBO} is proper, we
consider $S_{3}:=(N_{3},B_{T_{3}}),N_{3}:=\{a,b,c,d\}\cup\mathbb{Q}$\ and the
O-tree $T_{3}:=(N_{3},\prec)$ ordered such that:

\begin{quote}
- $a\prec b\prec i\prec j$ and $d\prec c\prec i\prec j$ for all $i,j\in
\mathbb{Q}$ such that $\sqrt{2}<i<j,$\ and

- $i\prec j$ if $i,j\in\mathbb{Q}$, $i<j.$
\end{quote}

It is shown in Figure 9(\emph{a}). The upper dotted line is isomorphic to
$\mathbb{Q}_{>}(\sqrt{2}):=\{i\in\mathbb{Q\mid}i>\sqrt{2}\}$ and the lower one
is isomorphic to 
$\mathbb{Q}_{<}(\sqrt{2})
:=\mathbb{Q}-\mathbb{Q}_{>}(\sqrt{2})$.

We let then $S_{4}:=S_{3}[\{a,b,c,d,1,2,3\}]$ with corresponding O-tree
$T_{4}$ (Figure 9(\emph{b})). The structure $S_{4}$ is in \textbf{IBO} but not
in \textbf{BO}.\ Otherwise, as it is finite, it would be a quasi-tree.\ But
$S_{4}$ does not satisfy A8 : we have $B_{T_{3}}(a,b,3)\wedge\lnot A_{T_{3}%
}(b,c,3)$ but $(a,b,c)\notin B_{T_{3}}$.\ For this reason, $S_{4}$ is not in
\textbf{IBQT} either.

Note that $S_{4}$ in \textbf{IBO}\ is finite but is not the induced
betweenness relation of a \emph{finite} O-tree.\ Otherwise, it would be in
\textbf{IBQT} because a finite O-tree is a join-tree.%

\begin{figure}
[ptb]
\begin{center}
\includegraphics[
height=2.0833in,
width=2.6515in
]%
{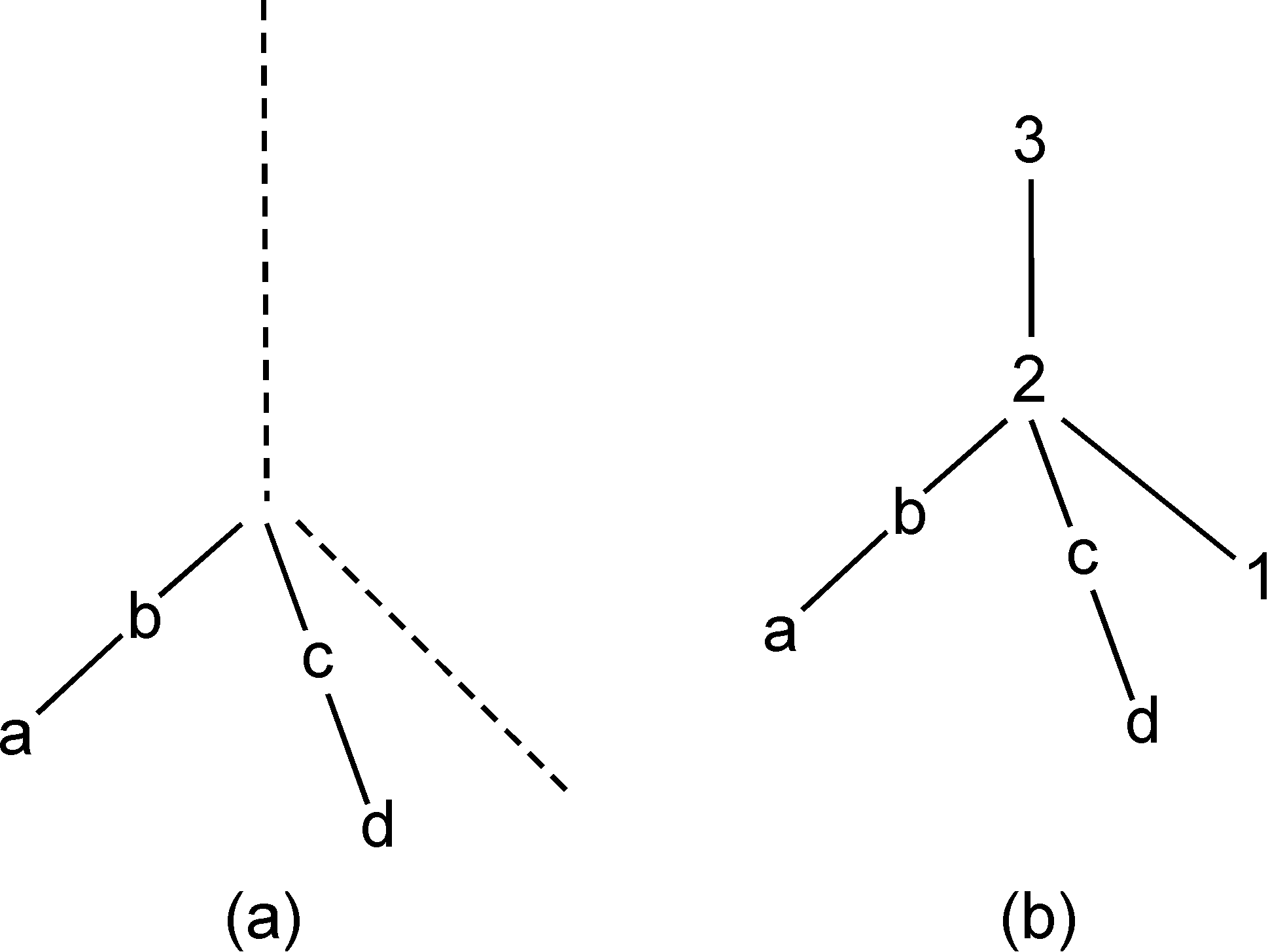}%
\caption{Part (\emph{a}) shows $T_{3}$ and (\emph{b}) shows $T_{4}$ of the
proof of Proposition 2.15, Part (3), and Example (3.6).}%
\end{center}
\end{figure}

\item Let $T_{5}$ be the O-tree $T_{2}[N_{5}]$\ where $N_{5}:=\mathbb{N}%
\cup\{b,c\}$ and $S_{5}:=(N_{5},B_{T_{5}}).$\ (Figure 8 shows $T_{2}$). Then
$S_{5}$ is in \textbf{BO}, and also in \textbf{IBQT}\ : just add to $T_{5}$ a
least upper-bound $m$ for $b$ and $c$ such that $m<\mathbb{N}$, one obtains a
join-tree.\ It is not a quasi-tree because A7 does not hold for the triple
$(b,c,3)$ (relative to $B_{5}$). Hence, we have \textbf{QT} $\subset$
\textbf{IBQT}$\cap$\textbf{BO}.\ 

Note that $S_{2}$ is not in \textbf{IBQT} but its induced substructure
$S_{5}\ $is.
\qedhere
\end{enumerate}
\end{proof}

Figure 7\ shows how these examples are located in the different classes of
betweenness relations. The structures\ $S_{1}$ and $S_{4}$ are finite, $S_{2}$
and $S_{5}$\ are infinite, which is necessary because the finite structures in
\textbf{BO} and \textbf{QT}\ are the same.\

\begin{Remark}{2.16} An alternative betweenness relation for an O-forest
$F=(N,\leq)$ could be defined by $B_{F}^{\prime}:=B_{\widehat{F}}[N]$ (see
Definition 1.3 for $\widehat{F}$).\ If $F$ is an O-tree, we have $(x,y,z)\in
B_{F}^{\prime}$ if and only if $\neq(x,y,z)$ and, either $x<y\leq m\geq z$ or
$z<y\leq m\geq x$ for some $m$ that need not be the join in $F$ of $x$ and
$z$.\ As $(N,B_{F}^{\prime})$ is an induced betweenness in a join-tree, this
definition does not bring anything new.
\end{Remark}

\section{Axiomatizations}

\subsection{First-order axiomatizations}

Our first main result is Theorem 3.1\ that provides a\ first-order
axiomatization of the class \textbf{IBQT}, among countable (finite or
countably infinite) structures.

All our constructions are relative to countable structures. The letter $B$
will always denote ternary relations. Writing $(x,y,z)\in B$ is equivalent to
stating that $B(x,y,z)$ holds.

\subsubsection{Induced betweenness in quasi-trees}

\begin{Theorem}{3.1} The class \textbf{IBQT} is axiomatized by the
first-order properties A1-A6 and A8.
\end{Theorem}

\bigskip

With $S=(N,B)$ and $r\in N,$ we associate the binary relation $\leq_{r}$ on
$N$\ such that $x\leq_{r}y:\Longleftrightarrow x=y\vee y=r\vee B(x,y,r).$

\begin{Lemma}{3.2} Let $S=(N,B)$ satisfy Axioms A1-A6 and $r\in N.$ Then :

(1)\textbf{ }$T(S,r):=(N,\leq_{r})$ is an O-tree,

(2) if $x<_{r}y<_{r}z$, then $(x,y,z)\in B$,

(3) if $(x,y,z)\in B$, $x<_{r}y$ and $z<_{r}y$, then $y=x\sqcup_{r}z$,

(4) if $x<_{r}w<_{r}y$ and $z<_{r}w$, then $(x,y,z)\notin B$.

\end{Lemma}

\begin{proof}
  \begin{enumerate}
  \item
    The relation $\leq_{r}$ is a partial order: antisymmetry
follows from A3 and transitivity from A5.\ The node $r$ is its largest
element. Axiom A6\ implies that, for any $x\in N$, the set $L_{\geq_{r}}(x)$
is linearly ordered. Hence, $T(S,r):=(N,\leq_{r})$ is an O-tree with root $r$.

\item This is clear if $z=r$ and follows from A5 otherwise.

\item Assume that $B(x,y,z)$ holds $x<_{r}y$ and $z<_{r}y$.\ We cannot have
$x<_{r}z$\ or $z<_{r}x$ because otherwise, we have by (2) $B(x,z,y)$ or
$B(z,x,y),$ contradicting $B(x,y,z)$ by A3.\ 

Assume for a contradiction, that $x<_{r}w<_{r}y$ and $z<_{r}w<_{r}y.$ Then, by
(2), we have $B(x,w,y)$ and $B(z,w,y).$ We get $B^{+}(x,w,y,z)$ by A5, which
gives $B(w,y,z)$, contradicting A3 since we have $B(z,w,y).$

\item From $x<_{r}w<_{r}y$ we get $B(x,w,y)$ by (2).\ With $B(x,y,z)$, A5 gives
$B^{+}(x,w,y,z)$, whence $B(w,y,z)$ by the definitions. From $z<_{r}w<_{r}y,$
we get $B(z,w,y)$ by (2), which is incompatible with $B(w,y,z)$ by
A3.
\qedhere
\end{enumerate}
\end{proof}

\begin{Lemma}{3.3} Let $S:=(N,B)$ satisfy A1-A6 and A8, and $r\in N$.
  \begin{enumerate}

\item Let $x$ and $y$ be incomparable with respect to $\leq_{r}$.\ If $z<_{r}y$,
then $(x,y,z)\in B$.

\item If $(x,y,z)\in B$, then $x<_{r}y$ or $z<_{r}y$.

\item We have $B\subseteq B_{T(S,r)}$ if $N$ is finite.
\end{enumerate}
\end{Lemma}

\begin{proof} In this proof, $<$, $\leq$\ and $\sqcup$ will denote
$<_{r},\leq_{r}$\ and $\sqcup_{r}.$

\begin{enumerate}
\item Let $x$ and $y$ are incomparable and $z<y$. The root $r$ is not any of
$x,y,z.$ If $B(x,r,y)$ holds, then, from $B(r,y,z)$ we have $B^{+}(x,r,y,z)$
by A4, whence $B(x,y,z)$. Otherwise, $A(x,y,r)$ does not hold, and as we have
$B(z,y,r)$, we get by A8 $B(z,y,x),$ \emph{i.e.},$B(x,y,z)$.

\item Let $(x,y,z)\in B$.\ We have several cases.

\emph{Case 1} : $x$ or $z$ is $r$.\ We get respectively $z<y$ or $x<y$.

\emph{Case 2:} $y=r$.\ Then $z<y$ and $x<y$.

\emph{Case 3:} $\neq(x,y,z,r)$ and $\lnot(A(y,z,r)$. We have $B(x,y,r)$ by A8,
hence, $x<y$.

\emph{Case 4:} $\neq(x,y,z,r)$ and $A(y,z,r)$ holds.\ If $B(y,z,r),$ we have
$B^{+}(x,y,z,r)$ by A4, hence, $x<y$. If $B(y,r,z),$ we have $B^{+}(x,y,r,z)$
by A5, hence, $x<y$. If $B(r,y,z),$ we have $z<y$.

\item Let $(x,y,z)\in B$. We have $x<y$ or $z<y$ by (2). As $N$ is finite, $x$
and $z$ have a join $x\sqcup z$\ in the rooted tree $T(S,r)$.\ \ Assume $x<y.$
If $y\leq x\sqcup z$, we have $(x,y,z)\in B_{T(S,r)}$ by Definition 2.13, as
desired.\ Otherwise, $x\sqcup z<y$, hence $x\leq x\sqcup z<y$. We cannot have
$x\sqcup z=z$ because then $(x,z,y)\in B$ by lemma 3.2(2), contradicting
$(x,y,z)\in B$ (by A3). Hence, $x\bot z$\ and then $x<x\sqcup z<y$ and
$z<x\sqcup z<y$. Lemma 3.2(3) yields $y=x\sqcup z$, contradicting the
assumption. Hence, we have $(x,y,z)\in B_{T(S,r)}$. The case $z<y$ is
similar.
\qedhere
\end{enumerate}
\end{proof}

\begin{Examples}{3.4}
  \begin{enumerate}[label=(\alph*)]

    \item In statement (3) above, we may have a proper
inclusion.\ Consider\ $S_{6}$\ defined as $(N_{6},B_{6})$ with $N_{6}%
:=\{0,1,2,a,c\}$, $B_{6}^{+}(0,1,2,a)$, $B_{6}^{+}(0,1,2,c)$ and $r:=0$.\ Then
$T(S_{6},0)=T[N_{6}]$ where $T$ is the join-tree at the left of Figure
1.\ \ We have $(a,2,c)$ in $B_{T(S_{6},0)}$ but not in $B_{6}$.

\item The inclusion $B\subseteq B_{T(S,r)}$ may be false if $S$ is
infinite.\ \ Consider $S_{7}=(\mathbb{N}\cup\{a,b,c\},B_{7})$ defined from
$S_{2}=(\mathbb{N}\cup\{a,b,c\},B_{2})$\ in the proof of Proposition 2.15 (see
Figure 8), where $B_{7}:=B_{2}\cup\{(a,b,c),(c,b,a)\}$. Then $T(S_{7}%
,0)=T_{2}$ of this proof, but $(a,b,c)\notin B_{T(S_{7},0)}.$

\item We give an example showing how we will prove Theorem 3.1. Let
$S_{8}:=(N_{8},B_{8})$ such that $N_{8}:=\{0,a,b,c,d,e,f,g,h\}$ and $B_{8}$ is
defined by the following properties :

\begin{quote}
(i) $B_{8}^{+}(b,a,c,d)$, $B_{8}^{+}(f,e,g,h)$,

(ii) $B_{8}^{+}(b,a,0,e,f)$, $B_{8}^{+}(d,c,0,e,f)$, $B_{8}^{+}(b,a,0,g,h)$,
$B_{8}^{+}(d,c,0,g,h)$.\ 
\end{quote}

Figure 10(\emph{a}) shows this structure drawn with the conventions of Figures
3 and 5\ (right\ part).\ It shows properties $B_{8}(b,a,0)$, $B_{8}(d,c,0)$,
$B_{8}(e,f,0)$ and $B_{8}(g,h,0)$.\ It does not show the four conditions of
type (ii) for the purpose of clarity. We have neither $B_{8}^{+}(b,a,0,c,d)$
nor $B_{8}^{+}(e,f,0,g,h)$.

By adding new nodes $1$ and 2 to $T(S_{8},0)$ such that $a<1<0,c<1<0,e<2<0$
and $g<2<0,$ we get the rooted tree $T_{8}$ of Figure 10(\emph{b}).\ \ Then
$B_{7}=B_{T_{8}}[N_{7}]$, hence, is in \textbf{IBQT}.

The proof of Theorem 3.1 will consist in adding new elements to trees $T(S,r)$
for such cases.

\item Let $S=(N,B)$ satisfies A1-A7\ (and thus A8 by Lemma 2.7).\ For each $r\in
N$, the O-tree $T(S,r)$ is a join-tree\ and $B=B_{T(S,r)}$ by Lemma 14 of
\cite{Cou14} and Proposition 5.6 of~\cite{CouLMCS}.
\end{enumerate}
\end{Examples}

\begin{figure}
[ptb]
\begin{center}
\includegraphics[
height=1.7417in,
width=3.1955in
]%
{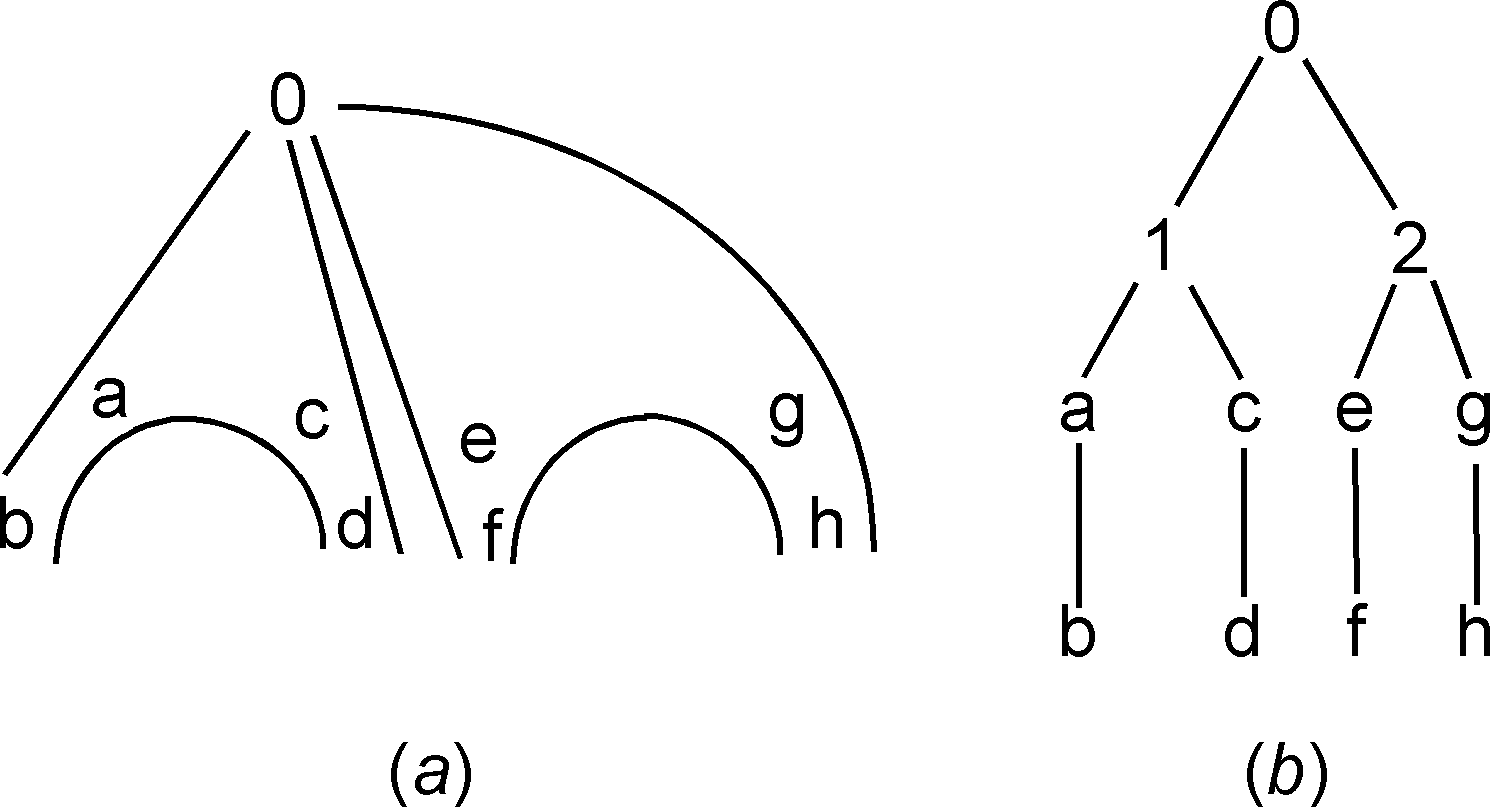}%
\caption{(\emph{a}) shows $S_{8}$ and (\emph{b}) shows $T_{8}$ of Example
3.4(c).\ See also Remark 3.16.}%
\end{center}
\end{figure}

\begin{Definitions}{3.5}[Directions in O-trees]
In a rooted tree $T$, each node that is not a leaf has sons $u_{1}%
,...,u_{p},...$ from which are issued subtrees whose sets of nodes are the
sets $N_{\leq}(u_{i})$. In O-trees directions replace such subtrees that need
not exist in O-trees because a node may have no son (for example node 2 of the
tree $T_{3}$ of Figure 9(a)).
\begin{enumerate}[label=(\alph*)]
\item Let $T=(N,\leq)$ be an O-tree\footnote{Or an O-forest, but we will use the
notion of direction only for O-trees.}. Let $L\subseteq N$\ be linearly
ordered and upwards closed\footnote{In particular, if $X\neq\emptyset$, the
set $L_{>}(X):=\{y\in N\mid y>X\}$ is linearly ordered and upwards closed.}.
It is a line according to Definition 1.1.\ Two nodes $x$ and $y$ in
$N_{<}(L):=\{w\in N\mid w<L\}$ are \emph{in the same direction w.r.t.} $L$ if
$x\leq u$ and $y\leq u$ for some $u\in N_{<}(L)$. This is an equivalence
relation that we denote by $\sim_{L}$. Clearly, $x\leq y$ implies $x\sim_{L}%
y$. Each equivalence class is called a\emph{ direction relative to} $L$.\ We
denote by $Dir_{L}(x)$ the direction relative to $L$\ that contains $x$ such
that $x<L$. The O-tree is \emph{binary} if each such line $L$\ has at most two directions.

\item Let $S=(N,B)$ satisfy Axioms A1-A6 (but not necessarily A8) and $r$ be any
node taken as root. Then $T(S,r).:=(N,\leq_{r})$ is an O-tree. If $x$ and $y$
in $N$\ are incomparable, the set $L_{>}(x,y)$ is an upwards closed line that
contains $r$, but not $x$ and $y$. We denote by $\mathfrak{L}$ the countable
set of such lines ($\mathfrak{L}$ $\subset\mathcal{K}$ defined in Definition 1.3).

\item For $L=L_{>}(x,y)\in\mathfrak{L}$, we denote by $\mathfrak{D}(L)$ the set
of directions relative to $L$. There are at least two different ones,
$Dir_{L}(x)$ and $Dir_{L}(x)$ . We have $L=L_{>}(N_{<}(L))$ and $N_{<}(L)$ is
the disjoint union of the directions in $\mathfrak{D}(L).$
\end{enumerate}
\end{Definitions}

\begin{Examples}{3.6}
  \begin{enumerate}

\item In the O-tree $T_{3}$ of Figure 9(a) (defined for
proving Proposition 2.15(3)), $L_{>}(b,c)$ is the set $\mathbb{Q}_{>}(\sqrt
{2})$ of rational numbers larger that $\sqrt{2}$ and the associated three
directions are $\{a,b\}$,$\{c,d\}$ and $\mathbb{Q}_{<}(\sqrt{2}):=\mathbb{Q}%
-\mathbb{Q}_{>}(\sqrt{2})$.

\item We consider again the join-tree $T:=(Seq_{+}(\mathbb{Q}),\preceq)$ of
Example 1.2(4) defined by Fra\"{\i}ss\'{e}. The partial ordered $\preceq$ is
defined as follows :

$\qquad(x_{1},...,x_{n})\preceq(y_{1},...,y_{m})$ if and only if

$\qquad n\geq m$, $(x_{1},...,x_{m-1})=(y_{1},...,y_{m-1})$ and $x_{m}\leq
y_{m}$.

The join of two incomparable nodes $\boldsymbol{x}:=(x_{1},...,x_{n})$ and
$\boldsymbol{y}:=(y_{1},...,y_{m})$ is $\boldsymbol{z}=(z_{1},...,z_{p})$ if
we have $p\leq n$, $\boldsymbol{x}=(z_{1},...,z_{p-1},x_{p},...,x_{n}%
),z_{p}>x_{p}$ and $\boldsymbol{y}=(z_{1},...,z_{p-1},z_{p},$ $y_{p+1}%
,...,y_{m})$. Then, the directions relative to $L=L_{>}(\boldsymbol{x}%
,\boldsymbol{y})=L_{>}(\boldsymbol{z})$ are :

\begin{quote}
$Dir_{L}(\boldsymbol{x})=\{(z_{1},...,z_{p-1},u_{p},u_{p+1},...,u_{q})\mid
u_{p},...,u_{q}\in\mathbb{Q},u_{p}\leq z_{p}\}$ and

$Dir_{L}(\boldsymbol{y})=\{(z_{1},...,z_{p},u_{p+1},...,u_{q})\mid
u_{p+1},...,u_{q}\in\mathbb{Q}\}$.
\end{quote}

This join-tree is binary.
\end{enumerate}
\end{Examples}

\begin{Lemma}{3.7} Let $S=(N,B),r$ and  $<_{r}$ be as in Lemma 3.2, and $L\in\mathfrak{L}$. Let $u,v\in D$ for some
direction $D$ in $\mathfrak{D}(L)$ (see Definition 3.5).\ Let $m\in L$ and
$w\in N$. Then $B(u,m,w)$ if and only if $B(v,m,w)$.
\end{Lemma}

\begin{proof} We will denote $\leq_{r}$ by $\leq$.\ Related notations are
$<$%
, $\sqcup$ and $\bot.$

We have $\{u,v\}<a<m$ for some $a\in D$, hence $B(u,a,m)$\ and $B(v,a,m)$ by
Lemma 3.2(2).\ If $B(u,m,w)$ we have $B^{+}(u,a,m,w)$ by A5, hence
$B(a,m,w)$.\ From this fact and $B(v,a,m),$ we get $B^{+}(v,a,m,w)$ by A4,
hence $B(v,m,w)$.\ $\square$

\bigskip

It follows that we can define, for $D,D^{\prime}\in\mathfrak{D}(L)$ and $m\in
L$:

\begin{quote}
$B(D,m,D^{\prime}):\Longleftrightarrow B(u,m,w)$ for some $u\in D$ and $w\in
D^{\prime}.$
\end{quote}

By Lemma 3.7, we have (directions are not empty by definition) :

\begin{quote}
$B(D,m,D^{\prime})\Longleftrightarrow B(u,m,w)$ for all $u\in D$ and $w\in
D^{\prime}.$
\end{quote}

In particular, we do not have $B(D,m,D).$
\end{proof}

\begin{Lemma}{3.8} Let $S=(N,B)$ satisfy A1-A6 and A8.\ Let $r\in N$,
$T(S,r):=(N,\leq_{r})$ and $m\in L\in\mathfrak{L}$. The binary relation $\lnot
B(D,m,D^{\prime})$ for $D,D^{\prime}\in\mathfrak{D}(L)$ is an equivalence relation.
\end{Lemma}

\begin{proof} Reflexivity and symmetry are clear. Assume that we have
$\lnot B(D,m,$ $D^{\prime})$\ and $\lnot B(D^{\prime},m,D^{\prime\prime})$ for
distinct directions $D,D^{\prime},D^{\prime\prime}$.\ Hence, by Lemma 3.7, we
have $\lnot B(u,m,v)$ and $\lnot B(v,m,w)$ for some $u,v,w$ respectively in
$D,D^{\prime},D^{\prime\prime}$.\ For a contradiction, we assume that
$B(u,m,w)$\ holds.

We have $\lnot B(u,m,v)$ as observed above. If $B(m,u,v)$ we have $m<u$ or
$v<u$ by Lemma 3.3(2), but we know that $u<m$ and\ $u\bot v.$ Hence, we
have\ $\lnot B(m,u,v)$ and similarly, $\lnot B(m,v,u)$. We have $\lnot
A(m,u,v)\wedge B(w,m,u)$, and A8 gives $B(w,m,v)$, contradicting an assumption.

Hence, $\lnot B(u,m,w)$ holds for all $u,w$ respectively in $D,D^{\prime
\prime}$ and we have $\lnot B(D,m,D^{\prime\prime})$.
\end{proof}

\begin{Definition}{3.9}[Independent directions]
Let $S=(N,B)$ satisfy A1-A6 and A8, $r\in N$, and $m\in L\in\mathfrak{L}$,
relative to $T(S,r):=(N,\leq_{r}).$

\begin{enumerate}[label=(\alph*)]

\item If $D$, $D^{\prime}\in\mathfrak{D}(L)$, we define $D\approx_{L}D^{\prime}$
if $B(D,m,D^{\prime})$ holds for no $m\in L$. By Lemma 3.2(2),
$B(D,m,D^{\prime})$ can hold only if $m$ is the smallest element of $L$.
Hence, $D\approx_{L}D^{\prime}$ holds if and only if, either $L$ has no
smallest element or $B(D,\min(L),D^{\prime})$ does not hold. Hence, by Lemma
3.8, $\approx_{L}$ is an equivalence relation\footnote{Not to be confused with
$\sim_{L}$\ of Definition 3.5(a), whose classes are the directions relative to
$L$.}. We say that $D$ and $D^{\prime}$ \ are \emph{independent} if
$D\approx_{L}D^{\prime}$ because they are not "linked" through any $m\in L$
such that $B(D,m,D^{\prime})$ holds.

\item For each $D\in\mathfrak{D}(L)$, we denote by $D_{\approx}$ the union of
the directions that are $\approx_{L}$-equivalent to $D$. The sets $D_{\approx
}$ form a partition of $N_{<}(L)$. We define $\mathcal{C}:=\mathcal{C}%
_{1}\uplus\mathcal{C}_{2}$ as the set of downward closed subsets of $N$ such
that:

\begin{quote}
$\mathcal{C}_{1}:=\{N_{\leq}(x)\mid x\in N\}$ (in particular $N=N_{\leq}(r)$) and

$\mathcal{C}_{2}:=\{D_{\approx}\mid D\in\mathfrak{D}(L)$, $L\in\mathfrak{L}$
and $D_{\approx}$ is the union of at least two directions\}.
\end{quote}
\end{enumerate}
\end{Definition}

\bigskip

We pause with technicalities for explaining how we will use these definitions
and lemmas.

Consider the structure $S_{8}=(N_{8},B_{8})$ of Figure 10(a), used in Example
3.4(c).\ The rooted tree $T(S_{8},0)$ is $T_{8}$ shown in Figure 10(b) minus
the nodes 1 and 2.\ There are in $T(S_{8},0)$ four directions relative to
$L:=\{0\}=L_{>}(a,c)=L_{>}(a,e)=L_{>}(g,c)$.\ They are $D(a)=\{a,b\}$, the
direction of $a$, and similarly, $D(c)=\{c,d\},D(e)=\{e,f\}$ and
$D(g)=\{g,h\}$.\ 

Let $B:=B_{8}$. We have $B(D(a),0,D(e))$, $B(D(c),0,D(e))$, $B(D(a),0,D(g))$
and $B(D(c),0,D(g))$ because $B$\ contains the triples $(a,0,e),$
$(c,0,e),(a,0,g)$ and $(c,0,g)$ (by clauses (ii) in Example 3.4(c)). But we
have neither $B(D(a),0,D(c))$ nor $B(D(e),0,D(g)).$\ The two equivalences in
$\mathfrak{D}(L)$ are $D(a)\approx_{L}D(c)$ and $D(e)\approx_{L}D(g).$

Since we have $B(b,a,c)$ we must have in any join-tree $R$ such that
$T(S_{8},0)\subseteq R$\ and $B=B_{R}[N_{8}]$ an element $x$ such that in
$B_{R}^{+}(b,a,x,c)$ holds with $b<_{R}a<_{R}x<_{R}0$ and $c<_{R}x<_{R}0$. To
build such a tree, we must add $x$, and similarly $y$ such that $f<_{R}%
e<_{R}y<_{R}0$ and $g<_{R}y<_{R}0$. They are the nodes 1 and 2 in Figure
10(b), formally defined as the two sets $D(a)\uplus D(c)=\{a,b,c,d\}$ and
$D(e)\uplus D(g)=\{e,f,g,h\}$ that form $\mathcal{C}_{2}$.

In the general construction, for each $\approx_{L}$-equivalence class $E$ of
independent directions, we introduce in $T(S,r)$ an element $x$ such that, for
each direction $D$ in $E$, we have $D<x<L$.\ Such an element is added only for
an equivalence class $E$ containing at least two different equivalent
directions. It is formally defined as the union of the directions in
$E$.\ These added elements correspond bijectively to the sets in
$\mathcal{C}_{2}$.

\begin{Lemma}{3.10} Let $S=(N,B)$ and $r\in N$ be as in Lemma 3.8, from
which we get $\mathcal{C}$\ by Definition 3.9(b).

(1) The family $\mathcal{C}$\ is not overlapping.

(2) It is first-order definable in $S$.
\end{Lemma}

\begin{proof}
  \begin{enumerate}
  \item
    Consider $E$ and $E^{\prime}$ in $\mathcal{C}$\ such that
$w\in E\cap E^{\prime}$.

There are three possible cases to consider.\ 

\emph{Case 1} : $E=N_{\leq}(x),E^{\prime}=N_{\leq}(y)$.\ Then $x\leq y$ or
$y\leq x$ because $w\leq x$ and $w\leq y$, which gives $E\subseteq E^{\prime}$
or $E^{\prime}\subseteq E$.

\emph{Case 2} : $E=N_{\leq}(x),w\leq x,E^{\prime}=D_{\approx}$, $D=Dir_{L}(w)$
where $L\in\mathfrak{L}$.\ Then $x<L$ (in particular if $x=w)$ or $x\in L$,
which gives $E\subseteq D\subseteq E^{\prime}$ or $E^{\prime}\subseteq E.$

\emph{Case 3}$\ :E=D_{\approx}$, $D\in\mathfrak{D}(L)$, and $E^{\prime
}=D_{\approx}^{\prime},D^{\prime}\in\mathfrak{D}(L^{\prime})$.\ Then $L\cup
L^{\prime}\subseteq L_{>}(w)$, hence $L^{\prime}\subset L$\ or $L\subset
L^{\prime}$ or $L=L^{\prime}$ .\ In the first case, we have $Dir_{L}%
(w)\subseteq E\subseteq N_{\leq}(x)$ for any $x\in L-L^{\prime}$.\ We have
$x<L^{\prime}$.\ Then, $N_{\leq}(x)\subseteq Dir_{L^{\prime}}(w)\subseteq
E^{\prime}$.$\ $The second case is similar and the last one gives
$Dir_{L}(w)=Dir_{L^{\prime}}(w)$, hence, $E=E^{\prime}.$

\bigskip

\item The set $\mathcal{C}$\ is relative to a rooted O-tree $T(S,r)$ where $r\in
N$. We will construct an FO formula $\varphi(X,r)$ (not depending on $S$) such
that for every $r$ and $X\subseteq N$,

\begin{quote}
$S=(N,B)\models\varphi(X,r)$ \ if and only if $X\in\mathcal{C}$.
\end{quote}

Since $\mathcal{C}$ is defined from $T(S,r)$, this formula will have the free
variable $r$.\ The partial order $\leq_{r}$ (denoted\ by $\leq$) is FO
definable in $S$ in terms of $r$, and so is incomparability, denoted
by\ $\bot$.

An FO formula $\varphi_{1}(X,r)$ can express that $X=N_{\leq}(x)$ for some
$x\in N$.

Next we define $\varphi_{2}(X,r)$ intended to characterize the sets
$D_{\approx}$. Let $x$ and $y$ be incomparable in $T(S,r)=(N,\leq).$\ Let
$L=L_{>}(x,y)$ and $u,v<L.$ The nodes $u$ and $v$ are in the direction
$Dir_{L}(u)\in\mathfrak{D}(L)$ if and only if :

\begin{quote}
$(N,B)\models\exists w[u<w\wedge v<w\wedge\forall z(z\in L\Longrightarrow
w<z)],\ $
\end{quote}

which can be expressed by an FO formula $\alpha(r,x,y,u,v)$ because $z\in L$
is FO expressible\footnote{This is a key point of the proof.\ In the proof of
Theorem 3.25, we will use an alternative description of sets $L$ in
$\mathfrak{L}$\ in which membership is still FO expressible.} in terms of
$r,x$ and $y$. Similarly, $u$ and $v$ are in a same set $D_{\approx}$ for some
set $D\in\mathfrak{D}(L)$ (then $D_{\approx}=Dir_{L}(u)_{\approx}$) if and
only if :

\begin{quote}
$(N,B)\models\forall z[z\in L\Longrightarrow\lnot B(u,z,v)],\ $
\end{quote}

which can be expressed by an FO formula $\sigma(r,x,y,u,v)$.

If $u<L$, the set $Dir_{L}(u)_{\approx}$ is the union of at least two
directions in $\mathfrak{D}(L)$ if and only if :

\begin{quote}
$(N,B)\models u<L\wedge\exists v[v<L\wedge\sigma(r,x,y,u,v)\wedge\lnot
\alpha(r,x,y,u,v)]$
\end{quote}

which is expressed by an FO\ formula $\delta(r,x,y,u)$ (for convenience, this
formula includes the condition $u<L$).

We let finally $\varphi_{2}(X,r)$ be the FO formula that :

\begin{quote}
$\exists x,y[x\bot y\wedge\exists u(u\in X\wedge\delta(r,x,y,u))\wedge
\forall u(u\in X\Longrightarrow\forall v[v\in
X\Longleftrightarrow\sigma(r,x,y,u,v)])].$
\end{quote}

It expresses that $X=$\ $Dir_{L_{>}(x,y)}(u)_{\approx}$\ for some incomparable
elements $x,y$, and that $X$\ is the union of at least two directions in
$\mathfrak{D}(L_{>}(x,y))$.

Hence, the formula $\varphi_{1}(X,r)\vee\varphi_{2}(X,r)$ expresses that
$X\in\mathcal{C}$.
\qedhere
\end{enumerate}
\end{proof}

We will use $\mathcal{C}$ to build a join-tree witnessing that $S$ is in
\textbf{IBQT}. With the notation of Lemma 3.10, we have the following obvious facts.

\begin{Lemma}{3.11} For all $x,y\in N$, $D\in\mathfrak{D}(L)$, $D^{\prime
}\in\mathfrak{D}(L^{\prime})$ and $L,L^{\prime}\in\mathfrak{L}$ we have:

\begin{enumerate}

\item $N_{\leq}(x)\subset N_{\leq}(y)$ if and only if $x<y$.

\item $N_{\leq}(x)\subset D_{\approx}\ $if and only if $x<L$ and $D_{\approx
}=Dir_{L}(x)_{\approx},$

\item $D_{\approx}\subset N_{\leq}(x)$ if and only if $x\in L$,

\item $D_{\approx}\subset D_{\approx}^{\prime}$ if and only if\ $L^{\prime
}\subset L$; if $\ D_{\approx}\subset D_{\approx}^{\prime}$, we
have$\ D_{\approx}\subseteq N_{\leq}(x)\subseteq D_{\approx}^{\prime}$ for
every $x$ in $L-L^{\prime}.$
\end{enumerate}
\end{Lemma}

In the next three lemmas, $S$ and the related objects are as in Lemma 3.10.
\begin{Lemma}{3.12} The structure $T(\mathcal{C}):=(\mathcal{C},\subseteq)$
is a join-tree.
\end{Lemma}

\begin{proof} First, $T(\mathcal{C}):=(\mathcal{C},\subseteq)$ is an O-tree
because if $E\subseteq E^{\prime}$ and $E\subseteq E^{\prime\prime}$, we have
$E^{\prime}\subseteq E^{\prime\prime}$ or $E^{\prime\prime}\subseteq
E^{\prime}$\ by Lemma 3.10(1). Next we consider $E$ and $E^{\prime}$,
incomparable in $T(\mathcal{C})$. They are disjoint.\ We will prove that they
have a join $E\sqcup_{T(\mathcal{C})}E^{\prime}$ in $T(\mathcal{C}).$ There
are three cases and several subcases.

\emph{Case 1} : $E=N_{\leq}(x),E^{\prime}=N_{\leq}(y)$ where $x\bot y$ .

\emph{Subcase 1.1} : $(x,m,y)\notin B$ for every $m$ in $L:=L_{>}(x,y)$.\ Then
$Dir_{L}(x)\approx_{L}Dir_{L}(y)$ and $E^{\prime\prime}:=Dir_{L}(x)_{\approx
}\supseteq E\uplus E^{\prime}$. We have $Dir_{L}(x)_{\approx}\in\mathcal{C}$
because $Dir_{L}(x)\neq Dir_{L}(y)$.

We prove that $E^{\prime\prime}=E\sqcup_{T(\mathcal{C})}E^{\prime}.$ If this
is not the case, we could have $E^{\prime\prime}\supset N_{\leq}(z)\supseteq
E\uplus E^{\prime}$. But then $x,y<z$ by Lemma 3.11(1), hence $z\in L$ and
$N_{\leq}(z)\supseteq N_{<}(L).$ So we cannot have $N_{\leq}(z)\subset
E^{\prime\prime}\subseteq N_{<}(L).$

Otherwise, we have $E^{\prime\prime}\supset D_{\approx}^{\prime}\supseteq
E\uplus E^{\prime}$.\ By Lemma 3.11(2), we have $D_{\approx}^{\prime
}=Dir_{L^{\prime}}(x)_{\approx}=Dir_{L^{\prime}}(y)_{\approx}$ and $L\subset
L^{\prime}$. Let $z\in L^{\prime}-L$. Then $x,y<z$, hence $z\in L$,
contradicting the choice of $z$. Hence, $Dir_{L}(x)_{\approx}=E\sqcup
_{T(\mathcal{C})}E^{\prime}.$

Note that $E^{\prime\prime}$ is not of the form $N_{\leq}(z)$ for any $z$
because it is the disjoint union of at least two directions in $\mathcal{D}%
(L)$. If $E^{\prime\prime}=N_{\leq}(z)$, then $z$ would belong to one
direction, say $F$, and all these directions, in particular $Dir_{L}(x)$\ and
$Dir_{L}(y),$ would be included in $F,$ hence equal to $F$ because directions
in $\mathcal{D}(L)$\ do not overlap.

\emph{Subcase 1.2} : $(x,m,y)\in B$ where $m=x\sqcup_{T}y=\min(L)$. Let
$E^{\prime\prime}:=N_{\leq}(m)\supset E\uplus E^{\prime}$.

We claim that $E^{\prime\prime}=E\sqcup_{T(\mathcal{C})}E^{\prime}.$ If this
is not the case, we could have $E^{\prime\prime}=N_{\leq}(m)\supset N_{\leq
}(z)\supseteq E\uplus E^{\prime}$. But then $\{x,y\}<z<m$, hence $m$ is not
the join of $x$ and $y$. Otherwise, $E^{\prime\prime}=N_{\leq}(m)\supset
D_{\approx}^{\prime}\supseteq E\uplus E^{\prime}$ where $D_{\approx}^{\prime
}=Dir_{L^{\prime}}(x)_{\approx}=Dir_{L^{\prime}}(y)_{\approx}$ and $L\subset
L^{\prime}$. Let $z\in L^{\prime}-L$. Then $\{x,y\}<z<m$, hence $m$ is not the
join of $x$ and $y$. Hence, $N_{\leq}(m)=E\sqcup_{T(\mathcal{C})}E^{\prime}.$

\emph{Case 2} : $E=N_{\leq}(x),E^{\prime}=Dir_{L}(y)_{\approx}$.\ Since
$N_{\leq}(x)\cap Dir_{L}(y)_{\approx}=\emptyset$, we do not have
$Dir_{L}(y)\approx_{L}Dir_{L}(y)$, hence we have $(x,m,y)\in B$ for some $m$
that must be $x\sqcup_{T(S,r)}y=\min(L).$ We have $N_{\leq}(m)=E\sqcup
_{T(\mathcal{C})}E^{\prime}$ as in Subcase 1.2.

\emph{Case 3}$\ :E=D_{\approx}$, $D\in\mathfrak{D}(L)$, and $E^{\prime
}=D_{\approx}^{\prime},D^{\prime}\in\mathfrak{D}(L^{\prime})$.\ If
$L=L^{\prime}$ then, as $D_{\approx}\neq D_{\approx}^{\prime}$, we have
$B(D,m,D^{\prime})$ where $m=\min(L)$, and then $E\sqcup_{T(\mathcal{C}%
)}E^{\prime}=N_{\leq}(m)$, as in Case 2.

Otherwise, $L$ and $L^{\prime}$ are incomparable by Lemma 3.11(4) since $E$
and $E^{\prime}$ are so, and $r\in L\cap L^{\prime}$.\ Hence, there are $w\in
L-L^{\prime}$ \ and $w^{\prime}\in L^{\prime}-L$.\ We have $L\cap L^{\prime
}=L_{>}(w,w^{\prime})$. If $(w,m,w^{\prime})\in B$ for some $m\in L\cap
L^{\prime}$ then $m=\min(L_{>}(w,w^{\prime}))$ and $N_{\leq}(m)=E\sqcup
_{T(\mathcal{C})}E^{\prime}$ as in Case 2.

If $(w,m,w^{\prime})\in B$ for no $m\in L\cap L^{\prime}$ then, $F\approx
_{L\cap L^{\prime}}F^{\prime}$ where $F=Dir_{L\cap L^{\prime}}(w)$ and
$F^{\prime}=Dir_{L\cap L^{\prime}}(w^{\prime})$.\ We claim that $F_{\approx}$
is $E\sqcup_{T(\mathcal{C})}E^{\prime}$ as in Subcase 1.1.
\end{proof}

\bigskip

The next two lemmas prove that the join-tree $T(\mathcal{C})$\ witnesses that
$S$ is in \textbf{IBQT}. We identify $N$ with the subset $\mathcal{C})_{1}$  of $N_{T(\mathcal{C})}$.

\begin{Lemma}{3.13} $B\subseteq B_{T(\mathcal{C})}[N]$.
\end{Lemma}

\begin{proof} We recall that
$<$
denotes $<_{r}=<_{T(S,r)}$ which is, by Fact (1) of Lemma 3.11, the
restriction of $<_{T(\mathcal{C})}$\ to $N$.\ The joins in $T(S,r)$ and
$T(\mathcal{C})$ are not always the same.\ 

Consider $(x,y,z)\in B$. By Lemma 3.3(2), we have $x<y$ or $z<y$. Assume
$x<y$. If $y<z$ then $x<_{T(\mathcal{C})}y<_{T(\mathcal{C})}z$, hence
$(x,y,z)\in B_{T(\mathcal{C})}[N]$. If $z<y$, then $y=x\sqcup_{T(S,r)}z$, by
Lemma 3.2(3).\ We are in Subcase 1.2\ of Lemma 3.12, hence, $y=x\sqcup
_{T(\mathcal{C})}z$ and $(x,y,z)\in B_{T(\mathcal{C})}$. The last case is
$y\bot z.$ Let $E:=y\sqcup_{T(\mathcal{C})}z$. We have $x<y<_{T(\mathcal{C}%
)}E$, hence $(x,y,E)\in B_{T(\mathcal{C})}$, and also $(y,E,z)\in
B_{T(\mathcal{C})}$, hence $(x,y,z)\in B_{T(\mathcal{C})}$.

The case $z<y$ is similar.
\end{proof}

\begin{Lemma}{3.14} $B_{T(\mathcal{C})}[N]\subseteq B$.
\end{Lemma}

\begin{proof} Let $x,y,z\in N$\ be such that $(x,y,z)\in
B_{T(\mathcal{C})}.$

If $x<_{T(\mathcal{C})}y<_{T(\mathcal{C})}z$, or $z<_{T(\mathcal{C}%
)}y<_{T(\mathcal{C})}x$, then $x<y<z$, or $z<y<x$ since
$<$
is the restriction of\ $<_{T(\mathcal{C})}$\ to $N$.\ Hence, $(x,y,z)\in B$ by
the definition of  $<$
as $<_{T(S,r)}.$

Otherwise $x<y\leq_{T(\mathcal{C})}E>_{T(\mathcal{C})}z$ or $x<_{T(\mathcal{C}%
)}E\geq_{T(\mathcal{C})}y>z$, where $x$ and $z$ are incomparable in
$T(\mathcal{C}),$ hence also in $T(S,r),$ and $E=x\sqcup_{T(\mathcal{C})}%
z$.\ We assume the first.

\emph{Case 1} : $y\bot z$ in $T(S,r)$.\ Then we have $(x,y,z)\in B$ by Lemma
3.3(1) since $x<y$.

\emph{Case 2}\ : If $y$ and $z$ are comparable, the case $y<z$ has been first
considered.\ Otherwise, $y>z$, hence $y\geq_{T(\mathcal{C})}E=x\sqcup
_{T(\mathcal{C})}z$.\ As $y\leq_{T(\mathcal{C})}E,$ we must have $y=E$.\ Hence
we are in Subcase 1.2 of Lemma 3.12, with $y=x\sqcup_{T(S,r)}z$ so that
$(x,y,z)\in B$.
\end{proof}

\bigskip

\begin{proof}[Proof of Theorem 3.1] From $(N,B)$\ satisfying A1-A6 and A8, we
have built a join-tree $T(\mathcal{C})$ whose nodes $\mathcal{C}$ contains
$N$\ (with $x\in N$ identified with $N_{\leq}(x)\in N_{T(\mathcal{C})}$) such
that, by Lemmas 3.13 and 3.14, the restriction of its betweenness relation
$B_{T(\mathcal{C})}$ to $N$\ is $B$.\ Hence, together with Theorem 2.9, a
structure $(N,B)$\ is in \textbf{IBQT}\ if and only if it satisfies A1-A6 and
A8.
\end{proof}

\bigskip

We know from Definition 10\ and Proposition 17\ of \cite{Cou14} \ that a
quasi-tree $(N,B)$ is the betweenness relation of a tree if and only if $B$ is
\emph{discrete}, \emph{i.e.}, that each set $[x,y]_{B}:=\{x,y\}\cup\{z\in
N\mid B(x,z,y)\}$ is finite (cf.\ Definition 2.8(a)).\ 

\bigskip

\begin{Corollary}{3.15} A structure $S=(N,B)$ is an induced betweenness
relation in a tree if and only if it satisfies axioms A1-A6, A8\ and is
discrete. These conditions are monadic second-order expressible.
\end{Corollary}

\begin{proof}  An induced substructure $S=(N,B)$ of a discrete one is
discrete, which gives the "only if" directions by Theorem 2.9. Conversely, if
$S=(N,B)$ satisfies axioms A1-A6, A8\ and is discrete, then for all $x,y\in N$
such that $x\leq_{T(S,r)}y,$ the set $\{z\in N\mid x\leq_{T(S,r)}%
z\leq_{T(S,r)}y\}=[x,y]_{B}$ is finite.\ Hence, $T(S,r)$ is a rooted tree.\ 

For all $x,y\in N_{T(\mathcal{C})}$ such that $x\leq_{T(\mathcal{C})}y,$ the
set $\{z\in N_{T(\mathcal{C})}\mid x\leq_{T(\mathcal{C})}z\leq_{T(\mathcal{C}%
)}y\}$ is finite because, by Lemma 3.11(4), its number of elements belonging
to $\mathcal{C}_{2}$\ is at most one plus its number of elements belonging to
$\mathcal{C}_{1}$, that is finite as observed above. Hence, $T(\mathcal{C})$
is a rooted tree.\ 

Recall from Section 1.4 that the finiteness of a linear order is MSO
expressible.\ On each set $[x,y]_{B}$ such that $x<_{T(S,r)}y,$ the linear
order $\leq_{T(S,r)}$ is FO definable.\ Hence, the finiteness of $[x,y]_{B}$
is MSO\ expressible.
\end{proof}

\begin{ThmEnv}{Examples and Remarks}{3.16}[About the proof of Theorem 3.1.]
\begin{enumerate}[beginpenalty=99]
\item Consider the structure $S_{8}$\ of Figure 10(a).\ The O-tree $T(S_{8},0)$
is $T_{8}$ (in Figure 10(b)) minus the nodes 1 and 2.\ As observed above,
there are four directions relative to $L:=\{0\}=L_{>}(a,c)$ : $D(a)$, the
direction of $a$, and similarly, $D(c),D(e)$ and $D(g)$.\ The two sets of
$\mathcal{C}_{2}$\ are $D(a)_{\approx}=D(a)\uplus D(c)=\{a,b,c,d\}$ and
$D(e)_{\approx}=D(e)\uplus D(g)=\{e,f,g,h\}.$ The nodes 1 and 2 of Figure
10(b) represent the two nodes $D(a)_{\approx}$ and $D(e)_{\approx}$ added to
$T(S_{8},0)$ to form the tree $T_{8}$ such that $S_{8}=B_{T_{8}}%
[\{0,a,...,h\}].$

\item Consider the O-tree of Figure 8 and its betweenness relation to which we
add the fact $B(a,b,c)$ (and of course $B(c,b,a)).$ Let $L:=\mathbb{N}$.\ This
new structure satisfies A1-A6 and A8.\ The two directions relative to $L$\ are
$\{a,b\}$ and $\{c\}$. They are $\approx_{L}$-equivalent.\ Only one node is
added : $\{a,b,c\}=D(a)\uplus D(c)$.

\item Let $T=(N,\leq)$ be a join-tree with root $r$.\ Let $S:=(N,B_{T})$.\ Then,
$T=T(S,r)$.\ Let us apply the construction of Theorem 3.1.$\ \ $%
Each$\ L\in\mathfrak{L}$ has a minimal element because $T$ is a join-tree.\ By
Definition 3.9(a), no two different directions relative to $L$ are
$\approx_{L}$-equivalent. Hence, The family $\mathcal{C}$\ consists only of
the sets $N_{\leq}(x)$ and so, $T(\mathcal{C})=T(S,r)=T$.\ 

\item If $S=(N,B)$ is an induced betweenness in a quasi-tree, then any node $r$
can be taken as root for defining an O-tree $T(S,r)$ and from it, a join-tree
$T(\mathcal{C})$. This fact generalizes the observation that the betweenness
in a tree $T$ does not dependent on any root. Informally, quasi-trees and
induced betweenness in quasi-trees are "undirected notions". This will not be
true for betweenness in O-trees. See the remark about $U$ in the proof of
Proposition 2.15, Part (2).

\item If $S=(N,\emptyset)$, then $T(S,r)$ consists of the root $r$ having sons
$u$ for all $u\in N-\{r\}$.\ These sons are in pairwise independent directions
relative to $\{r\}$. The rooted tree $T(\mathcal{C})$ is $T(S,r)$ augmented
with a unique new node $x$ corresponding to $N-\{r\}=D_{\approx}$ where $D$ is
$\{u\}$ for any $u\in N-\{r\}$. We have $u<_{T(\mathcal{C})}x<_{T(\mathcal{C}%
)}r$ for each $u\in N-\{r\}$.
\end{enumerate}
\end{ThmEnv}

\subsubsection{Betweenness in rooted O-trees}

We let \textbf{BO}$_{\mathbf{root}}$ be the class of betweenness relations of
rooted O-trees. These relations satisfy A1-A6.\ 

\bigskip

\begin{Proposition}{3.17}The class \textbf{BO}$_{\mathbf{root}}$ is
axiomatized by a first-order sentence.
\end{Proposition}

\begin{proof}Consider $S=(N,B)$. If $B$ is the betweenness relation of an
O-tree $(N,\leq)$ with root $r$, then, $\leq$ is nothing but $\leq_{r}$
defined\ before Lemma 3.2\ from $B$ and $r$. Let $\varphi$\ be the FO sentence
that expresses properties A1-A6\ (relative to $B$) together with the following
one:

\begin{quote}
A9\ : there exists $r\in N$ such that the O-tree $T(S,r):=(N,\leq_{r})$ whose
partial order is defined by $x\leq_{r}y:\Longleftrightarrow x=y\vee y=r\vee
B(x,y,r)$ has a betweenness relation $B_{T(S,r)}$ equal to $B$.\ 
\end{quote}

That $S$ satisfies A1-A6\ insures that $(N,\leq_{r})$ is an O-tree with root
$r$. The sentence $\varphi$\ holds if and only if $S$ is in \textbf{BO}%
$_{\mathbf{root}}$. When it holds, the found node $r$ defines via $\leq_{r}%
$ the relevant O-tree.
\end{proof}

\bigskip

The following counter-example shows that we do not obtain an
FO\ axiomatization of the class \textbf{BO}.

\begin{Example}{3.18}[\textbf{BO}$_{\mathbf{root}}$ is properly
included in \textbf{BO}]
Let $T$ be the O-tree with set of nodes $\mathbb{Q}$ and defining partial
order $\preceq$ such that $x\preceq y:\Longleftrightarrow x\leq y\wedge
y\in\mathbb{Q}-\mathbb{Z}$ (see Figure 11). Any two elements of $\mathbb{Z}$
are incomparable and no two incomparable elements have a join. We claim that
$B_{T}$ is not in \textbf{BO}$_{\mathbf{root}}$.\ We have $B_{T}%
=\{(i,j,k),(k,j,i)\mid i,j,k\in\mathbb{Q},j,k\notin\mathbb{Z}$ and $i<j<k\}$.

Assume that $B_{T}\ =B_{U}$ for some O-tree $U$ with root $r\in\mathbb{Q}$. We
will derive a contradiction.

If $r\in\mathbb{Z}$ we take, without loss of generality, $r=0$.\ Let $a=-1/2$
and $b=-3/2$.\ These two nodes are incomparable in $U$ otherwise, we would
have $(0,a,b)$ or $(0,b,a)$ in $B_{U}=B_{T}$\ which is false. Hence
$(a,0,b)\in B_{U}$, but $(a,0,b)\notin B_{T}.$

If $r\in\mathbb{Q}-\mathbb{Z}$ we take, without loss of generality,
$r=1/2$.\ Let $a=1$ and $b=2$.\ These two nodes are incomparable in $U$
otherwise, we would have $(1/2,a,b)$ or $(1/2,b,a)$ in $B_{U}=B_{T}$\ which is
false. Hence $(a,1/2,b)\in B_{U}$, but $(a,1/2,b)\notin B_{T}.$%
\end{Example}

\begin{figure}
[ptb]
\begin{center}
\includegraphics[
height=2.0833in,
width=0.8233in
]%
{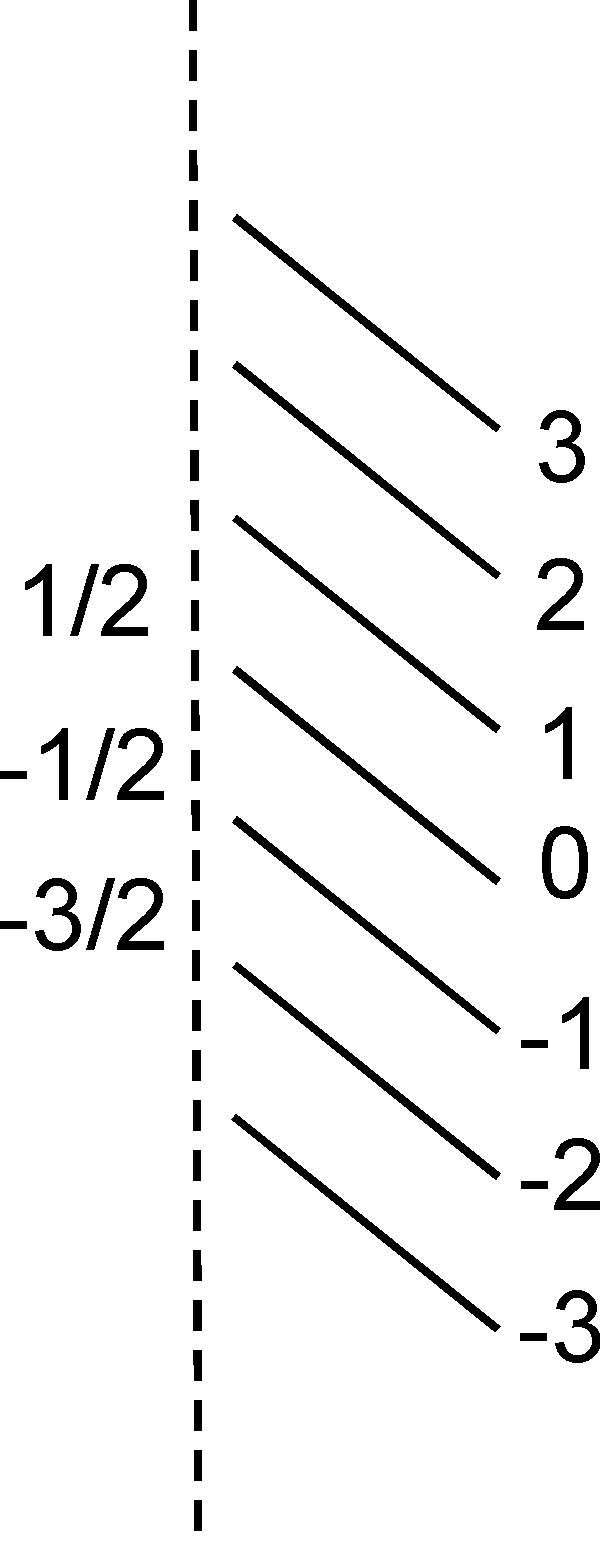}%
\caption{The O-tree of Example 3.18.}%
\end{center}
\end{figure}

\subsection{Monadic second-order axiomatizations}

\subsubsection{Betweenness in O-trees.}

\bigskip

We will prove that the class \textbf{BO} is axiomatized by a monadic
second-order sentence. In the proof of Proposition 3.17, we have defined from
$S=(N,B)$ satisfying A1-A6 and $r\in N$ a candidate partial order $\leq_{r}$
for $(N,\leq_{r})$ to be an O-tree with root $r$ whose betweenness relation
would be $B$. The order $\leq_{r}$ being expressible by a first-order
sentence, we finally obtained a first-order characterization of \textbf{BO}%
$_{\mathbf{root}}$. For \textbf{BO}, a candidate order will be defined from a
line, not from a single node.\ It follows that we will need for our
construction a set quantification.

\bigskip

The next lemma is Proposition 5.3\ of \cite{CouLMCS}.

\begin{Lemma}{3.19} Let $(L,B)$ satisfy properties A1-A7'.\ Let $a,b$ be
distinct elements of $L$.\ There exist a unique linear order $\leq$\ on $L$
such that $a<b$ and $B_{(L,\leq)}=B$.\ This order is quantifier-free
definable, in terms of $a$ and $b,$ in the relational structure $(L,B)$.
\end{Lemma}

\bigskip

We will denote this order by $\leq_{L,B,a,b}$.\ There is a quantifier-free
formula $\lambda,$ written with the ternary relation symbol $B,$\ such that,
for all $a,b,u,v$ in $L$, $(L,B)\models\lambda(a,b,u,v)$ if and only if
$u\leq_{L,B,a,b}v$. We recall from Definition 1.1(b) that a line $L$ in an
O-tree $T$ is a linearly ordered set that is convex, \emph{i.e}., $y\in L$ if
$x,z\in L$ and $x\leq_{T}y\leq_{T}z$.

\begin{Lemma}{3.20} Let $T=(N,\leq_{T})$ be an O-tree and $L$ a
maximal\footnote{Maximality of $L$ is for set inclusion.\ } line in $T$\ that
has no largest node.\ Let $a,b\in L$, such that $a<_{L}b$, where $<_{L}$ is
the restriction of $<_{T}$ to $L$.
\begin{enumerate}
\item The partial order $\leq_{T}$ is first-order definable in a unique way in
the structure ($N,B_{T})$ in terms of $L,\leq_{L},a$ and $b$.

\item It is first-order definable in ($N,B_{T})$ in terms of $L,a$ and
$b$.$\square$
\end{enumerate}
\end{Lemma}

\begin{proof} The line $L$\ is upwards closed and infinite.

Let $x,y\in N$. We first prove the following facts.

\emph{Fact 1 }: If $x,y\in L$, then $x<_{T}y$ if and only if $x<_{L}y.$

\emph{Fact 2 : }If $x\notin L,y\in L,$ then $x<_{T}y$ if and only if
$B_{T}(x,y,z)$ holds for some $z\in L$ such that $z>_{L}y.$

\emph{Fact 3 :} If $x,y\notin L,$ then $x<_{T}y$ if and only if $B_{T}%
^{+}(x,y,z,u)$ holds for some $z,u$ in $L$, such that $u>_{L}z.$

Fact 1 is clear from the definitions.

For Fact 2, we have some $z>_{L}y$ because $L$\ has no largest element. If
$x<_{T}y<_{L}z$, then $B_{T}(x,y,z)$ holds.

Assume now that $B_{T}(x,y,z)$ holds for some $z>_{L}y.$ By the definition of
$B_{T}$, we have $x<_{T}y\leq_{T}x\sqcup_{T}z$ or $z<_{T}y\leq_{T}x\sqcup
_{T}z$. Since $z>_{L}y,$ we cannot have $z<_{T}y.$ Hence, $x<_{T}y.$ \ (We
have actually $B_{T}(x,y,z)$ for every $z>_{L}y).$

For Fact 3, we note that for every $y\notin L$, we have some $z\in L$,
$z>_{T}y$ : take for $z$ any upper-bound of $y$ and some element of $L$, then
\ $z\in L$ because $T$ is an O-tree.\ Hence, we have $z,u\in L$ such that
$y<_{T}z<_{L}u$ because $L$\ has no largest element, hence $B_{T}(y,z,u)$
holds by Fact 2.

If $x<_{T}y$, we have $x<_{T}y<_{T}z$ hence $B_{T}^{+}\ (x,y,z,u)$ hold (by
A4) since we have $B_{T}(x,y,z)$ and $B_{T}(y,z,u).$

Assume now for the converse that $B_{T}^{+}(x,y,z,u)$ holds for $z,u\in L$
such that $z<_{L}u$.\ We have $B_{T}(x,y,z)$ and $z>_{T}y$ by Fact 2 (since we
have $B_{T}(y,z,u)$).\ By the definition of $B_{T}$, we have $x<_{T}y\leq
x\sqcup_{T}z$ $\ $or $\ z<_{T}y\leq x\sqcup_{T}z$. Since $z>_{T}y,$ we cannot
have $z<_{T}y$, hence, $x<_{T}y.$

\bigskip

We now prove the two assertions of the statement.

(1) The above four facts show that $\leq_{T}$ is first-order definable in
$(N,B_{T})$ in terms of $L,\leq_{L},a$\ and $b$. \ More precisely, Facts 1,2
and 3 can be expressed as a first-order formula $\theta$ written with the
relation symbols $L,B$ and $R$ of respective arities 1,3 and 2, such that, if
$L$ is a maximal line in $T$\ that has no largest node, $a,b\in L$ and
$a<_{L}b$, then, for all $u,v\in N$, $(N,L,B_{T},\leq_{L})\models
\theta(a,b,u,v)$ if and only if $u\leq_{T}v.$ For the validity of
$\theta(a,b,u,v)$, $B_{T}$ is the value of $B,$ and $\leq_{L}$ is that of $R$.\ 

(2) However, $\leq_{L}$ is FO definable in $(L,B_{T}[L])$ by Lemma 3.20.\ By
replacing the atomic formulas $R(x,y)$ by $\lambda(a,b,x,y),$ we ensure that
$R$ is $\leq_{L}$, hence, we obtain a first-order formula $\psi(a,b,u,v),$
written with $L$ and $B$\ such that, for $u,v\in N$ $\ $we have $(N,B_{T}%
)\models\psi(a,b,u,v)$ if and only if $u<_{T}v$ where $B_{T}$ is the value of
$B$.
\end{proof}

A \emph{line} in a structure $S=(N,B)$ that satisfies A1-A6 is a set
$L\subseteq N$\ of at least 3 elements in which any 3 different elements are
aligned (cf.\ Section 2.1) and that is convex, \emph{i.e., }is such that
$[x,y]_{B}$ $\subseteq L$ for\ all $x,y$ in $L$.

\begin{Theorem}{3.21} The class \textbf{BO} is axiomatized by a monadic
second-order sentence.
\end{Theorem}

\begin{proof} Let $\varphi(L)$ be the monadic second-order formula
expressing the following properties of a structure $S=(N,B)$ and a set
$L\subseteq N$:

\begin{enumerate}[label=(\roman*)]
\item $S$ satisfies A1-A6,

\item $L$ is a maximal line in $S$,

\item there are $a,b\in L$ such that the formula $\psi(a,b,u,v)$\ of Lemma
3.20\ defines a partial order $\leq$ on $N$\ such that $a<b$,

\item $(N,\leq)$ is an O-tree $U$, in which $L$ is a maximal line without
largest element, and

\item $B_{U}=B.$
\end{enumerate}

We need a set quantification to express the maximality of $L$.\ All other
conditions are first-order expressible.

If $S=(N,B_{T})$ is the betweenness relation of an O-tree $T=(N,\leq)$ without
root, and $L$ is a maximal line in $T$, then $L$ is also a maximal line in
$S$. As $T$ has no root, $L$ has no largest element. Then $\varphi(L)$ holds
where $a,b\in L$ are such that $a<_{L}b$.\ Hence, $S\models\exists
L.\varphi(L)$.

Conversely, if $S=(N,B)$ satisfies $\exists L.\varphi(L)$, then, conditions
(iv) and (v) show that $S$ is in the class \textbf{BO}.

Together with Proposition 3.17, we can express by an MSO\ sentence that
$(S,N)$ is the betweenness relation of an O-tree, with or without root.

A structure $S=(N,B)$ is the betweenness relation of an O-forest if and only
if its connected components (cf.\ Remark 2.14) are the betweenness relations
of O-trees.\ Hence, we get a monadic second-order sentence expressing that a
structure $S$ is the betweenness relation of an O-forest.
\end{proof}

\subsubsection{Induced betweenness in O-trees.}

Next we examine in a similar way the class \textbf{IBO}. It is easy to see
that \textbf{IBO}\ = \textbf{IBO}$_{\mathbf{root}}$.\ 

\begin{Proposition}{3.22} Every structure in the class \textbf{IBO}%
$\ $satisfies Properties A1-A6 but these properties do not characterize this class.
\end{Proposition}

\begin{proof} Every structure $S$ in the class \textbf{IBO} is an induced
substructure of some $S^{\prime}$\ in \textbf{BO}, that thus satisfies
Properties A1-A6.\ Hence, $S$ satisfies also these properties as they are
expressed by universal sentences.%

\begin{figure}
[ptb]
\begin{center}
\includegraphics[
height=1.7435in,
width=2.9516in
]%
{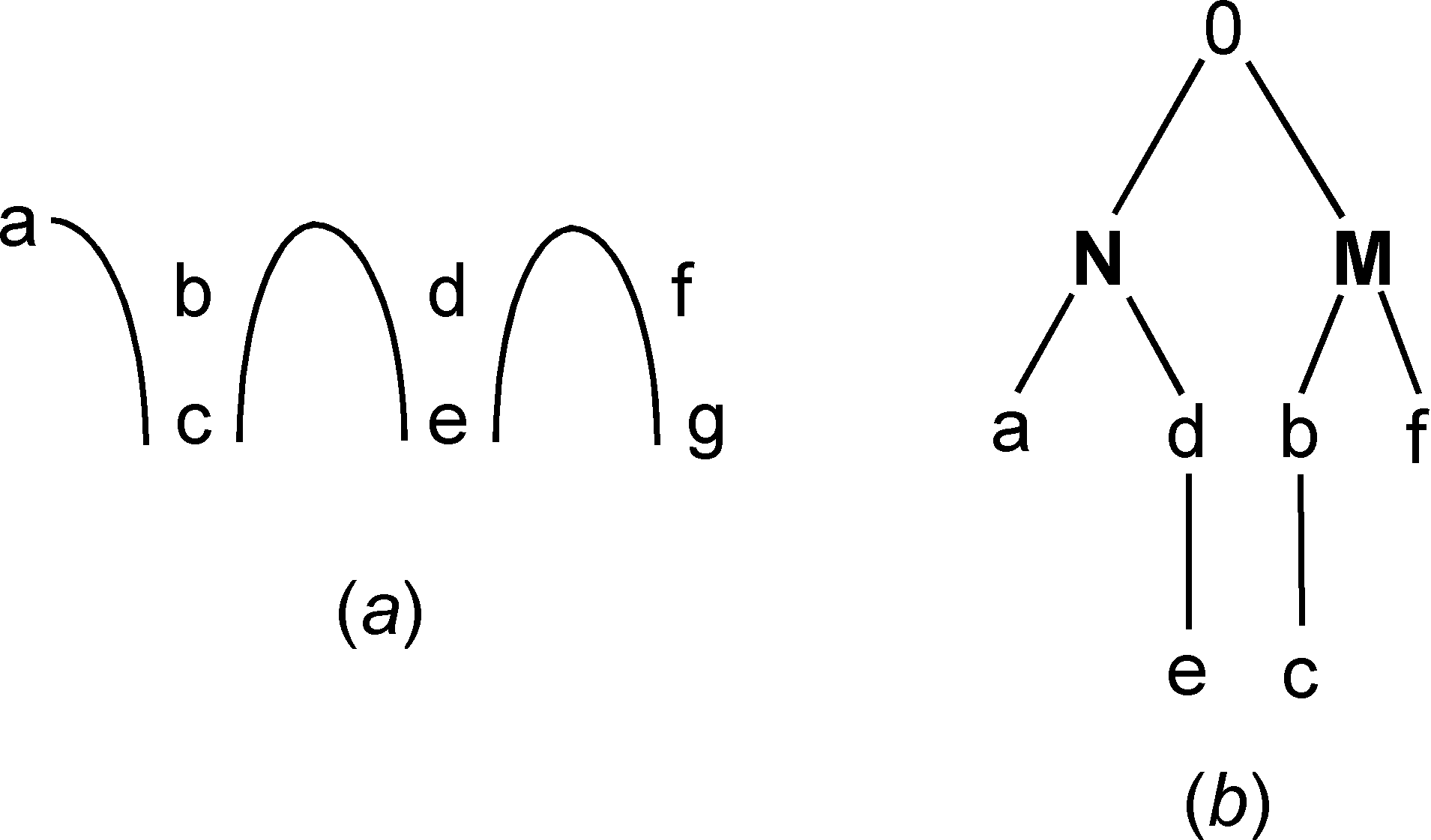}%
\caption{The structure $U$ of Proposition 3.22 (the counter-example) and the
O-tree $T$\ of Remark 3.23.}%
\end{center}
\end{figure}

Now, we give an example of a structure $U=(N,B)$ that satisfies Properties
A1-A6 but is not in \textbf{IBO}$_{\mathbf{root}}$.

We let $N:=\{a,b,c,d,e,f,g\}$ and $B$\ such that\footnote{And also $B(c,b,a)$
to satisfy Axiom A2.} $B(a,b,c)$, $B^{+}(c,b,d,e),$\ $B^{+}(e,d,f,g)$ hold,
and nothing else.\ See Figure 12(a),\ using the conventions of Figures 3 and
5. Assume that $B=B_{T}[N]$ where $T$ is an O-tree $(M,\leq)$ such that
$N\subseteq M$. We will consider several cases leading each to $B\subset
B_{T}[N]$, hence to a contradiction. The relations $<,\leq,\bot$ \ \ and
$\sqcup$\ refer to $T$.

(1) We first assume that $a,c,e,g$ are pairwise incomparable.

The joins $a\sqcup c$, $c\sqcup e$ and $e\sqcup g$ must be defined (because
$(a,b,c),(c,b,e)$ and $(e,f,g)$ are in $B_{T}$) and furthermore $b\leq a\sqcup
c$, $b\leq c\sqcup e$, $d\leq c\sqcup e$, $d\leq e\sqcup g$ and\ $f\leq
e\sqcup g.$ The joins $a\sqcup c$ and $c\sqcup e$ must be comparable (because
$c<\{a\sqcup c,c\sqcup e\})$ and so must be $c\sqcup e$ and $e\sqcup g$.

(1.1) These three joins are pairwise distinct, otherwise $B_{T}[N]$ contains
triples not in $B$, as we now prove.

(1.1.1) Assume $a\sqcup c=c\sqcup e=e\sqcup g=\alpha.$ At least one of
$a\sqcup e,c\sqcup g$ and $a\sqcup g$ is defined and equal to $\alpha.$

If $a\sqcup e=\alpha=a\sqcup c=c\sqcup e,$ then either $c<d\leq\alpha$ or
$e<d\leq\alpha$ \ because $(c,d,e)\in B_{T}$. Hence, we have $(a,d,c)$ or
$(a,d,e)$ in $B_{T}[N]$ but these triples do not belong to $B$. All other
proofs will be of this type.

If $c\sqcup g=\alpha=c\sqcup e=e\sqcup g,$ then $(c,f,e)$ or $(c,f,g)$ is in
$B_{T}[N]-B$ if, respectively, $e<f\leq\alpha$ or $g<f\leq\alpha$ \ (because
$(e,f,g)\in B_{T}$).

If $a\sqcup g=\alpha=c\sqcup e=e\sqcup g,$ then $(a,f,g)$ or $(c,f,e)$ is in
$B_{T}[N]-B$, if, respectively, $g<f\leq\alpha$ \ or $e<f\leq\alpha$
\ (because $(e,f,g)\in B_{T}$).

(1.1.2) We now consider the cases where only two of $a\sqcup c$, $c\sqcup e$
and $e\sqcup g$ are equal.

Assume $a\sqcup c=c\sqcup e=\alpha$. If $\alpha<e\sqcup g,$ then $(a,b,g)$ or
$(c,b,g)$ is in $B_{T}[N]-B$ (because $(a,b,c)\in B_{T}$); if $e\sqcup
g<\alpha,$ then $(c,f,e)$ or $(c,f,g)$ is in $B_{T}[N]-B$ because
$\alpha=c\sqcup e=c\sqcup g.$

If $c\sqcup e=e\sqcup g=\alpha$ and $a\sqcup c<\alpha$, then $e<d\leq\alpha$
or $c<d\leq\alpha$ which gives $(a,d,e)$\ or $(a,d,c)$ in $B_{T}[N]-B$; if
$\alpha<a\sqcup c$, then $(a,f,g)$ or $(a,f,e)$ is in $B_{T}[N]-B$.

If $a\sqcup c=e\sqcup g=\alpha,$ then we have $c\sqcup e<\alpha$ and $a\sqcup
e=\alpha.$ Hence, $(a,d,c)$ or $(a,d,e)$ is in $B_{T}[N]-B$. We cannot have
$\alpha<$ $c\sqcup e$ because then $c,e<\alpha<c\sqcup e.$

(1.2) If $a\sqcup c$ and $e\sqcup g$ are incomparable, then $a\sqcup c<c\sqcup
e$ and $e\sqcup g<c\sqcup e$. We have then $c\sqcup e=c\sqcup g=a\sqcup
g$.\ Hence, we get that $(a,b,g)$ or $(c,b,g)$ is in $B_{T}[N]-B.$

(1.3) Hence, $a\sqcup c$, $c\sqcup e$ and $e\sqcup g$ are pairwise different
but comparable.\ We have six cases to consider : $a\sqcup c<c\sqcup e<e\sqcup
g$ and five other ones, corresponding to the six sequences of three objects.

If $a\sqcup c<c\sqcup e<e\sqcup g$ then, $a<b<a\sqcup c$ or $c<b<a\sqcup c$
and $(a,b,g)$ or $(c,b,g)\in B_{T}[N]-B$.

The verifications are similar in the five other cases.

\bigskip

(2) We consider cases where $a,c,e,g$ are not pairwise incomparable.

\emph{Observation} : If $u<x,(x,y,z)\in B_{T}$ and we do not have $x>z$, then
$B_{T}^{+}(u,x,y,z)$ holds.\ (If $x>z$, then $x$ may not be the join of $u$
and $z$).

If $a>c$, then we have $a>b>c$ and $c\sqcup e>c$.\ Hence $c\sqcup e\geq b$, or
$b>c\sqcup e$.\ We get triples $(e,b,c)$ or $(a,b,e)$ in $B_{T}[N]-B$.

If $a<c$, then we have $a<b<c\leq c\sqcup e$. Hence $(a,c,e)\in B_{T}[N]-B$.

Hence $a\bot c$.\ By the observation, we cannot have $e<c,$ $g<c,$ $e<a$ or
$g<a$.

If $c<e$, then, if $a\sqcup c\leq e$ we have $(e,b,c)$ or $(e,b,a)$ in
$B_{T}[N]-B$; if $e<a\sqcup c$, then $(a,e,c)\in B_{T}[N]-B$.

Hence, $c\bot e$. By the observation, we cannot have $a<c,$ $a<e,$ or $g<e$.

If $e<g$, then, either $c\sqcup e\leq g$ or $g<c\sqcup e$ which gives
$(g,b,c)$, $(g,b,e)$ or $(c,g,e)$ in $B_{T}[N]-B$.

Hence, $e\bot g$. By the observation, we cannot have $a<g$ or $c<g$.

All cases yield $B\subset B_{T}[N]$. \ Hence, $S$ is not in \textbf{IBO}%
.
\end{proof}

\bigskip

\begin{Remarks}{3.23}
  \begin{enumerate}
  \item If we modify $U$ of the previous proof by
replacing $B^{+}(c,b,d,e)$ by $B^{+}(c,d,e)$ (but we keep $b$ in the set of
nodes), we get a modified structure $U^{\prime}$ for which the same result
holds, by a similar proof.

\item If we delete $g$ from $U$, we get a structure $W$ that is in
\textbf{IBO}$_{\mathbf{root}}$. A witnessing O-tree $T$ is shown in Figure
12(b)\ where \textbf{N}\ and \textbf{M} represent two copies of $\mathbb{N}%
$\ ordered top-down as in the O-tree $T_{2}$\ of Figure 8 (cf. the proof of
Proposition 2.15).

\item For every finite structure $H=(N_{H},B_{H})$, let $\varphi_{H}$ be a
first-order sentence expressing that a given structure $(N,B)$ has an induced
substructure isomorphic to $H$. Hence, every structure in \textbf{
IBO}\ satisfies properties A1-A6\ and $\lnot\varphi_{U}\wedge\lnot
\varphi_{U^{\prime}}$.
\end{enumerate}
\end{Remarks}

We do not know whether this first-order sentence axiomatizes the class
\textbf{IBO}, and more generally, whether there exists a finite set of
"excluded" finite induced structures like $U$ and $U^{\prime}$, that would
characterize the class \textbf{IBO. }The existence of such a set would give a
first-order axiomatization of \textbf{IBO. }

The construction of Theorem 3.21 does not extend to \textbf{IBO }because, as
we noted in the proof of Proposition 2.15 (point (3)), a \emph{finite}
structure in \textbf{IBO}\ may not be an induced betweenness relation of any
\emph{finite }O-tree. No construction like that of $T(\mathcal{C})$ in the
proof of Theorem 3.1\ can produce an infinite structure from a finite one.
Nevertheless:

\begin{Conjecture}{3.24} The class \textbf{IBO}\ is characterized by a
monadic second-order sentence.
\end{Conjecture}

\subsection{Logically defined transformations of structures}

Each betweenness relation is a structure $S=(N,B)$ defined from a \emph{marked
O-tree}, \emph{i.e.}, a structure $T=(P,\leq,N)$ where $(P,\leq)$ is an O-tree
and $N\subseteq P$ , the set of \emph{marked} nodes, is handled as a unary
relation. The different cases are shown in Table 1.\ In each case a
first-order formula can check whether the structure $(P,\leq,N)$ is of the
appropriate type, and another one can define the relation $B$\ in $(P,\leq
,N)$. Hence, the transformation of $(P,\leq,N)$ into $(N,B)$ is a first-order
transduction (Definition 1.7).%

\bigskip
\begin{tabular}
[c]{|c|c|c|c|}\hline
\emph{Structure } & \emph{Axiomatization} & \emph{Source } & \emph{From}
$(N,B)$ \emph{to a }\\
$(N,B)$ &  & \emph{structure} & \emph{source structure}\\\hline\hline
\multicolumn{1}{|l|}{\textbf{QT}} & \multicolumn{1}{|l|}{FO : A1-A7, Thm 2.9}
& \multicolumn{1}{|l|}{join-tree $(N,\leq,N)$} & \multicolumn{1}{|l|}{FOT}%
\\\hline
\multicolumn{1}{|l|}{\textbf{IBQT}} & \multicolumn{1}{|l|}{FO : A1-A6, A8, Thm
3.1} & \multicolumn{1}{|l|}{join-tree $(P,\leq,N)$} &
\multicolumn{1}{|l|}{MSOT}\\\hline
\multicolumn{1}{|l|}{\textbf{BO}} & \multicolumn{1}{|l|}{MSO : Theorem 3.21} &
\multicolumn{1}{|l|}{O-tree $(N,\leq,N)$} & \multicolumn{1}{|l|}{MSOT}\\\hline
\multicolumn{1}{|l|}{\textbf{IBO}} & \multicolumn{1}{|l|}{MSO\ ? : Conjecture
3.24} & \multicolumn{1}{|l|}{O-tree $(P,\leq,N)$} & \multicolumn{1}{|l|}{not
MSOT}\\\hline
\end{tabular}

\begin{center}
Table 1
\end{center}

The last colomun indicates which type of transduction, FO transduction
(\emph{FOT}) or MSO transduction (\emph{MSOT}) can produce, from a structure
$(N,B),$ a relevant marked O-tree $(P,\leq,N)$. For \textbf{QT}, this follows
from the proof of Theorem 2.9(1) : if $S=(N,B)$ satisfies A1-A7 and $r\in N$,
then, the O-tree $T(S,r)=(N,\leq_{r})$\ is a join-tree and $B=B_{T(S,r)}$. For
\textbf{BO}, the MSO\ sentence that axiomatizes the class constructs a
relevant O-tree (it guesses one and checks that the guess is correct).\ For
\textbf{IBO}, we observed that the source tree may need to be infinite for
defining a finite betweenness structure, which excludes the existence of an
MSO transduction, because these transformations produce structures whose
domain size is linear in that of the input structure. (cf.\ Definition 1.7,
and Chapter 7 of \cite{CouEng}).\ 

It remains to prove that the transformation of $S\in$ \textbf{IBQT}\ into a
witnessing marked O-tree $(P,\leq,N)$ is a monadic second-order transduction.
This is the content of the following statement.

\begin{Theorem}{3.25} A marked join-tree witnessing that a given structure
$S$ is in \textbf{IBQT} can be defined from $S$\ by MSO formulas.
\end{Theorem}

\bigskip

We first describe the proof strategy.\ We want to prove that, for a given
structure $S=(N,B)$ that satisfies Axioms A1-A6 and A8, the tree
$T(\mathcal{C})$ of the proof of Theorem 3.1 can be constructed by
MSO\ formulas (of course independent of $S$).\ 

The first step is the construction of $T(S,r)=(N,\leq_{r})$ : one chooses a
node $r$ from which the partial order $\leq_{r}$\ is FO definable in $S$ by
using $r$ as value of a variable. The nodes of $T(\mathcal{C})$ (constructed
from $T(S,r)$) are the sets in $\mathcal{C}$\ (cf.\ the proof of Theorem 3.1)
and they are of two types :

either $N_{\leq}(z)$, they are in $\mathcal{C}_{1}$,

or $Dir_{L}(u)_{\approx}$ for $u<L$ and $L\in\mathfrak{L}$ such that
$Dir_{L}(u)_{\approx}$ is the union of at least two directions
(cf.\ Definition 3.9); they are in $\mathcal{C}_{2}$.

A set $N_{\leq}(z)$ is represented by its maximal element $z$ in a natural
way, $T(S,r)$ embeds into $T(\mathcal{C})$ (cf.\ Section 1.1), and the
order between such sets in $T(\mathcal{C})$ is as in $T(S,r)$ by Lemma 3.11(1).\ A
set $Dir_{L}(u)_{\approx}$ is a new node added to $T(S,r)$. In order to make
the transformation of $S\longmapsto T(\mathcal{C})$ into a transduction as
in\ Definition 1.7(b), we define $N_{T(\mathcal{C})}$\ in bijection with
($N\times\{1\})\uplus(M\times\{2\})$ where $(x,1)$ encodes $N_{\leq}(x)$ and
each $w\in M\subseteq N$ encodes (bijectively) some set $Dir_{L}(u)_{\approx
}\in\mathcal{C}_{2}$. An MSO\ formula will express that a node $z$ encodes
$U=Dir_{L}(u)_{\approx}$ for some $L$ and $u$.

Lemma 3.10(2)\ has shown that each set $Dir_{L}(u)_{\approx}$ in
$\mathcal{C}_{2}$ can be defined by FO formulas from three nodes $x,y$ and
$u$.\ We need a definition by a single node, in order to obtain a monadic
second-order transduction. The sets $U$\ in $\mathcal{C}_{2}$ are FO definable
but not pairwise disjoint.\ Hence, one cannot select arbitrarily an element of
$U$ to represent it.\ We will use a notion of structuring of O-trees, that
generalizes the one defined in \cite{CouLMCS} for join-trees, and that we will
also use in Section 4. We will also have to prove that the partial order
$\leq_{T(\mathcal{C})}$ is defined by MSO formulas, but this will be
straightforward by Lemma 3.11, by means of the formula expressing that a node
$z$ encodes a set in $\mathcal{C}_{2}$.

\begin{Definition}{3.26}[Strict upper-bounds]
Let $(N,\leq)$ be a partial order and $X\subseteq N.$ A \emph{strict
upper-bound} of $X$\ is an element $y$ such that $y>X$, that is, $y\in
N_{>}(X)$. We denote by $lsub(X)$ the \emph{least strict upper-bound} of
$X$\ if it exists.\ If $X$ has no maximum element but has a least upper-bound
$m$, then $lsub(X)=m.$ If $X$ has a maximum element $m$, its least strict
upper-bound if it does exist \emph{covers} $m,$ that is, $lsub(X)>m$ and there
is no $p$ such that $lsub(X)>p>m$.
\end{Definition}

\begin{Definition}{3.27}[Structurings of O-trees.]
In the following definitions, $T=(N,\leq)$ is an O-tree.
\begin{enumerate}[label=(\alph*)]
\item If $U$ and $W$ are two lines (convex and linearly ordered subsets of $N$),
we say that $W$ \emph{covers} $U$, denoted\footnote{The relation $\prec$\ is
not an order.\ It is not transitive.} by $U\prec W$, if $U<w$ for some $w$ in
$W$ and, for such $w$ and any $x\in N$, if $U<x<w,$ then $x\in W$. (See
Examples 3.28 below). Note that $lsub(U)$ may not exist, but if it does, it is
in $W$.

\item A \emph{structuring} of $T$ is a set $\mathcal{U}$ of nonempty lines that
forms a partition of $N$ and satisfies the following conditions:

\begin{enumerate}[label=(\arabic*)]

\item One distinguished line called the \emph{axis} is upwards closed.\ 

\item There are no two lines $U,U^{\prime}\in\mathcal{U}$ such that $U<U^{\prime
}$.

\item For each $x$ in $N$, $L_{\geq}(x)=I_{k}\uplus I_{k-1}\uplus...\uplus I_{0}$
for nonempty intervals $I_{0},...,I_{k}$ of $(L_{\geq}(x),\leq)$ such that:

\begin{enumerate}[label=(3.\arabic*)]
\item $x=\min(I_{k})$ \ and\ $I_{k}<I_{k-1}<...<I_{0}$,

\item for each $j$, there is a line $U\in\mathcal{U}$ such that $I_{j}\subseteq
U,$ and it is denoted by $U_{j}$; $U_{0}$\ is the axis,

\item each $I_{j}$ is upwards closed in $U_{j}$, that is, if $x\in U_{j}$ and
$x>y\in I_{j}$ then $x\in I_{j}$.
\end{enumerate}
\end{enumerate}

Hence, $U_{j}\neq U_{j^{\prime}}$ if $j\neq j^{\prime}$, and $U_{j}\prec
U_{j-1}$ for $j=1,...,k$.\ The sequence $I_{0},I_{1},...,I_{k}$ is unique for
each $x$, and $k$ is called the \emph{depth }of $x$ and also of $U_{k}$.\ \ We
denote by $U(x)$ the unique line that contains $x\in N.$
We say that $T=(N,\leq,\mathcal{U})$ is a \emph{structured O-tree}.
\end{enumerate}
\end{Definition}

\begin{Examples}{3.28}[On using structurings]
The notion of structuring will be used as follows.\ Consider in an O-tree $T$
a line $L_{>}(x,y)$ defined from incomparable nodes $x$ and $y$.\ For the
construction of MSO\ transductions, it is essential to define it from a single
node. If it has a minimal element $z$, then it is $L_{\geq}(z)$. Otherwise, we
can use a structuring $\mathcal{U}$.\ Assume that the line $U(x)$ is at depth
$k$, the line $U(y)$ is at depth $k+1$, $U(y)\prec U(x)$ and $L_{>}(x,y)=$
$U(x)\cap L_{>}(y)$.\ This latter line is defined in a unique way from $y$
(equivalently, from any $y^{\prime}$ in $U(y)$), and will be denoted by
$L^{+}(y).\ $Every line in $\mathcal{L}$\ is $L^{+}(z)\ $for some $z$ (not on
the axis). We give examples.

\begin{enumerate}
\item The tree $T_{3}$ of Figure 9(a) described in the proof of Proposition
2.15\ has several structurings. Its upper part consists of the line
$\mathbb{Q}_{>}(\sqrt{2}).$ A first structuring consists of the axis
$\mathbb{Q}$ and the two lines $\{a,b\}$ and $\{c,d\}$ at depth 1.\ Then
$\mathbb{Q}_{>}(\sqrt{2})=L_{>}(a,c)=L_{>}(b,c)=L_{>}(c,1)=L^{+}%
(a)=L^{+}(b)=L^{+}(c)$. A second one consists of $\mathbb{Q}_{>}(\sqrt{2}%
)\cup\{a,b\}$ and the two lines 
$\mathbb{Q}_{<}(\sqrt{2}):=$ 
$\mathbb{Q}-\mathbb{Q}_{>}(\sqrt{2})$ and $\{c,d\}$ at depth 1.\ Then 
$\mathbb{Q}_{>}(\sqrt{2})=L^{+}(c)=L^{+}(d)=L^{+}(1)$.

\item The rooted tree of Figure 10(b), has a structuring consisting of the axis
$\{0,1,a,b\},$ of $\{c,d\}$ and $\{2,e,f\}$ at depth 1 and $\{g,h\}$ at depth
2. We have $\{0\}=L_{>}(a,h)=L^{+}(2)=L^{+}(e)$ and $\{0,2\}=L^{+}(h).$

\item Consider again the join-tree $T:=(Seq_{+}(\mathbb{Q}),$ \ $\preceq)$ of
Examples 1.2(4) and 3.6(2). It has a structuring consisting of the axis
$\{(x)\mid x\in\mathbb{Q}\}$ and the lines $\{(x_{1},...,x_{n},z)\mid
z\in\mathbb{Q}\}$ for all $x_{1},...,x_{n}\in\mathbb{Q}$.\ A node
$(x_{1},...,x_{n})$ is at depth $n-1$. Then $L^{+}((x_{1},...,x_{n}))$, where
$n>1$ is $\{(x_{1},...,x_{n-1},z)\mid z\leq x_{n}\}$.

\item Figure 13\ shows a structuring of a join-tree with axis $U_{0}$ and lines
$U_{0}$,..., $U_{6}$ such that $U_{1}\prec U_{0},U_{3}\prec U_{2}\prec U_{0}$
$U_{6}\prec U_{2}$ and $U_{5}\prec U_{4}\prec U_{0}$. We have $L_{\geq
}(i)=I_{2}\uplus I_{1}\uplus I_{0}$ where $I_{2}=U_{3}\cap L_{\geq}%
(i),I_{1}=U_{2}\cap L_{\geq}(g),$ $I_{0}=U_{0}\cap L_{\geq}(e).$

We have $L_{>}(n,m)=L_{>}(g,j)=L^{+}(j).$%
\end{enumerate}
\end{Examples}

\begin{figure}
[ptb]
\begin{center}
\includegraphics[
height=2.6654in,
width=2.2727in
]%
{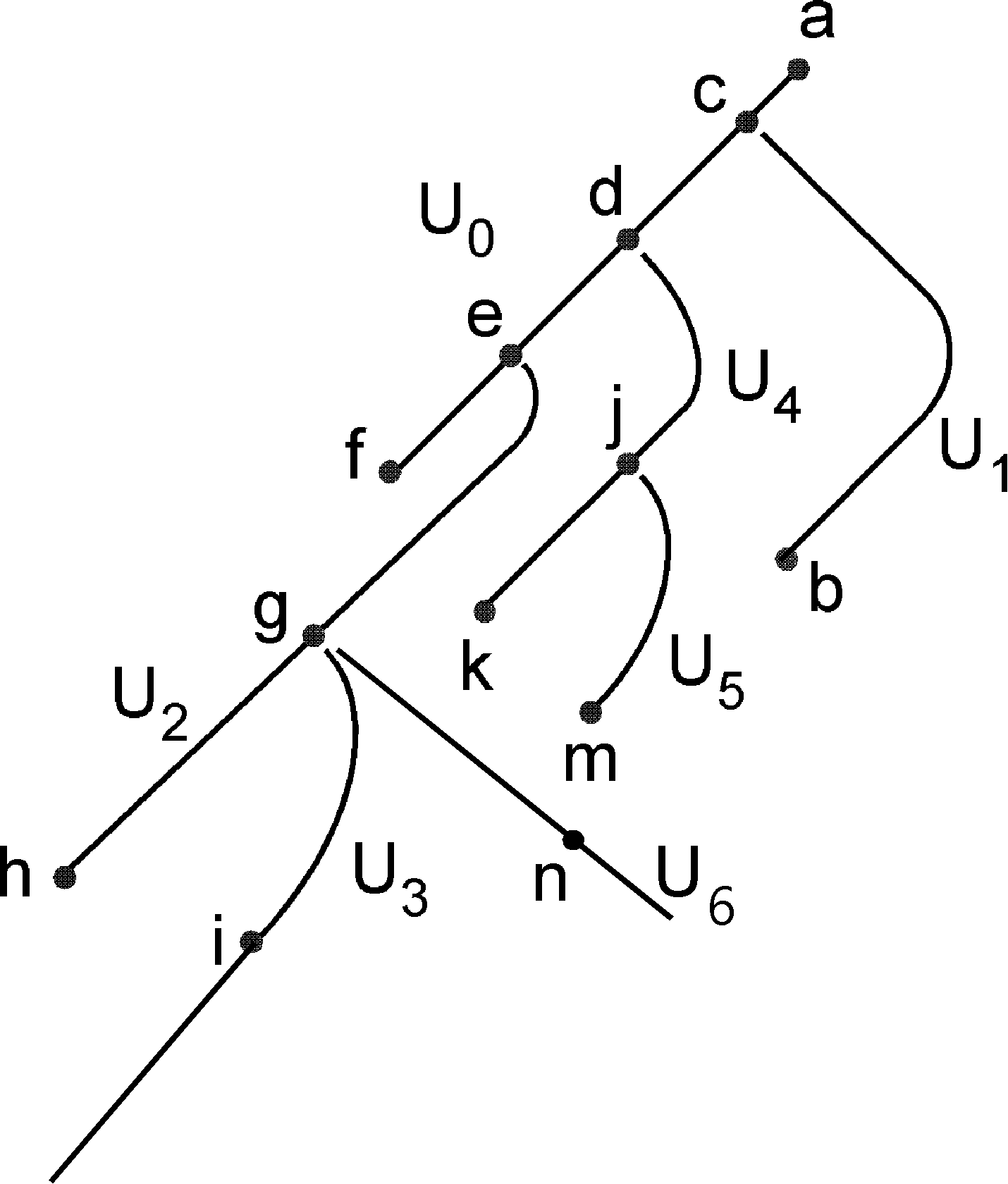}%
\caption{The structuring of Example 3.28(4).}%
\end{center}
\end{figure}

\begin{Proposition}{3.29} Let $\mathcal{U}$ be a structuring of an O-tree
$T=(N,\leq)$.\ Then, $T$ is a join-tree if and only if each $U\in$
$\mathcal{U}$ that is not the axis has a least strict upper-bound, and
$lsub(U)\in W$ \ where $W$ is the line in $\mathcal{U}$ that covers $U$.
\end{Proposition}

\begin{proof} Clear from Definition 3.27.
\end{proof}

\begin{Proposition}{3.30} Every O-tree has a structuring.
\end{Proposition}

\begin{proof} The proof is similar to that of \cite{CouLMCS}\ establishing
that every join-tree has a structuring. We give it for completeness.\ \ Let
$T=(N,\leq)$ be an O-tree.\ We choose an enumeration $x_{0},x_{1}%
,...,x_{n},...$\ of $N$ and a maximal line $B_{0}$; it is thus upwards
closed.\ We define $U_{0}:=B_{0}$.\ For each $i>0$, we choose a maximal line
$B_{i}$ containing the first node not in $B_{i-1}\cup...\cup B_{0}$, and we
define $U_{i}:=B_{i}-(U_{i-1}\uplus...\uplus U_{0})$\ \ \ $=B_{i}-(B_{i-1}%
\cup...\cup B_{0})$. We define $\mathcal{U}$ as the set of lines $U_{i}$. It
is a structuring of $J$. The axis is $U_{0}$. Condition 2) is guaranteed
because we choose a maximal line $B_{i}$ at each step.
\end{proof}

\begin{Lemma}{3.31} If $(N,\leq,\mathcal{U})$ is a structured O-tree, we
define $S(N,\leq,\mathcal{U})$ as the relational structure $(N,\leq
,N_{0},N_{1})$\ such that $N_{0}$ is the set of nodes at even depth and
$N_{1}:=N-N_{0}$.

\begin{enumerate}

\item The class of structures $(N,\leq,N_{0},N_{1})$\ that represent a
structured O-tree is MSO definable.

\item There is a first-order formula $\nu(X,N_{0},N_{1})$\ expressing in every
structure $S(N,\leq,\mathcal{U})$ representing a structured O-tree that a set
$X$\ belongs to $\mathcal{U}$.
\end{enumerate}
\end{Lemma}

\begin{proof} \begin{enumerate} \item
 The proof is, up to minor details, that Proposition
3.7(1) in \cite{CouLMCS}. We let $\sigma(N_{0},N_{1})$\ be the corresponding MSO\ formula.

\item We let $\nu(X,N_{0},N_{1})$\ express that:

\begin{quote}
(i) $X$ is nonempty, linearly ordered and convex,

(ii) $X\subseteq N_{0}$\ or $X\subseteq N_{1}$,

(iii) if $x\in N_{0}\cap X$ , $y\in N$ and, $[x,y]\subseteq N_{0}$ or
$[y,x]\subseteq N_{0}$, then $y\in X,$

(iv) the same holds for $N_{1}$ instead of $N_{0}.$
\end{quote}

Let $X\in\mathcal{U}$. Condition 3) of Definition 3.27\ yields that, if $x<y$,
then $[x,y]\subseteq N_{0}$ or $[x,y]\subseteq N_{1}$ if and only if $x$ and
$y$ belong to the same line in $\mathcal{U}$ (in particular because if
$[x,y]\subseteq N_{0}$ or $[x,y]\subseteq N_{1}$, then $[x,y]\subseteq
I_{k}\subseteq U_{k}).$ Conditions (i)-(iv) hold.

Conversely, assume that $\nu(X,N_{0},N_{1})$ holds. Let $x\in X.$\ We have
$X\subseteq U(x)$: let $y\in X$; if $x<y$, then $[x,y]\subseteq N_{0}\cap
X\ $or $[x,y]\subseteq N_{1}\cap X$.\ Hence, $y\in U(x)$ by the above remark ;
if $y<x$, then, $x\in U(y)$ and so $y\in U(x)$ (because $U(z)$ is the unique
line of the structuring that contains $z$).

If there is $z\in U(x)-X$, then, as $X$\ is an interval, we have $z<X$ or
$X<z$. The intervals $[z,x]$ (or $[x,z])$ is contained in $N_{0}$ or in
$N_{1}$, hence, $z\in X$\ by (iii) and (iv). Contradiction.\ Hence,
$X=U(x).$\ The formula $x\in X\wedge\nu(X)$ expresses that $X=U(x)$%
.
\qedhere
\end{enumerate}
\end{proof}

\bigskip

\subsection*{Some more notation} Let $T=(N,\leq,\mathcal{U})$ be a structured
O-tree with axis $A$.\ Let $x\in N-A$\ and $L_{\geq}(x)=I_{k}\uplus
I_{k-1}\uplus...\uplus I_{0}$ as in Definition 3.27(b).\ We define
$L^{+}(x):=I_{k-1}\uplus...\uplus I_{0}.$ We have $U_{k-1}=W_{k-1}\uplus
I_{k-1}$ for some interval $W_{k-1}$ of $U_{k-1}$ such that $W_{k-1}<I_{k-1}$.
With these hypotheses and notation:

\begin{Lemma}{3.32}
  \begin{enumerate}
 \item The interval $W_{k-1}$ is not empty.

\item For every $y\in\downarrow(W_{k-1})$, we have $L_{>}(x,y)=L^{+}(x).$

\item Every set $L\in\mathfrak{L}$ is of the form $L^{+}(z)$ for some $z$.
\end{enumerate}
\end{Lemma}

\begin{proof}
  \begin{enumerate}
  \item
    If $W_{k-1}$ is empty, then $U_{k}<I_{k-1}=U_{k-1}$,
contradiction with Condition 2) of Definition 3.27(b).

\item Clear from Condition 2) of Definition 3.27(b).

\item Let $L=L_{>}(x,y)$. We have $L_{\geq}(x)=I_{k}\uplus I_{k-1}%
\uplus...\uplus I_{0}$ and $L_{\geq}(y)=J_{\ell}\uplus J_{\ell-1}%
\uplus...\uplus J_{0}$ (cf.\ Condition 3) of Definition 3.27(b)). \ We have
three cases:

\emph{Case 1}: $I_{m-1}\uplus...\uplus I_{0}=J_{m-1}\uplus...\uplus J_{0}$ for
some $m\leq\min(k$,$\ell)$ such that $I_{m}\cap J_{m}=\emptyset.$

Then $L_{>}(x,y)=L^{+}(z)$ for any $z$ in $I_{m}\cup J_{m}$ (or even in
$U_{m}\cup U_{m}^{\prime}$, where $J_{m}\subseteq U_{m}^{\prime}\in
\mathcal{L}$). We have also :

\begin{quote}
$L_{>}(x,y)=$ $L_{>}(x^{\prime},y^{\prime})=$ $L_{>}(x^{\prime},u)=L_{>}%
(y^{\prime},u)$
for every $x^{\prime}\in\downarrow(I_{m})$,$y^{\prime}\in\downarrow(J_{m})$
and $u\in\downarrow(U_{m-1}-I_{m-1}))=\downarrow(W_{m-1})$, (cf.\ (1) and (2)).
\end{quote}

\emph{Case 2} : $I_{m-1}\subset J_{m-1}$ and $I_{p}=J_{p}$ for every $p<m-1$.

Then $L_{>}(x,y)=L^{+}(z)$ for any $z$ in $I_{m}$ (or even in $U_{m}$). We
have also

\begin{quote}
$L_{>}(x,y)=$ $L_{>}(x^{\prime},u)$ for every $x^{\prime}\in\downarrow(I_{m}%
)$, and
$u\in\downarrow(U_{m-1}-I_{m-1})=\downarrow(W_{m-1})$.
\end{quote}

\emph{Case 3}\ : Similar to Case 2 by exchanging $x$ and $y$.
\qedhere
\end{enumerate}
\end{proof}

\begin{ThmEnv}{Example and Remarks}{3.33}{}
  \begin{enumerate}
  \item
    In Case 1, the sets $\downarrow
(I_{m}),$ $\downarrow(J_{m})$ and $\downarrow(U_{m-1}-I_{m-1})$ are three
different directions relative to $L$. In Case 2, $\downarrow(I_{m})$ and
$\downarrow(U_{m-1}-I_{m-1})$ are similarly different directions.

\item In the example of Figure 13, we have :

$L_{>}(i,n)=L_{>}(h,n)=L^{+}(i)=L^{+}(n)=L_{\geq}(g)$ illustrating Cases 1 and 2,

$L_{>}(g,m)=L_{>}(h,m)=L^{+}(j)=L^{+}(k)=L_{\geq}(d)$ and

$L_{>}(k,m)=L^{+}(m)=L_{\geq}(j)$ illustrating Case 2.
\end{enumerate}
\end{ThmEnv}

\begin{Lemma}{3.34} There exist FO formulas $\alpha(N_{0},N_{1},r,x,z)$ and
$\beta(N_{0},N_{1},r,x,$ $z)$\ that express the following properties in a
structure $(N,B,N_{0},N_{1},r)$ that satisfies A1-A6 and A8 and defines a
structuring of the O-tree $T((N,B),r)$; the corresponding set $\mathcal{C}%
_{2}$ is as in Definition 3.9.

\begin{enumerate}
  \item
The formula $\alpha(N_{0},N_{1},r,x,z)$ expresses that $x\in L^{+}(z).$

\item The formula $\beta(N_{0},N_{1},r,X,z)$ expresses that $X=Dir_{L^{+}%
(z)}(z)_{\approx}$ and $X\in\mathcal{C}_{2}$.
\end{enumerate}
\end{Lemma}

\begin{proof}
  \begin{enumerate}
  \item
  The property $x\in L^{+}(z)$ is expressed by the
following FO\ formula $\alpha(N_{0},N_{1},r,x,z)$ \ defined as :

\begin{quote}
$[z\in N_{0}\wedge\exists y(z<y\leq x\wedge y\in N_{1})]\vee
\lbrack z\in N_{1}\wedge\exists y(z<y\leq x\wedge y\in
N_{0})].$
\end{quote}

\item Lemma 3.10(2) shows that the property $X=Dir_{L}(z)_{\approx}\wedge
X\in\mathcal{C}_{2}$ is FO expressible provided $x\in L$ is.\ Assertion (1)
shows precisely that $x\in L^{+}(z)$ is FO expressible.
\qedhere
\end{enumerate}
\end{proof}
\bigskip

\begin{proof}[Proof of Theorem 3.25] By using the previous lemmas, we now prove
the existence of MSO\ formulas that define in a structure $S=(N,B)$ that
satisfies A1-A6 and A8, a marked join-tree $T$\ such that $N_{T}\supseteq N$
and $B=B_{T}[N]$. In the technical terms of \cite{CouEng}\ there is a monadic
second-order transduction that transforms a structure $S=(N,B)$ into such a
marked join-tree ($N_{T},\leq_{T},N$).

The formulas implement the following steps, assuming that $S$ that satisfies
A1-A6 and A8.

\emph{First step}: One chooses $r\in N$, there is no constraint on this
choice. One obtains an O-tree $T(S,r)$.

\emph{Second step}: One guesses a partition $(N_{0},N_{1})$ of $N$\ that
defines a structuring of $T(S,r)$, according to Lemma 3.31. As the order on
$T(S,r)$ depends on $r$, the formula $\sigma(N_{0},N_{1})$\ of Lemma 3.31 can
be transformed into $\sigma^{\prime}(N_{0},N_{1},r),$\ written with $r$ to
define $\leq_{r}$.

\emph{Third step} : All this yields the set $\mathcal{C}=\mathcal{C}_{1}%
\uplus\mathcal{C}_{2}$\ and the associated notions of Definition 3.9 and Lemma
3.32.\ \ We will \emph{encod}e each set in $\mathcal{C}_{2}$ by a unique node
$z$ that defines a unique set $Dir_{L^{+}(z)}(z)_{\approx}\in\mathcal{C}_{2}$.
We may have $Dir_{L^{+}(z)}(z)_{\approx}=Dir_{L^{+}(w)}(w)_{\approx}$ where
$z\neq w$, but we wish to have each set in $\mathcal{C}_{2}$ encoded by a
unique node.\ For insuring this, we choose a set $M$ of nodes such that each
set in $\mathcal{C}_{2}$ is $Dir_{L^{+}(z)}(z)_{\approx}\ $for a unique node
$z\in M$. That a set $M$\ is correctly chosen can be checked by using the
formula $\beta$\ of Lemma 3.34.

We now have the set of nodes of $T(\mathcal{C})$ in bijection with\ 
$(N\times\{1\})\uplus$ \ $(M\times\{2\})$\ where $(x,1)$ encodes $N_{\leq}(x)$ and each
$w\in M$ in a pair $(w,2)$ encodes a unique set in $\mathcal{C}_{2}$. Hence, we
have constructed a structure isomorphic to $T(\mathcal{C})=(N_{T(\mathcal{C})},\leq)$ 
where $\leq$ is the inclusion of the sets encoded by the pairs in
$N_{T(\mathcal{C})}$. This partial order is easy to define by means of the
formula $\beta$.

To sum up, the formulas will use the parameters $r,N_{0}$ and $M$ and check
they are correctly chosen by existential quantifications :

\begin{quote}
  \noindent
$r$ to be the root of the O-tree $T(S,r)=(N,\leq_{r})$,

  \noindent
$N_{0}\subseteq N$\ such that the structure $(N,\leq_{r},N_{0},N-N_{0})$
represents a structured O-tree,

  \noindent
$M$ intended to be in bijection with $\mathcal{C}_{2}$.
\end{quote}

First-order\ formulas can check that these parameters are correctly chosen.
However, the choices of $N_{0}$ and $M$ need set quantifications.

We obtain a join-tree $T^{\prime}$ with set of nodes $N_{T^{\prime}}%
=(N\times\{1\})\uplus(M\times\{2\})$.\ Then $S=(N,B)$ is isomorphic to
$(N\times\{1\},B_{T^{\prime}}[N\times\{1\}])$ where $(x,1)$ corresponds to
$x\in N$. Hence, $S$ is defined by $(N_{T^{\prime}},\leq_{T^{\prime}}%
,N\times\{1\})$ constructed by MSO formulas.
\end{proof}

\begin{Remark}{3.35}[About join-completion]
The join-completion builds an O-tree $T$\ from the sets 
$L_{>}(x,y)$ for incomparable $x$ and $y$,
cf.\ Definition 1.3(b).\ By means of a structuring of $T$, such a set is of
the form $L^{+}(z)$, hence can be encoded by a single node
 $z$.\ The technique
of Theorem 3.25\ is applicable to prove that join-completion is an
MSO\ transduction. The join-completion is built with the set of nodes
$(N_{T}\times\{1\})\uplus(M\times\{2\})$ where $M$ contains a single node $z$
for each set $L_{>}(x,y)$, where $L^{+}(z)=L_{>}(x,y)$.
\end{Remark}

\section{Embeddings in the plane}

In order to give a geometric characterization of join-trees and of induced
betweenness in quasi-trees (equivalently, in join-trees), we show how a
structured join-tree can be embedded in portions of straight lines in the
plane that form a \emph{topological tree}.

\begin{Definition}{4.1}[Trees of lines in the plane]
  \begin{enumerate}[label=(\alph*)]

\item In the Euclidian plane, let $\mathcal{L}=(L_{i})_{i\in\mathbb{N}}$ be a
family of straight half-lines\footnote{One could equivalently use bounded
segments of straight lines because on each such segment, one can designate
countably many points.} (simply called \emph{lines} below) with respective
origins $o(L_{i})$, that satisfies the following conditions :

(i) if $i>0,$ then $o(L_{i})\in L_{j}$ for some $j<i$,

(ii) for all $i,j\in\mathbb{N}$, $i\neq j$, the set $L_{i}\cap L_{j}$ is
$\{o(L_{i})\}$ or $\{o(L_{j})\}$ or is empty.\ (We may have $o(L_{i}%
)=o(L_{j}))$.

We call $\mathcal{L}$ a \emph{tree of lines : }the union of the lines $L_{i}$
is a connected set $\mathcal{L}^{\#}$ in the plane. A \emph{path} from $x$ to
$y\neq x$ in $\mathcal{L}^{\#}$ is a homeomorphism $h$ of the interval $[0,1]$
of real numbers into $\mathcal{L}^{\#}$ such that $h(0)=x$ and $h(1)=y.$ A
\emph{cycle} is a homeomorphism of the circle $S^{1}$ into $\mathcal{L}^{\#}$.

For any two distinct $x,y\in\mathcal{L}^{\#}$, there is a unique path from $x$
to $y$ (it "follows the lines"), and consequently, there is no cycle.\ This
path goes through lines $L_{k}$ such that $k\leq\max\{i,j\}$ where $x\in
L_{i}$ and $y\in L_{j}$, hence, through finitely many of them.\ This path uses
a single interval of each line it goes through, otherwise, there would be a cycle.

\item We define the ternary \emph{betweenness} relation:

\begin{quote}
$B_{\mathcal{L}}(x,y,z):\Longleftrightarrow\neq(x,y,z)$ and $y$ is on the path
between $x$ and $z$.\ 
\end{quote}

\item On each line $L_{i}$, we define a linear order as follows:

\begin{quote}
$x\preceq_{i}y$ if and only if $y=x$ or $y=o(L_{i})$ or $y$ is between $x$ and
$o(L_{i})$.
\end{quote}

On $\mathcal{L}^{\#}$, we define a partial order by:

\begin{quote}
$x\preceq y$ if and only if $x=y$ or

$x\prec_{i_{k}}o(L_{i_{k}})\prec_{i_{k-1}}o(L_{i_{k-1}})\prec_{i_{k-2}%
}...\prec_{i_{1}}o(L_{i_{1}})\prec_{i_{0}}y$

for some $i_{0}<i_{1}<...<i_{k}.$ If $k=0$, then $x\prec_{i_{0}}y$.
\end{quote}

It is clear that $(\mathcal{L}^{\#},\preceq)$ is an uncountable rooted O-tree
: for each $x$ in $\mathcal{L}^{\#}$, the set $\{y\in\mathcal{L}^{\#}\mid
x\preceq y\}$ is linearly ordered with greatest element $o(L_{0})$.
\end{enumerate}
\end{Definition}

\begin{Definition}{4.2}[Embeddings of join-trees in trees of lines]

Let $T=(N,\leq,\mathcal{U})$ be a structured join-tree (cf.\ Definition
3.27).\ An\emph{ embedding }of $T$\ into a tree of lines $\mathcal{L}$ is an
injective mapping $m:N\rightarrow\mathcal{L}^{\#}$ such that:

\begin{quote}
for each $U\in\mathcal{U}$, $m$ is order preserving : $(U,\leq)\rightarrow
(L_{i},\preceq_{i})$ for some $i\in\mathbb{N}$, and if $U$ is not the axis,
then\footnote{See Definition 3.26\ for $lsub(U)$.} $m(lsub(U))=o(L_{i}).$
\end{quote}
\end{Definition}

\bigskip

\begin{Lemma}{4.3} If $T$ is a structured join-tree embedded by $m$ into a
tree of lines $\mathcal{L}$, then, its betweenness satisfies:

\begin{quote}
$B_{T}(x,y,z)\Longleftrightarrow\lbrack\neq(x,y,z)\wedge B_{\mathcal{L}%
}(m(x),m(y),m(z))].$
\end{quote}
\end{Lemma}

\begin{proof}[Proof Sketch] Let $(x,y,z$)$\in B_{T}$. Assume that $x<y<x\sqcup z$
and let us compare $L_{\geq}(x)=I_{k}\uplus I_{k-1}\uplus...\uplus I_{0}$
\ and $L_{\geq}(z)=J_{\ell}\uplus J_{\ell-1}\uplus...\uplus J_{0}$ (as in the
proof of Lemma 3.32(3)).\ There are three cases.\ In each of them, we have a
path in $T$ between $x$ and $z$, that goes through $y$ and is a concatenation
of intervals of lines of the structuring of $T$.\ By concatenating the
corresponding segments of the lines in $\mathcal{L}$, we get a (topological)
path between $m(x)\ $and $m(z)$ that contains $m(y).$ Hence, we have
$(m(x),m(y),m(z))$\ in $B_{\mathcal{L}}.$ The proof is similar in the other
direction.
\end{proof}

\begin{Theorem}{4.4} If $\mathcal{L}$ is a tree of lines and $N$ is a
countable subset of $\mathcal{L}^{\#}$, then $S:=(N,B_{\mathcal{L}}[N])$ is in
\textbf{IBQT}, \emph{i.e.}\ is an induced betweenness in a
quasi-tree.\ Conversely, every structure in \textbf{IBQT} is isomorphic to
some $S=(N,B_{\mathcal{L}}[N])$ of the above form.
\end{Theorem}

\begin{proof}If $\mathcal{L}$ is a tree of lines and $N\subset
\mathcal{L}^{\#}$\ is countable, then $S:=(N,B_{\mathcal{L}}[N])$\ is in
\textbf{IBQT}. A witnessing join-tree $T$\ is defined as follows.\ Its set of
nodes is $N\cup O$\ where $O$ is the set of origins of all lines in
$\mathcal{L}$.\ Its partial order is the restriction to $N\cup O$\ of the
partial order $\preceq$\ on $\mathcal{L}^{\#}$. Then $(N,B_{\mathcal{L}%
}[N])=(N,B_{T}[N])$ hence belongs to \textbf{IBQT}.\ 

Conversely, let $S=(N,B_{T}[N])$ such that $T$ is a structured join-tree.\ It
is isomorphic to $(N,B_{\mathcal{L}}[N])$ for some tree of lines $\mathcal{L}$
by the following proposition.
\end{proof}

\begin{Proposition}{4.5} Every structured join-tree $T$ embeds into a tree
of lines $\mathcal{L}$.
\end{Proposition}
\bigskip

The proof will use some notions of geometry relative to positions of lines in
the plane.

\begin{Definitions}{4.6}[Angles and line drawings]
An orientation of the plane, say the trigonometric one is fixed.
\begin{enumerate}[label=(\alph*)]

\item Let $L,K$ be two lines with same origin. Their \emph{angle}
$L\bigtriangleup K$ is the real number $\alpha,$ $0\leq\alpha<2\pi,$ such that
$L$ becomes $K$\ by a rotation of angle $\alpha$.

If $o(K)$ is in $L-\{o(L)\}$, we define $L\bigtriangleup K:=L^{\prime
}\bigtriangleup K$ where $L^{\prime}$ is the unbounded half-line included in
$L$ with origin $o(K).$

\item For a line $L$, an angle $\alpha$ such that $0<\alpha<\pi$ and $O\in L$,
we define $S(L,O,\alpha)$ as the union of the lines $K$\ with origin $O$ such
that $0\leq L\bigtriangleup K<\alpha$. We call \emph{sector} such a set.
\end{enumerate}
\end{Definitions}

\begin{Lemma}{4.7} For given $L$ and $\alpha$ as above, one can draw
countably many lines with origin $o(L)$ inside the sector $S(L,o(L),\alpha).$
\end{Lemma}

\begin{proof} We draw $K_{1},K_{2},...,K_{i},...$ \ such that
$L\bigtriangleup K_{1}=\alpha/2$ and $K_{i}\bigtriangleup K_{i+1}%
=\alpha/2^{i+1}$\ for each $i$.
\end{proof}

\begin{Lemma}{4.8} Let $L$, $\alpha$ be as above and $X$ be a countable set
enumerated as $\{x_{1},x_{2},...,x_{i},...\}\subseteq L-\{o(L)\}.$ One can
draw lines $K_{1},K_{2},...,K_{i},...$ in the sector $S(L,o(L),\alpha)$ in
such a way that $o(K_{i})=x_{i}$ for each $i$, no two lines are parallel or
meet except at their origins, and none is included in $L$.
\end{Lemma}

\begin{proof} We must have $0<L\bigtriangleup K_{i}<\alpha$ for each $i$.
For each $i$, we let $\gamma_{i}:=\alpha/2^{i+1}$ and $\beta_{i}%
:=\Sigma\{\gamma_{j}\mid x_{j}\prec x_{i}\}<\alpha$ where $x_{j}\prec x_{i}$
means that $x_{i}$ is between $o(L)$ and $x_{j}$. Then, we draw $K_{1}%
,K_{2},...,K_{i},...$ \ with respective origins $x_{1},x_{2},...,x_{i},...$
such that $L\bigtriangleup K_{i}=\beta_{i}$.
\end{proof}

\bigskip

For each $i$, the sector $S(K_{i},x_{i},\gamma_{i})$ contains nothing else
than $K_{i}$. By Lemma 3.8, one can draw inside $S(K_{i},x_{i},\gamma_{i})$
countably many lines with origin $x_{i}$.

\bigskip%

\begin{figure}
[ptb]
\begin{center}
\includegraphics[
height=1.817in,
width=3.403in
]%
{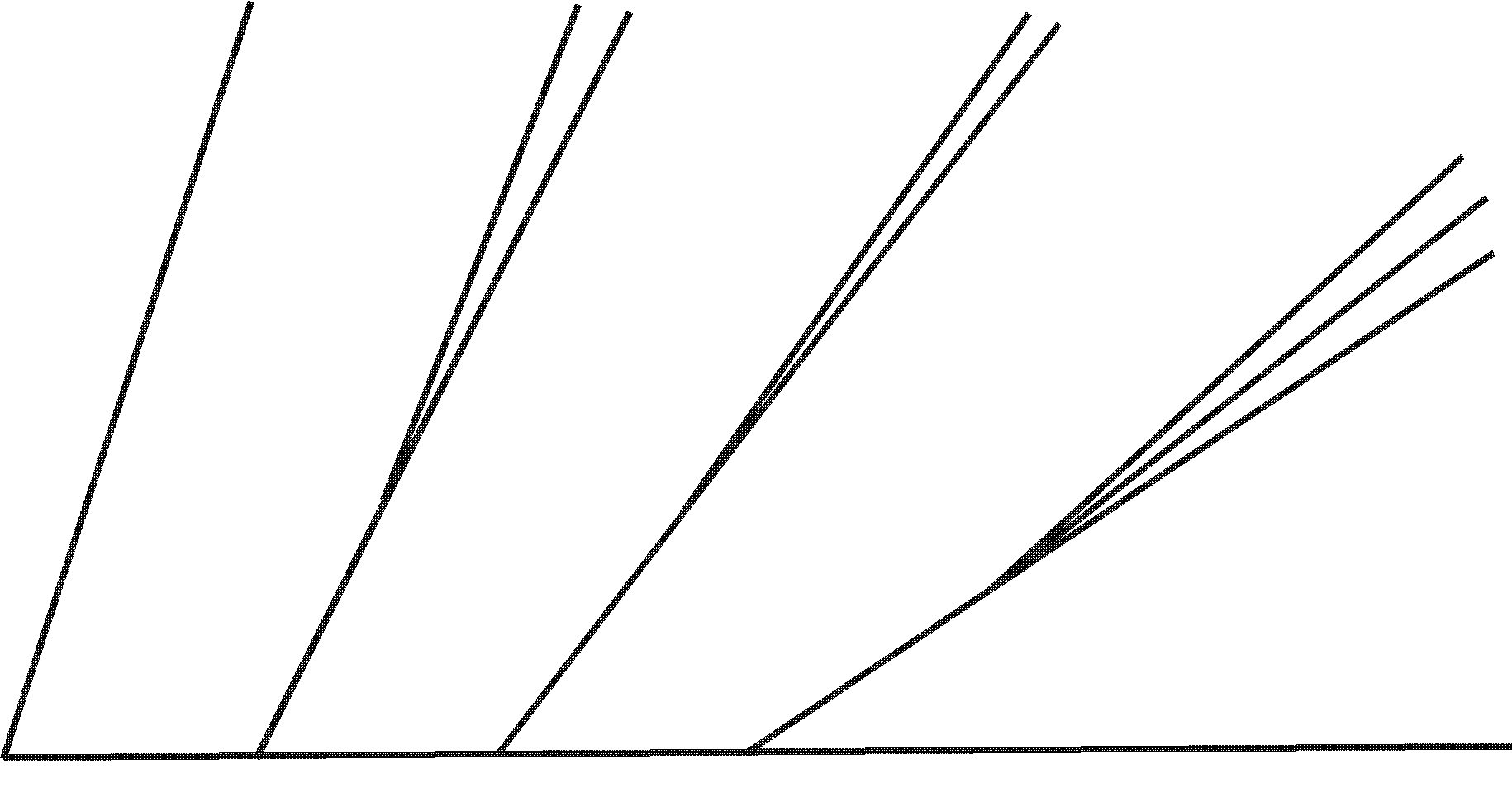}%
\caption{For the proof of Lemma 4.8.}%
\end{center}
\end{figure}

\begin{proof}[Proof of Proposition 4.5] Let $\mathcal{U}$ be a structuring of a
join-tree $T$.\ Let $A$ be the axis.\ Hence, $lsub(A)$ is undefined.\ 

The depth $\partial(U)$ of $U\in\mathcal{U}$ is defined in Definition 3.27 for
O-trees.\ It satisfies the following induction :

\begin{itemize}
\item $\partial(A)=0$,

\item $\partial(U)=\partial(U^{\prime})+1$ if $U^{\prime}$\ has the minimal depth
  such that $lsub(U)\in U^{\prime}$.
\item (Hence, $lsub(U)\neq lsub(U^{\prime})).$
\end{itemize}

We draw lines $L_{0},L_{1},...$\ and define an embedding $m$ such that the
conditions of Definition 4.2 hold. We first draw $L_{0}$ and define $m$\ on
$A$, as required. We choose $\alpha$ such that $0<\alpha<\pi$.\ All further
constructions will be inside the sector $S(L_{0},o(L_{0}),\alpha).$By Lemmas
4.7\ and 4.8, we can draw the lines of depth 1.\ There is space for drawing
the lines of depth 2.\ We continue in this way.
\end{proof}

\section{Conclusion}

We have defined betweenness relations in different types of generalized trees,
and obtained first-order or monadic second-order axiomatizations. In Section
4, we have given a geometric characterization of join-trees and the associated
betweenness relations.

We have proved that the class \textbf{IBQT}\ of induced substructures of the
first-order class \textbf{QT}\ of quasi-trees is first-order
axiomatizable.\ This is not an immediate consequence of the FO\ axiomatization
of \textbf{QT} as shown in the appendix.

We conjecture that betweenness in O-trees is \emph{not }first-order definable
(although the class of O-trees is).\ We also conjecture that the class
\textbf{IBO}\ of induced betweenness relations in O-trees has a monadic
second-order axiomatization.

In \cite{CouLMCS}, we have defined quasi-trees and join-trees of different
kinds from regular infinite terms, and proved they are equivalently the unique
models of monadic second-order sentences.\ Both types of characterizations
yield finitary descriptions and decidability results, in particular for
deciding isomorphism.\ In a future work, we will extend these results to
O-trees and to their betweenness relations.

\section{Appendix : Induced relational structures}

The following example shows that the FO characterization of \textbf{IBQT} does
not follow from the FO characterization of the class \textbf{QT}.

\begin{ThmEnv}{Counter-Example}{6.1}{}[Taking induced substructures does not
preserve first-order axiomatizability]
We prove a little more. We define an FO class $\mathcal{C}$ of relational
structures such that $Ind(\mathcal{C})$, the class of induced substructures of
those in $\mathcal{C}$, is not MSO axiomatizable.

Let $R$ be a binary relation symbol and $A,B,C$ be unary ones.\ We let
$\mathcal{C}$ be the class of structures $S=(V,R,A,B,C)$ that satisfy the
following conditions (i) to (iv)~:

\begin{enumerate}[label=(\roman*)]
\item The sets defined by $A,B,C$ form a partition of $V$,

\item $\forall x,y.(\lnot R(x,x)\wedge\lbrack R(x,y)\Longrightarrow\lnot
R(y,x)]).$
\end{enumerate}
Hence $S$ can be considered as a directed graph whose vertex set is $V$ and
vertices are \emph{colored} by $A,B$ or $C$. Further conditions are as follows:

\begin{enumerate}[resume*]
\item each infinite connected component of $S$ is a "horizontal ladder" that
is infinite in both directions,\ and a portion of which is shown in Figure 15;
the sets of $A$- and $C$-colored vertices\ form two biinfinite horizontal
directed paths.
\end{enumerate}

\begin{figure}
[ptb]
\begin{center}
\includegraphics[
width=3.6752in
]%
{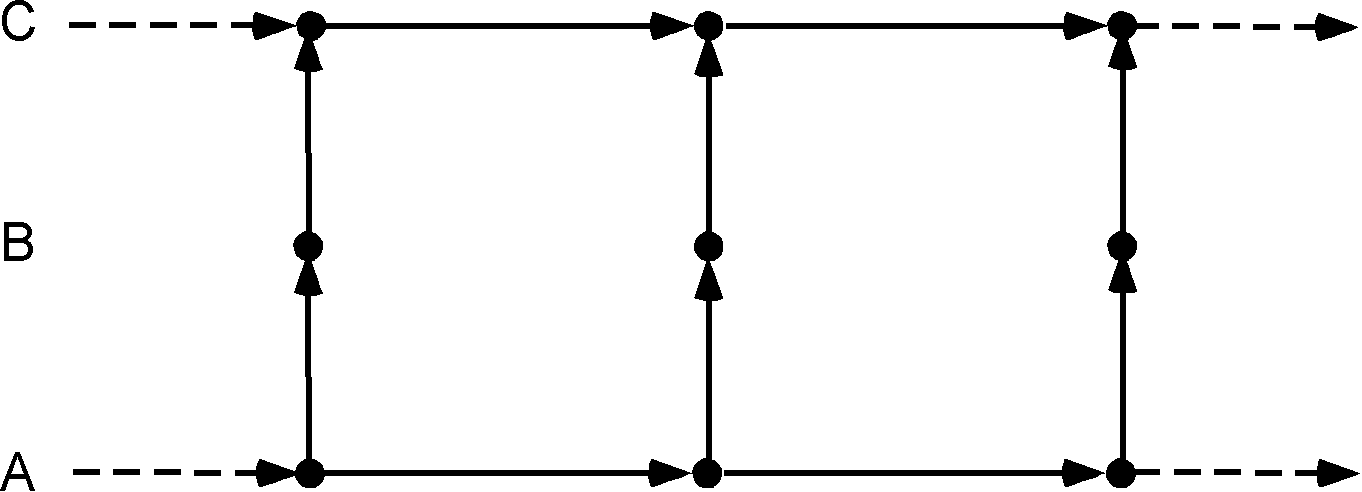}%
\caption{The ladder of Example 6.1.}%
\end{center}
\end{figure}

\begin{enumerate}[resume*]
\item Each finite connected component is a closed "ring", with two directed
cycles of $A$- and $C$-colored vertices~; Figure 15 shows a portion of such a ring.
\end{enumerate}

By a \emph{successor} (or \emph{predecessor}) of $x$, we mean a vertex $y$
such that $(x,y)\in R$ (or $(y,x)\in R$ respectively).
Conditions (iii) and (iv) can be expressed by an FO sentence saying in particular:
\begin{enumerate}[label=(\alph*)]
\item Every vertex $x_{A}$ in $A$ has a unique successor $y_{A}$ in $A$ and a
unique successor $x_{B}$ in $B$ ; this vertex $x_{B}$ has a unique successor
$x_{C}$ in $C$ ; $y_{A}$ has a unique successor in $y_{B}$ in $B$; $y_{B}$ has
a unique successor $y_{C}$ in $C$ that is also the unique successor of $x_{C}$
in $C$.

\item Every vertex in $A$ has a unique predecessor in $A$ and every vertex in
$C$ has a unique predecessor in $C$ \ 

\item There are no other edges than those specified by (a) and (b).
\end{enumerate}
Let us assume that $Ind(\mathcal{C})$ is characterized by an MSO sentence
$\psi$. We will derive a contradiction.

Let $\theta$ be an MSO sentence expressing that a structure $S=(V,R,A,B,C)$
consists of six vertices $x_{A}$, $z_{A}$, $x_{B}$, $z_{B}$, $x_{C}$, $z_{C}$,
of directed edges $x_{A}\rightarrow x_{B}$, $x_{B}\rightarrow x_{C}$,
$z_{A}\rightarrow z_{B}$ and $z_{B}\rightarrow z_{C}$, of a directed path
$p_{A}$ of $A$-colored vertices from $x_{A}$ to $z_{A}$ and of a directed path
$p_{C}$ of $C$-colored vertices from $x_{C}$ to $z_{C}$. These conditions
imply that $V$\ is finite.\ The construction of $\theta$ is routine.\ In
particular, the existence of paths $p_{A}$ and $p_{C}$ can be expressed in MSO
logic with set quantifications. (First-order logic cannot express transitive
closures.\ cf.\ \cite{CouEng}.)\ 

Then, the structures that satisfy $\theta\wedge\psi$ are exactly those that
satisfy $\theta$ and have paths $p_{A}$ and $p_{C}$ of equal lengths. But such
an equality is not MSO expressible (cf.\ \cite{CouEng}).\ Hence, no MSO
sentence $\psi$\ can characterize $Ind(\mathcal{C})$. \hfill $\qedsymbol$
\end{ThmEnv}

\bigskip

This example shows that the first-order axiomatization of the class
\textbf{IBQT}\ (Theorem 3.1) is not an immediate consequence of the
first-order axiomatization of the class \textbf{QT}\ of quasi-trees. To the
opposite, the proof of Proposition 2.12\ has used an argument based on the
structure of logical formulas.

\section*{Acknowledgment}
  \noindent I thank the referees for their useful questions and observations.

\end{document}